\documentclass[]{aa} 
\usepackage{rotating, graphicx}
\usepackage{lscape}
\usepackage{txfonts}
\usepackage{rotating,threeparttable,booktabs,caption,dcolumn} 
\usepackage[normalem]{ulem}
\usepackage{xcolor}
\usepackage[]{hyperref}
\usepackage{float}
\usepackage{amssymb}
\usepackage{multirow}
\usepackage{float}
\usepackage{gensymb}
\usepackage{verbatim} 
\usepackage{sidecap}

\hypersetup{backref=true, pagebackref=true,hyperindex=true,breaklinks=true,colorlinks=true,urlcolor=blue,linkcolor=blue,citecolor=blue,pagecolor=red,bookmarks=true,bookmarksopen=true}

\begin{document} 
   \title{
   Chemical survey of Class I protostars with the IRAM-30m}
   \author{S. Mercimek\inst{1,2} \and C. Codella \inst{1,3}
\and L. Podio\inst{1} \and E. Bianchi\inst{3} 
\and L. Chahine\inst{4,5} \and M. Bouvier\inst{3} \and  A. L\'{o}pez-Sepulcre\inst{3,5} \and R. Neri\inst{5}  \and C. Ceccarelli\inst{3}
}

\institute{
INAF, Osservatorio Astrofisico di Arcetri, Largo E. Fermi 5,
50125 Firenze, Italy
\and
Universit\`a degli Studi di Firenze, Dipartimento di Fisica e Astronomia, Via G. Sansone 1, 50019 Sesto Fiorentino, Italy
\and
Univ. Grenoble Alpes, CNRS, Institut de
Plan\'etologie et d'Astrophysique de Grenoble (IPAG), 38000 Grenoble, France
\and
\ \'Ecole doctorale de Physique, Universit\'e Grenoble Alpes, 110 Rue de la Chimie, 38400 Saint-Martin-d'H\`eres, France 
\and
Institut de Radioastronomie Millimétrique, 38406 Saint-Martin d’Hères, France
}
\offprints{S. Mercimek, \email{seyma.mercimek@inaf.it}}
\date{Received date; accepted date}
\authorrunning{Mercimek et al.}
\titlerunning{Chemical survey of Class I protostars}

\abstract
% CONTEXT
{Class I protostars are a bridge between Class 0 protostars ($\leq$ 10$^{5}$ yr old), and Class II ($\geq$ 10$^6$ yr) protoplanetary disks. Recent studies show gaps and rings in the dust distribution of disks younger than 1 Myr, suggesting that planet formation may start already at the Class I stage. To understand what chemistry planets will inherit, it is crucial to characterize the chemistry of Class I sources and to investigate how chemical complexity evolves from Class 0 protostars to protoplanetary disks.}
% AIMS
{The goal is twofold: (i) to obtain a census of the molecular complexity in a sample of four Class I protostars, and (ii) to compare it with
the chemical compositions of earlier
and later phases of the Sun-like star formation process.}
% methods
{We performed IRAM-30 m observations at 1.3 mm towards 4 Class I objects
(L1489-IRS, B5-IRS1, L1455-IRS1, and L1551-IRS5). The column densities of the detected species are derived assuming Local Thermodynamic  Equilibrium (LTE) or Large Velocity Gradient (LVG).}
%RESULTS
{We detected  27 species:
C-chains, N-bearing species, S-bearing species, Si-bearing species, deuterated molecules, and interstellar Complex Organic Molecules (iCOMs; CH$_{3}$OH, CH$_{3}$CN, CH$_{3}$CHO, and HCOOCH$_{3}$).  
Among the observed sample, L1551-IRS5 is the most chemically rich source.
Different spectral profiles are observed: (i) narrow ($\sim$1 km s$^{-1}$)
lines towards all the sources, (ii) broader ($\sim$4 km s$^{-1}$) lines
towards L1551-IRS5, and (iii) line wings due to outflows (in B5-IRS1,  L1455-IRS1, and L1551-IRS5). 
Narrow c-C$_3$H$_2$ emission originates from the envelope with temperatures of 5 -- 25 K and sizes of $\sim2\arcsec-10\arcsec$. 
The iCOMs in L1551-IRS5 reveal the occurrence of hot corino chemistry,
with CH$_3$OH and CH$_3$CN lines originating from a compact ($\sim$ 0$\farcs$15) and warm ($T>50$ K) region.
Finally, OCS and H$_{2}$S seem to probe the circumbinary disks in the L1455-IRS1 and L1551-IRS5 binary systems. The deuteration in terms of elemental D/H in the molecular envelopes is: $\sim 10-70\%$ (D$_{2}$CO/H$_{2}$CO), $\sim 5-15\%$ (HDCS/H$_{2}$CS), and $\sim 1-23\%$ (CH$_{2}$DOH/CH$_{3}$OH). For the L1551-IRS5 hot corino, we derive D/H $\sim$ 2\% (CH$_{2}$DOH/CH$_{3}$OH).} 
%Conclusions
{Carbon chain chemistry in extended envelopes is revealed towards all the sources. In addition, B5-IRS1, L1455-IRS1 and L1551-IRS5 show a low excitation methanol line which is narrow and centered at systemic velocity, suggesting an origin from an extended structure, plausibly UV illuminated. The abundance ratios of CH$_{3}$CN, CH$_{3}$CHO, and HCOOCH$_{3}$ with respect CH$_{3}$OH measured towards the L1551-IRS5 hot corino are comparable to that estimated at earlier stages (prestellar cores, Class 0 protostars), as well as to that found in comets. 
The deuteration in our sample is also consistent with the values estimated for sources at earlier stages. These findings support the inheritance scenario from prestellar cores to 
the Class I phase when planets start forming.}
\keywords{astrochemistry - stars: formation - ISM: molecules - ISM: individual objects: L1489-IRS, B5-IRS1, L1455-IRS1, L1551-IRS5}
\maketitle

%%%%%%%%%%%%%%%%%
%%%% 
%%%% INTRODUCTION
%%%%
%%%%%%%%%%%%%%%%%
\section{Introduction}
\label{Intro}
After just two decades from the first discovered exoplanet \citep{Wolszczan1992}, the field of exoplanets has reached maturity \citep[e.g.,][]{Doyle2011}.
The two most striking results have been the almost ubiquitous presence of planetary systems around Main Sequence stars and the overwhelming diversity of system architectures. Both of these findings strongly motivate the quests of the origins of such diversity. 
Having said that, what is the next step? The future is enlightened by a breakthrough discovery: planets start to form already during the Class I phase ($\geq$10$^{5}$ yr)  \citep[e.g.,][]{Sheehan2017, Fedele2018}. It is then mandatory to investigate the physical and chemical properties of the first stages of a Sun-like star, and to compare them with what has been found in our Solar System to unveil the chemical origin of planets.

%%%%%%%%%%%%%%%%%%%%%%%%%%%%%%%%%%%%%%%%%%%%%%%%%%%%%%%
%%  SOURCE COORDINATES AND SOME LITERATURE FEATURES   %%
%%%%%%%%%%%%%%%%%%%%%%%%%%%%%%%%%%%%%%%%%%%%%%%%%%%%%%%
\begin{table*}[ht]
\caption{The observed sample of Class I sources}
\centering
     \begin{tabular}{lcccccc}
        \hline
           Name & Region & $\alpha$ $_{\rm J2000}$ $^a$  & $\delta$ $_{\rm J2000}$ $^a$  & $d$ $^b$ & $V_{\rm sys}$  $^c$ & $L_{\rm bol}$ $^d$ \\
            & & ($^h$ $^m$ $^s$) & ($^{\circ}$ $\arcmin$ $\arcsec$) & (pc) & (km s$^{-1}$)& ($L_{\odot}$) \\
           \hline
           L1489-IRS & Taurus & 04:04:43.1 & +26:18:56.4 & 141 & 7.3 & 3.5 \\
            B5-IRS1  & Perseus  & 03:47:41.6  & +32:51:43.5 & 294 & 10.2 & 5.0 \\
            L1455-IRS1 & Perseus & 03:27:39.0 & +30:12:59.3 & 294 & 4.7 & 3.6\\
            L1551-IRS5  & Taurus  & 04:31:34.1  & +18:08:05.1 & 141 & 6.4 & 30 -- 40 \\
             
        \hline
        \end{tabular}
        \tablefoot{
            ($^a$) Coordinates: L1489-IRS from \citet{Jorgensen2009}, B5-IRS1 and L1455-IRS1 from \citet{Hatchell2007} and \citet{Bergner2017}, L1551-IRS5 from \citet{Froebric2005}. 
            ($^b$) Distances from \citet{Zucker2019}.
            ($^c$) Present work. 
            ($^d$) Bolometric lumonosities: L1489-IRS from \citet{Green2013}, B5-IRS1 from \citet{Evans2009}, L1455-IRS1 from \citet{Dunham2013}, L1551-IRS5 from \citet{Liseau2005}. 
            }
\label{Sources}
    \end{table*}
    
%%%%%%%%%%%%%%%%%%%%%%%%%%%%
%% OBSERVATIONAL SETTINGS %%
%%%%%%%%%%%%%%%%%%%%%%%%%%%%
\begin{table}[ht]
\small
 \caption{Observational IRAM-30m settings: the spectral windows are 214.5 -- 222.2 GHz and 230.2 -- 238.0 GHz. Forward efficiency ($F_{\rm eff}$) and beam efficiency ($B_{\rm eff}$) are 0.93 and 0.60, respectively. }
  \begin{tabular}{lccccc}
   
  \hline
 Source& $T_{\rm sys}$ & $t_{\rm on}$ & $\delta\nu$ & $\delta V$  & $HPBW$\\ 
  & (K) & (hr) & (MHz) & (km s$^{-1}$) & $(\arcsec)$\\
  &     &     &      &              & (au) \\
  \hline
L1489-IRS & 450 &26.3 & 0.20 &0.25 -- 0.27 & 10 -- 11\\
& & & & & 1400--1550\\
B5-IRS1  &350& 23.2  & 0.20 & 0.25 -- 0.27 & 10 -- 11\\
& & & & & 2900--3200\\
L1455-IRS1 & 540 &43.4 & 0.20 &0.25 -- 0.27 & 10 -- 11\\
& & & & & 2900--3200\\
L1551-IRS5 &351& 27.6 & 0.20 &0.25 -- 0.27 & 10 -- 11\\
& & & & & 1400--1550\\

   \hline
   \end{tabular}
   \tablefoot{
            $T_{\rm sys}$, $t_{\rm on}$, $\delta\nu$ and $\delta V$, $HPBW$ are system temperature, on-source observing time, spectral resolution at the unit of MHz and km s$^{-1}$, the half power beam width respectively.
            }
\label{observation}

\end{table}  

Class I protostars, with a typical age of 10$^{5}$ yr are a bridge between Class 0 protostars ($\sim$10$^{4}$ yr), where the bulk of the material that forms the protostar is still in the envelope and the Class II protoplanetary disks (10$^{6}$ yr). Class I sources have recently begun to be chemically characterized through spectral surveys at millimeter wavelengths to search for interstellar complex organic molecules, small organics and deuterated species \citep{Oberg2014,Bergner2017,Bianchi2017, Bergner2019, Bianchi2-2019, Bianchi2020, Legal2020}. Therefore, we are still far from concluding whether Class I protostars are also a bridge from a chemical point of view.

Interstellar Complex Organic Molecules (iCOMs\footnote{
Note that we added “i” to the commonly used COMs acronym in order to make clear that these molecules are only complex in the interstellar
context, contrary to what chemists would consider complex in the terrestrial
context.}), C-bearing molecules containing at least six atoms \citep{Herbst2009, Ceccarelli2017}, are the 
building blocks of pre-biotic molecules. Hence, it is very significant to understand how iCOM abundances vary in the evolutionary path from prestellar core to the Solar System small bodies.
Hot corinos around Class 0 protostars are chemically enriched due to the release of molecules, including iCOMs, from the surfaces of dust grains heated by the protostar up to temperatures larger than 100 K \citep{Ceccarelli2007}. They have been well studied so far \citep[e.g.][and references therein]{Bottinelli2007, Codella2016, Jorgensen2016, Belloche2020, Jorgensen2020,Yang2021}. Recently, ALMA observations also started to unveil the chemical content of protoplanetary disks (PPDs) with the detection of a few iCOMs \citep{Oberg2015, Walsh2016, Bergner2018,Favre2018disk, LeeJeong2019, Podio2020,Booth2021}. However, to our knowledge, there are only a few studies on the detection of a hot corino towards Class I protostars \citep{Bergner2019, Bianchi2019, Bianchi2020, Yang2021} and very little has been done in terms of comparison of the chemical complexity at the different stages along the star formation process from prestellar cores to comets \citep[see, e.g.,][]{Bianchi2019, Drozdovskaya2019, Podio2020,Drozdovskaya2021,Booth2021}. Conversely, protostars can be also enriched with carbon-chain molecules and are named as Warm Carbon Chain Chemistry sources \citep[e.g.][]{Sakai2008,Sakai2010,Sakai2013} which are known to be deprived of iCOMs. The origin of the chemical diversity is still unknown.

Deuterium bearing species are also a powerful key to investigate the chemical evolution from prestellar cores to our Solar System \citep{Ceccarelli2014}. The deuteration of molecules, i.e. the enhancement of the D/H abundance ratio in a given molecule with respect to the cosmic elemental deuterium abundance  \citep[D/H $=1.6 \times 10^{-5}$,][]{Linsky2007} occurs in cold and dense environments, such as prestellar cores, and is stored onto dust mantles. Deuterated molecules are then released into the gas phase when either the temperature is high enough to evaporate the grains mantles \citep[e.g.,][]{Ceccarelli2007, Parise2004, Parise2006} or when protostellar shocks sputter the grains \citep{Codella2012}. 
Therefore the D/H ratios are believed to be "fossil" of the Solar-like star forming process \citep[e.g.][]{Taquet2012, Jaber2017}. In this perspective, several prestellar cores and Class 0 protostars are studied by focusing on deuterium fractionations of  H$_{2}$CO, H$_{2}$CS, and CH$_{3}$OH \citep[e.g.,][and references therein]{Bacmann2003, Marcelino2005,Bizzocchi2014, Vastel2018, Parise2006, Bianchi2-2017, Drozdovskaya2018, Manigand2020}. On the other hand, there is only one Class I protostar for which the deuterium fractionation of these species was measured, SVS13-A \citep{Bianchi2017, Bianchi2-2019}.

In this context, we present a chemical survey of four Class I protostars by examining their molecular complexity, molecular deuteration, and their physical characterization. The paper is organized as follows: In Sect. \ref{sample} we present the source sample; in Sect. \ref{observations} we show the observations;  in Sect. \ref{results} we present the analysis of the detected molecular lines; in Sect. \ref{diversity} we discuss the molecular diversity among the targeted protostars, depending on the profiles, abundance, and deuteration of the detected molecules, and we put it in the context of the sources evolutionary stage; in Sect. \ref{summary} we summarize our findings.

%%%%%%%%%%%%%%%%%
%%%% 
%%%% SAMPLE
%%%%
%%%%%%%%%%%%%%%%%
\section{The Sample}
\label{sample}

Four Class I (see Table \ref{Sources}) sources have been selected according to the following criteria: 1) They are located in two different nearby star forming regions; Taurus at $d=141$ pc and Perseus at $d=294$ pc \citep{Zucker2019}; 2) They are classified as Class I sources, having a bolometric temperature $T_{\rm bol}$ $>$ 70 K \citep{Chen1995,Andre2000}; 3) They are associated with emission in methanol lines with upper level energies, $E_{\rm up}$, up to $\sim80$ K detected with IRAM-30m and ALMA observations suggesting the occurrence of a hot corino activity. The exception is L1489-IRS, where only methanol lines with low upper level energies were detected ($E_{\rm up} \leq 12$ K), pointing to emission from the surrounding envelope \citep[e.g.,][]{Oberg2014, Graninger2016, Bianchi2020}. \\

We ordered the sources in the same way that we use in the table and throughout all the paper, i.e. from the chemically poorest to the chemically richest one. The coordinates of the sources as well as distances, systemic velocities, and bolometric luminosities are presented in Table \ref{Sources}.

%%%%%%%%%%%%%%%%%%%%%%%%%%%%%% 
%% WHOLE RANGE OF 4 SPECTRA %%
%%%%%%%%%%%%%%%%%%%%%%%%%%%%%%
\begin{figure*}
  
 \includegraphics[width=19cm]{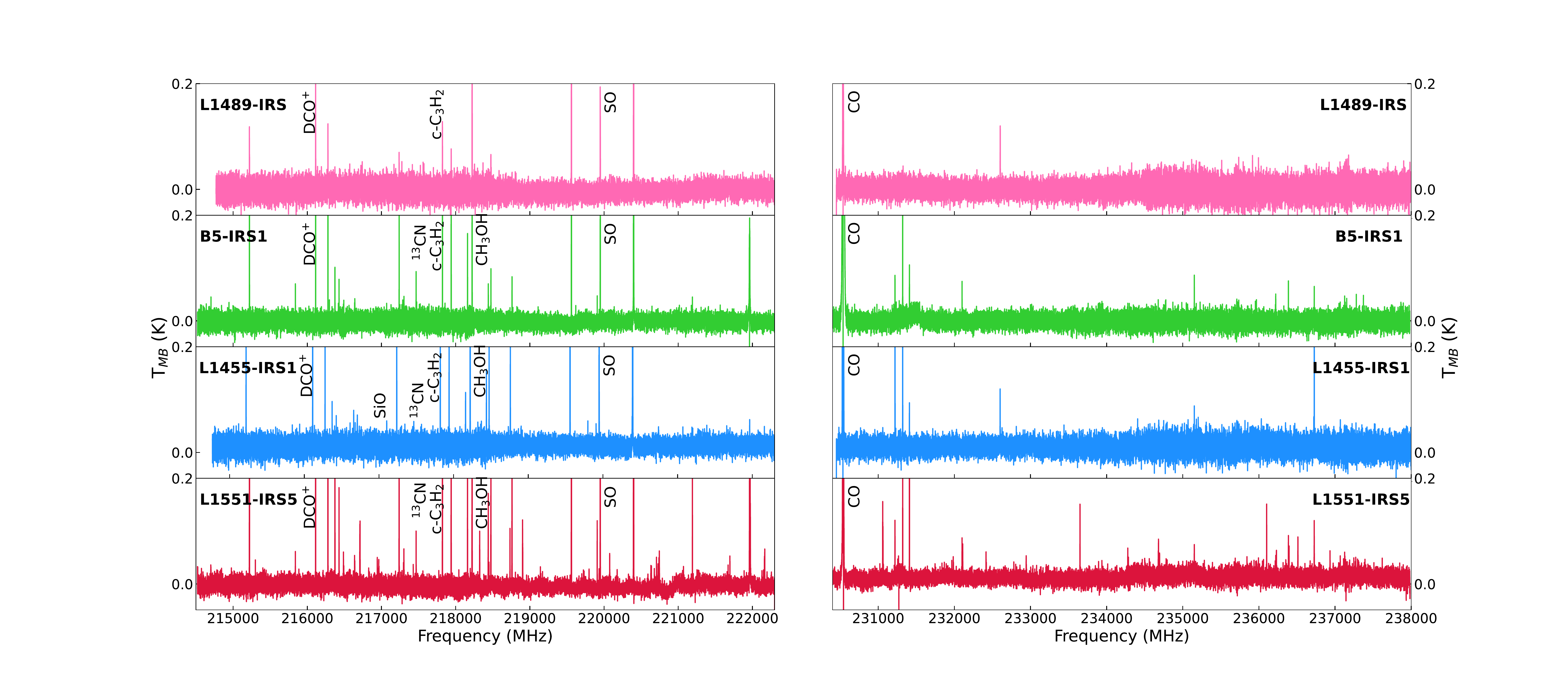}
     \caption{The spectra observed in the 214500 -- 222200 MHz and 230200 -- 238000 MHz range towards 
     L1489--IRS, B5--IRS1, L1455--IRS1, and L1551--IRS5,
     ordered from top to bottom according to the number of detected lines (see Table \ref{species}). Labels of selected detected species
     (see Sect. \ref{ID}, Fig. \ref{L1489_Spectra}--\ref{L1551_Spectra} and Tables \ref{L1489-list}--\ref{L1551-list}) are reported.} 
 \label{fig:4source}
 \end{figure*}
 
\subsection{L1489-IRS}
L1489 IRS (also named as IRAS 04016+2610) is located in the Taurus star forming region. The bolometric luminosity of the source is 3.5 $L_{\rm \odot}$ \citep{Green2013}. Interferometric observations of C$^{18}$O, $^{13}$CO, SO, HCO$^{+}$, and HCN showed infalling flows \citep{Hogerheijde2001, Yen2014}, a faint bipolar outflow  \citep{Myers1988, Hogerheijde1998, Yen2014}, and a large Keplerian disk, with $r$ $\sim$ 700 au \citep{Yen2014}. IRAM-30 m observations revealed emission from several molecules, i.e. sulfur- and nitrogen- bearing species, molecular ions, and deuterated molecules,
\citep[e.g.,][and references therein]{Law2018,Legal2020}. \citet{Oberg2014} detected CH$_{3}$OH lines with $E_{\rm up}\sim$ 6 -- 12 K.

\subsection{B5-IRS1}
B5-IRS1 (Barnard 5 IRS1, also called as IRAS 03445+3242 and Per-emb 53) is located in the B5 region of the Perseus molecular cloud. It has a bolometric luminosity of 5 $L_{\rm \odot}$ \citep{Beichman1984, Evans2009, Pineda2011} and it is embedded in a core within a filamentary structure observed in NH$_{3}$ emission. In addition \citet{Yu1999} reported a bright jet using H$_{2}$ and H$\alpha$ emission at optical, near-infrared wavelengths. A bipolar outflow has been also revealed \citep[e.g.,][and references therein]{Bally1996, Yu1999, Zapata2014}. More specifically, \citet{Zapata2014} resolved the outflow structure by using Submillimeter Array observations of CO emission. Their images showed a spider-like structure, with a high velocity outflow component nested inside a slower wide-angle component. Narrow (0.8 km s$^{-1}$) emission in CH$_{3}$OH, as well as in CH$_3$CN and,
tentatively, in CH$_3$CHO, has been detected using the IRAM-30m telescope \citep{Oberg2014}. In addition, emission from nitrogen-, sulfur-bearing molecules, and carbon chains was detected 
\citep{Oberg2014,Law2018}. \citet{Yang2021} detected CH$_{3}$OH emission using ALMA observations.

\subsection{L1455-IRS1}
L1455-IRS1 (also known as IRAS 03245+3002 and Per-emb 17) is located in the Perseus molecular cloud. This protostar is one of the brightest Class I sources in the L1455 region with a bolometric luminosity of 3.6 $L_{\rm \odot}$  \citep{Dunham2013}. \citet{Goldsmith1984} and  \citet{Chou2016} showed that the source is associated with a Keplerian disk ($r<200$ au) and drives an high velocity outflow mapped in CO and C$^{18}$O  out to distances of $\sim$ 6000 au. This source is also associated with three knots detected in H$_2$ lines at near-infrared wavelengths \citep{Davis1997}, which are located symmetrically with respect to the driving source in agreement with the collimated outflow structure imaged by \citet{Curtis2010}.
IRAM-30m studies showed that the source is associated with emission of both carbon chains and complex organic molecules: \citet{Graninger2016} detected CH$_{3}$OH lines with $E_{\rm up}$ up to $\sim$ 30 K; \citet{Law2018} reported several S-bearing and N-bearing species; \citet{Bergner2017} detected CH$_{3}$CHO at low E$_{up}$ $\sim$ 15 K as well as nitrogen-bearing molecules.
ALMA and VLA observations show that L1455-IRS1 is a binary system \citep{Tobin2018, Yang2021}: \citet{Tobin2018} reported the surrounding circumbinary disks around the two sources and an elongated outflow structure.
Using ALMA, \citet{Yang2021} reported a rich molecular complexity detecting iCOMs such as CH$_{3}$OH, CH$_{2}$DOH, HCOOCH$_{3}$, CH$_{3}$OCH$_{3}$, NH$_2$CHO, CH$_{3}$CN, CH$_{2}$DCN.

\subsection{L1551-IRS5}
L1551-IRS5 is located in the Taurus star forming region and is classified as a Class I protostar \citep{Adams1987,Looney1997} and as a FU Ori-like object \citep{Connelley2018} with bolometric luminosity between 30 -- 40 $L_{\rm \odot}$ \citep{Liseau2005}.
L1551-IRS5 is a binary system as shown by VLA \citep{Bieging1985} and BIMA observations \citep{Looney1997} and consists of a  Northern component of $0.8$ $M_{\rm \odot}$ and a Southern component of 0.3 $M_{\rm \odot}$ \citep{Liseau2005}, enclosed by a circumbinary disk \citep{Cruz2019,Takakuwa2020}. Both protostars drive jets as showed by VLA observations in the continuum image \citep{Rodridguez2003}. In additon, a recent ALMA study showed that the two circumstellar disks are associated with the two binary components, and are detected in CO and several S-bearing species \citep{Takakuwa2020}.
ALMA observations also allowed to constrain the systemic velocities of the Southern and Northern components as +4.5 km s$^{-1}$ and +7.5 km s$^{-1}$, respectively \citep{Bianchi2020}. They revealed a hot corino associated with the Northern protostar, thanks to CH$_{3}$OH, HCOOCH$_{3}$ and CH$_{3}$CH$_{2}$OH emission lines. 

%%%%%%%%%%%%%%%%%%%%%
%%% Hyperfine fits %%
%%%%%%%%%%%%%%%%%%%%%
%\centering
\begin{table}[ht]
  \caption{Results of the spectral fit of the hyperfine components performed using the GILDAS-CLASS tool: sum of the opacities, linewidth, and peak velocity. The errors are reported in parenthesis.
}

\centering
 { \begin{tabular}{lccc}
\hline
Species & $\tau$  & $FWHM$ & $V_{\rm peak}$ \\
 & & (km s$^{-1}$) & (km s$^{-1}$) \\
\hline
\hline
\multicolumn{4}{c}{L1489-IRS}\\
\hline
C$^{15}$N  &$\leq$0.1  &1.1 (0.3) & +7.59 (0.16)\\
CCD & $\leq$0.1  & 1.7 (0.4) & +7.75 (0.21) \\
DCN & 1.0 (0.1)  & 1.5 (0.3)& +7.75 (0.12)\\
\hline
\hline
\multicolumn{4}{c}{B5-IRS1}\\
\hline
$^{13}$CN& $\leq$0.1 &0.8 (0.1) & +10.10 (0.01)\\
$^{13}$CS &2.5 (2.3) &0.6 (0.1) & +9.94 (0.01) \\
CCD & $\leq$ 0.4& 0.8 (0.1)&+10.10 (0.01) \\
DCN&  $\leq$0.1  &0.9 (0.1) & +10.20 (0.01)\\
N$_{2}$D$^{+}$ &$\leq$0.1 &0.6 (0.1) &+10.20 (0.01) \\
\hline
\hline
\multicolumn{4}{c}{L1455-IRS1}\\
\hline
$^{13}$CN  &4.4 (2.9) &0.8 (0.1) &+4.83 (0.01) \\
$^{13}$CS &$\leq$0.1  &1.6 (0.1) &+4.66 (0.01) \\
CCD & $\leq$0.1&0.6 (0.1) &+4.74 (0.01) \\
DCN &$\leq$0.1  &1.4 (0.1) & +4.43 (0.01) \\
N$_{2}$D$^{+}$ & $\leq$ 0.2 & 0.6 (0.1) & +4.79 (0.01) \\
\hline
\hline
\multicolumn{4}{c}{L1551-IRS5}\\
\hline
$^{13}$CN  & 2.8 (1.0)&1.0 (0.1) & +6.33 (0.01) \\
C$^{15}$N & 1.0 (0.5) & 0.6 (0.1) & +5.97 (0.02)\\
$^{13}$CS & $\leq$ 0.1 & 1.4 (0.1) & +6.51 (0.03)\\
CCD & $\leq$ 0.1 & 1.1 (0.1) &+6.38 (0.01)  \\
DCN &  $\leq$ 0.1  &1.3 (0.1) & +6.29 (0.01)\\
N$_{2}$D$^{+}$ & 2.6 (0.6)& 0.5 (0.1) &+6.13 (0.01) \\
H$_{2}$C$^{33}$S & $\leq$ 0.1 & 0.8 (0.2) & +6.78 (0.09)\\
c-C$_{3}$H & $\leq$ 0.8 &0.9 (0.1) &+6.89 (0.01) \\

%10.3 (9.4)
\hline
\end{tabular}}
\label{Hyp}
\end{table}

%%%%%%%%%%%%%%%%%
%%%% 
%%%% OBSERVATION
%%%%
%%%%%%%%%%%%%%%%%
\section{Observations}
\label{observations}
Observations were carried-out at the IRAM-30m telescope located on Pico Veleta, Spain, during several sessions in July 2018 and May 2019. 
All sources were observed in band E2 (1.3 mm) in order to minimize beam dilution. To maximize the number of iCOMs and deuterated lines, the frequency ranges at 214.5 -- 222.2 GHz and 230.2 -- 238.0 GHz were selected with the Eight MIxer Receivers (EMIR) and the FTS200 backend, in wobbler switching mode. 
The spectral resolution is 0.2 MHz, corresponding to $\sim$0.26 km s$^{-1}$.
In order to increase the S/N ratio,
the spectral resolution of the weakest lines 
was degraded up to $\sim$1.0 km s$^{-1}$. 
Precipitable water vapour (pwv) is $\sim$3 -- 5 mm.  We estimate, using the W3(OH), 2251+158, 0430+052, and 0316+413 sources, the calibration uncertainty to be $\sim$20\%, while the error on pointing is $\leq$3$\arcsec$. The Half Power Beam Width ($HPBW$) of the telescope varies from $\sim$11$\arcsec$ (at 214 GHz) to $\sim$10$\arcsec$ (at 238 GHz). This corresponds to $\sim$1500 au for L1489-IRS and L1551-IRS5 and $\sim$3000 au for B5-IRS1 and L1455-IRS1.

GILDAS-CLASS\footnote{\label{note5}\url{http://www.iram.fr/IRAMFR/GILDAS}} package was used to perform the data reduction. Antenna temperature values ($T_{A}$) were converted to main beam temperature values ($T_{MB}$), according to beam efficiency of 0.60 and forward efficiency of 0.93, given in the IRAM-30m website\footnote{\label{note2}\url{http://www.iram.es/IRAMES/mainWiki/Iram30mEfficiencies}}. 
A summary of the observing settings is reported in 
Table \ref{observation}, which reports system temperature, on-source observing time, spectral resolution, and $HPBW$. The average root mean square noise (rms) in the 0.26 km s$^{-1}$ channel is $\sim$10 mK (in $T_{MB}$ scale).

%%%%%%%%%%%%%%%%%
%%%% 
%%%% RESULTS
%%%%
%%%%%%%%%%%%%%%%%
\section{Results} \label{results}

%%%%%%%%%%%%%%%%%
%%%% 
%%%% LINE IDENTIFICATION
%%%%
%%%%%%%%%%%%%%%%%
\subsection{Line Identification} \label{ID}

We analyzed the 1.3 mm spectra obtained for the four Class I sources presented in Sect. \ref{sample} and searched for molecular emission.
For the line identification, we used the JPL\footnote{\label{note3}\url{https://spec.jpl.nasa.gov}} and CDMS\footnote{\label{note4}\url{https://cdms.astro.uni-koeln.de}} spectral catalogs. 
The lines have been identified by fitting the profile with a Gaussian function in GILDAS-CLASS, which allowed us to retrieve the integrated line intensity, $I_{\rm int}$, the line full width at half maximum, $FWHM$, and the intensity and velocity of the line peak, $T_{\rm peak}$ and $V_{\rm peak}$. 

We claim a line detection if the following three criteria are fulfilled: (1) the integrated intensity ($I_{\rm int}$) is above $3 \sigma$, i.e. the line is detected with a signal-to-noise (S/N) higher than 3; (2) the line peaks at $|V_{\rm peak} - V_{\rm sys}| < 0.6$ km s$^{-1}$, where  $V_{\rm sys}$ is the source systemic velocity; (3) the linewidths of different transitions due the same species is the same within uncertainties. 
The results of the Local Thermodynamic
Equilibrium (LTE) and non-LTE analysis (see below) have been used as a-posteriori check to spot false identifications.
Clearly, diatomic and triatomic molecules (such as CO, CS, DCN) are identified even if only one line is detected. The identification of larger molecules ($\ge 4$ atoms) has been done by (i) verifying that in the observed frequency range the detected line(s) of one molecular species are expected to be the brightest one(s) assuming the typical conditions of protostellar regions; (ii) performing a careful comparison with the 1--3 mm unbiased spectral survey ASAI of chemical rich Class 0 and I sources \citep{Lefloch2018}. Future observations over a wider frequency range would be instructive.

The detected lines are shown 
%(in $T_{MB}$ scale) 
in Fig. \ref{L1489_Spectra}, \ref{B5_Spectra}, \ref{L1455_Spectra}, \ref{L1551_Spectra}. Tables \ref{L1489-list}, \ref{B5-list}, \ref{L1455-list}, \ref{L1551-list} report the spectral line parameters as well as the results of the Gaussian fit for all the detected lines, namely: frequency ($\nu$), 
telescope half power beam width ($HPBW$)
, upper level energy ($E_{\rm up}$), 
line strength (S$\mu^{2}$), 
root mean square noise (rms), 
channel width ($\delta V$), 
peak temperature ($T_{\rm peak}$), 
line peak velocity ($V_{\rm peak}$), 
$FWHM$, 
velocity integrated line intensity ($I_{\rm int}$) 
, and spectral catalog used. 

For the transitions associated with multiple hyperfine components, we fit the spectral patterns taking these components into account. In this case, the fit provides,
again, estimates of $V_{\rm peak}$, the $FWHM$ linewidth, as well as the sum of the line opacities. The results, which are obtained with the CLASS tool, are reported in Table \ref{Hyp}. Most of the species are optically thin (i.e. $\tau << 1$), while DCN, $^{13}$CS, $^{13}$CN, N$_{2}$D$^{+}$ and C$^{15}$N are moderately optically thick, 
with $\tau$ $\sim$ 1 -- 4. The $J$ = 2 -- 1 transition of the CO, $^{13}$CO, and C$^{18}$O has been observed towards all the sources. 

In addition,
taking into account all the sources, we detected the following species
(see Tables \ref{L1489-list}--\ref{L1551-list}):

\begin{itemize}

    \item C-chains (c-C$_{3}$H, c-C$_{3}$H$_{2}$, and CH$_{3}$CCH);
    \item iCOMs (CH$_{3}$OH, CH$_{3}$CN, CH$_{3}$CHO, and HCOOCH$_{3}$);
    \item N-bearing species ($^{13}$CN, C$^{15}$N, and HNCO);
    \item S-bearing molecules (SO, $^{34}$SO, SO$_{2}$, $^{13}$CS, OCS, O$^{13}$CS, CCS, H$_{2}$S, H$_{2}$CS, and H$_{2}$C$^{33}$S);
    \item Si-bearing species (SiO);
    \item Deuterated molecules (DCO$^{+}$, N$_{2}$D$^{+}$, CCD, DCN, HDCS, D$_{2}$CO, and CH$_{2}$DOH).
\end{itemize}

\begin{figure*}

  \centering
 \includegraphics[width=18cm]{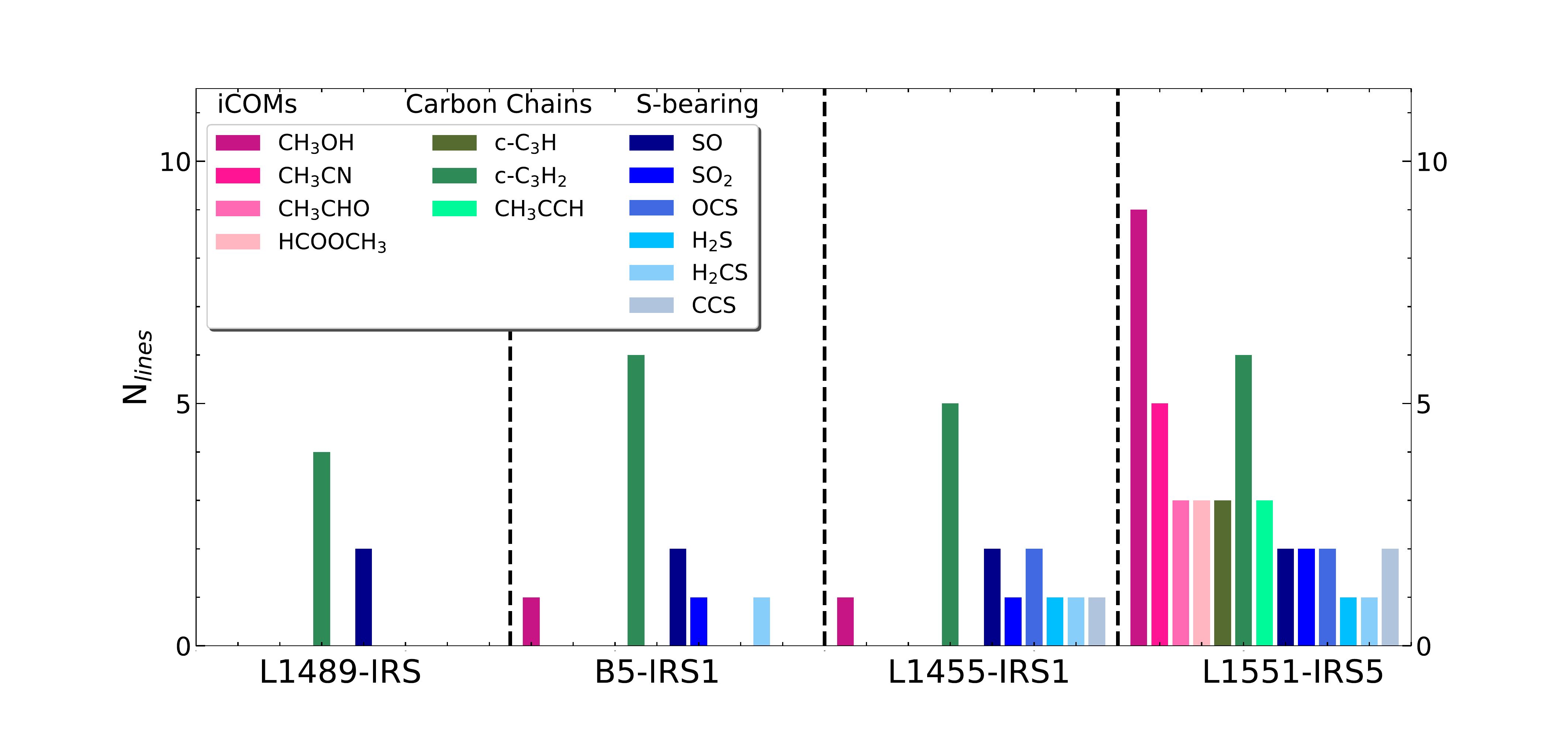}
 \caption{Histogram of the numbers of detected lines for iCOMs (pink), C-chains (green), S-bearing (blue) molecules for the four protostars in our sample (the protostars are ordered by increasing number of detected species).
 }
 \label{histogram}
   \end{figure*}

Figure \ref{fig:4source} shows the 1.3 mm spectra of the four observed Class I protostars. The labels indicate the brightest detected lines towards the four sources, in particular CO and the brightest line for each category of molecules listed above.
More specifically:  c-C$_{3}$H$_{2}$ for C-chains, CH$_{3}$OH for iCOMs, $^{13}$CN for N-bearing species, SO for S-bearing
molecules, SiO for Si-bearing species, and DCO$^{+}$ for deuterated isotopologues. The sources are ordered by increasing number of line detections (L1489-IRS, B5-IRS1, L1455-IRS1, and L1551-IRS5, respectively). Figure \ref{histogram} summarizes the number of detected lines per species and per source showing main isotopologues of iCOMs, C-chains, and S-bearing molecules. As listed in Table \ref{species}, we detected: 17 transitions due to 10 species in L1489-IRS, 29 transitions due to 15 species in B5-IRS1, and 36 transitions due to 21 species in L1455-IRS1. L1551-IRS5 has the richest spectra in these observations in terms of detected number of molecules with 75 transitions due to 27 species. 
 
To summarize, CO, SO, and H$_2$CO emission have been detected in all four of the sources; also the C-chain c-C$_3$H$_2$ as well as several deuterated species (DCO$^+$, DCN, CCD, and D$_2$CO) have been revealed towards all the targets; CH$_3$OH and CH$_2$DOH  have been detected in all the sources apart from L1489-IRS; L1455-IRS1 and L1551-IRS5 show H$_2$S, OCS, and H$_2$CS emission; SiO emission has been observed only towards L1455-IRS1; CH$_3$CN and CH$_3$CHO have been detected towards L1551-IRS5  supporting the occurrence of an hot corino, first detected through   CH$_3$OH and HCOOCH$_3$ emission by \citet{Bianchi2020}.

The observed chemical differentiation will be discussed in Sect. \ref{diversity},
in light of gas properties and molecular column densities inferred through
the line analysis presented in the following subsections.

\subsection{Constraints on line optical depth}
\label{iso}
To 
Given the large number of isotopologues (12 lines due to 5 species) detected in the present survey, we derive constraints on the opacity of the detected emission lines by assuming  interstellar isotopic ratios.
Figures \ref{CO1} and \ref{CO2} show the $^{12}$C/$^{13}$C from $J$ = 2 -- 1 profiles of $^{12}$CO, $^{13}$CO, and C$^{18}$O.
The spectra of the rarer isotopologues have been scaled assuming the isotopic ratios of $^{12}$C/$^{13}$C = 77 and $^{16}$O/$^{18}$O = 560 \citep{Milam2005}.
The line emission at velocities close to the systemic one are affected by absorption. In addition, in the case of L1551-IRS5, there is also foreground emission producing an absorbing dip at blueshifted velocity of $\sim$8 km s$^{-1}$. On the other hand, the line intensities at the highest velocities are consistent with the isotopic ratios, indicating optically thin emission ($\tau$ $\leq$ 0.1). 

Regarding SO 6$_{\rm 5}$ -- 5$_{\rm 4}$, we compare the $^{32}$SO and $^{34}$SO line emission towards  B5-IRS1 and L1551-IRS5. The line ratio is
$\sim$ 12 -- 20, which assuming an isotopic ratio $^{32}$S/$^{34}$S = 22 \citep{Wilson1994}, leads to an opacity of $^{32}$SO  6$_{\rm 5}$ -- 5$_{\rm 4}$ of $\sim 1$.

We also estimate the opacity of the H$_{2}$C$^{32}$S 7$_{\rm 1,7}$ -- 6$_{\rm 1,6}$ emission by comparing with H$_{2}$C$^{33}$S 7$_{\rm 1,7}$ -- 6$_{\rm 1,6}$. Assuming an isotopic ratio $^{32}$S/$^{33}$S = 138 \citep{Wilson1994}, the measured line ratio of $\sim$ 4 implies optically thick H$_{2}$C$^{32}$S emission ($\tau$ $\simeq$ 35).
Similarly the opacity of OCS $J$ = 19 -- 18 emission is inferred from the line ratio between O$^{12}$CS $19-18$ and O$^{13}$CS $19-18$ assuming $^{12}$C/$^{13}$C = 77 \citep{Milam2005}. We found that the opacity
of the O$^{12}$CS 19 -- 18 line is around 20. 

\begin{table}[ht]
    \caption{Detected molecular species in the observed Class I sources. 
    The number of lines are reported in parenthesis.}
    \label{species}
    \centering
     \begin{threeparttable}
    
     \begin{tabular}{lcccc}
        \hline
            Species$^{a}$ & L1489 & B5 &  L1455& L1551 \\
            & IRS & IRS1 & IRS1 & IRS5\\
        \hline
             CO & Y(1) &Y(1) &Y(1) & Y(1)  \\
             $^{13}$CO &Y(1) &Y(1) &Y(1) & Y(1)   \\
             C$^{18}$O &Y(1) &Y(1) &Y(1) & Y(1)   \\
             $^{13}$CN & - & - &Y(2) & Y(6) \\
             C$^{15}$N &Y(1) &Y(1) & - & Y(2) \\
             $^{13}$CS &- &Y(1) &Y(1) & Y(1) \\
             SO &Y(2) &Y(2) &Y(2) & Y(2) \\
             $^{34}$SO & - & Y(1) & - & Y(1) \\
             SiO &- &- & Y(1)& - \\
             CCD &Y(1) &Y(2) &Y(2) &Y(2)  \\ 
             DCN &Y(1) &Y(1) &Y(1) &Y(1)  \\
             N$_{2}$D$^{+}$&- &Y(1) &Y(1) & Y(1) \\
             DCO$^{+}$ &Y(1) &Y(1) &Y(1) & Y(1) \\
             SO$_{2}$ &- &Y(1) &Y(1) & Y(2) \\
             OCS &-&-&Y(2)&Y(2) \\
             O$^{13}$CS &- &- &-&Y(1) \\
             H$_{2}$S & & & Y(1) &Y(1)\\
             CCS &- &- &Y(1) &Y(2)  \\
             H$_{2}$CO &Y(3) &Y(4) &Y(4) & Y(4)  \\
             D$_{2}$CO &Y(1) &Y(2) &Y(2) & Y(4) \\
             H$_{2}$CS &- &Y(1) &Y(1) & Y(1) \\
             H$_{2}$C$^{33}$S &- &- &-&Y(1) \\
             HDCS&-&-&Y(2)& Y(1) \\
             HNCO&-&-&Y(1)&Y(1)\\
             c-C$_{3}$H &- &- &- & Y(3) \\
             c-C$_{3}$H$_{2}$ &Y(4)& Y(6) &Y(5) & Y(6) \\
             H$_{2}$CCO&- &- &-&Y(2)\\
             CH$_{3}$OH & -&Y(1) &Y(1) & Y(9)   \\
             CH$_{2}$DOH &- &Y(1) &Y(1)&Y(1)  \\
             CH$_{3}$CN&- &- &-&Y(5) \\ 
             CH$_{3}$CCH&- &- &-&Y(3) \\
             CH$_{3}$CHO&- &- &-&Y(3) \\
             HCOOCH$_{3}$&- &-&-&Y(3)\\
         \hline
         Total & 17 & 29 & 36 & 75 \\
         \hline
        \end{tabular}
\tablefoot{($^a$) Some species have been detected using less than 3 lines, according to the criteria explained in Sect.~\ref{ID}. 
  }   
     \end{threeparttable}
    \end{table}     
%%%%%%%%%%%%%%%%%
%%%% 
%%%% LTE ANALYSIS
%%%%
%%%%%%%%%%%%%%%%%
\subsection{LTE Analysis}

For the molecules detected in at least two lines, we used the standard rotational diagram (RD) approach to determine the rotational temperature and the column density \citep{Goldsmith1999}. We assumed optically thin line emission and Local Thermodynamic Equilibrium (LTE) conditions. For the species for which we estimated optically thick emission (see Sect. \ref{iso}) we apply an a posteriori correction to the estimated column density, as explained below. The upper level column density of the $u$ $\to$ $l$ transition
is calculated using: 

\begin{equation} 
N_{u} = \frac{8 \pi k \nu^{2}}{h c^{3} A_{ul}} \frac{1}{\eta_{bf}}  \int T_{mb} dV.
\label{eq:col_dens}
\end{equation}

Here $N_{\rm u}$ is the upper level column density (cm$^{-2}$), $k$ and $h$ are
respectively
the Boltzmann (erg K$^{-1}$) and Planck (erg s) constants, $A_{\rm ul}$ is the Einstein coefficient (s$^{-1}$), $\eta_{bf}$\footnote{ $\eta_{bf} = \theta_{s}^{2} \times (\theta_{s}^{2}+\theta_{b}^{2})^{-1}$; $\theta_{s}$ and $\theta_{b}$ are the source and the beam sizes (assumed to be both a circular Gaussian).} is the beam-filling factor,
$c$ is speed of the light (cm s$^{-1}$), $\nu$ is the rest frequency (GHz), and $\int T_{mb} dV$ is the integrated emission (K km s$^{-1}$) in main beam temperature scale. As the beam is roughly the same for all the observed transitions ($10\arcsec$--$11\arcsec$) and we do not know a priori the source size, we do not correct the line intensities for the beam filling factor (i.e. $\eta_{bf} = 1$ in Eq. (\ref{eq:col_dens})). Hence, the derived $N_{\rm u}$ is the beam-averaged column density.

According to the LTE assumption, we can describe $N_{\rm tot}$ and $T_{\rm rot}$ with an equation as follow:
\begin{equation} 
ln \frac{N_{u}}{g_{u}} = ln N_{tot} - ln Q(T_{rot}) - \frac{E_{up}}{k T_{rot}} 
\end{equation}
where $N_{\rm tot}$ is the total beam-averaged column density of the molecule, $g_{\rm up}$ is the degeneracy of the upper level, $E_{\rm up}$ is the energy of the upper level, and Q($T_{\rm rot}$) is the partition function depending on the rotational  temperature $T_{\rm rot}$.

The $T_{\rm rot}$ and beam averaged $N_{\rm tot}$ (cm$^{-2}$) values derived from the rotational diagram analysis for each molecule and for each source are listed in Table \ref{RD}. In the table we also report the number of lines used for the RD, and the interval of upper level energies (K) of the detected lines. The rotational diagrams for the examined molecules towards the four protostars are shown in Figs. \ref{RD-L1489} - \ref{RD-L1551}. 
In Section \ref{LVG} we applied the non-LTE analysis in the Large Velocity Gradient (LVG) approximation for the molecules where more than three lines are detected and collisional rates are available. In these cases we  derived the source size, hence the column densities are source-averaged.

In the case of H$_{2}$CS and OCS, whose
emission is optically thick, we correct the column density values (see Table \ref{L1551-CD}) 
for the optical depth derived in Sect. \ref{iso}, using the following equation: 
\begin{equation} 
N = N_{thin}(\frac{\tau}{1 - \exp^{-\tau}}). 
\end{equation}

In L1489-IRS, B5-IRS1, and L1455-IRS1 the rotational temperatures 
of all the analyzed species are quite low, less than about 30 K, with the exception of H$_{2}$CO in L1455-IRS1 for which we find $T_{\rm rot}$ = (68 $\pm$ 25) K. On the other hand, L1551-IRS5 is associated with iCOMs emission for which we find higher $T_{\rm rot}$, up to (148 $\pm$ 30) K. The beam averaged column densities (as derived using the RD analysis) are in the range of 10$^{11}$ -- 10$^{13}$ cm$^{-2}$, with the exception of SO and CH$_{3}$OH for which we find $N_{\rm tot}$ of a few 10$^{14}$ cm$^{-2}$.

\subsection{Line Profiles}
\label{LineProfiles}

The observed line profiles are very complex and
differ from source to source (see Fig. \ref{L1489_Spectra}--\ref{L1551_Spectra}),
with high-velocity wings, secondary peaks, and different linewidths. The CO and $^{13}$CO profiles are affected by deep absorption at systemic velocity as well as by absorption features due to CO molecules along the line of sight. Broad line wings up to about $\pm$10 km s$^{-1}$ with respect to $V_{\rm sys}$ have been detected in CO, $^{13}$CO, C$^{18}$O, as well as in other species. 
Lines with $FWHM$ $\sim$ 1 -- 2 km s$^{-1}$, centered at $V_{\rm sys}$, are expected to be emitted by the molecular envelope, while iCOMs emission with $\sim$4 km s$^{-1}$ linewidth is plausibly tracing the compact (less than 1$\arcsec$) hot corino. A special case is represented by L1551-IRS5, where the multi-peak profiles allow us to disentangle the emission  coming from both the envelope and the hot corino (see Sect. \ref{L1551Case}).

By selecting different velocity ranges, we were able to derive the column densities of the different kinematical components, i.e. the envelope
(Sect. \ref{envelope}), the outflow (Sect. \ref{outflow}), and the hot corino
(Sect. \ref{hot corino}). In some cases, the need to have S/N of at least 3 (for the intensities in selected velocity ranges) forced us to use less lines and, as a consequence, to assume the rotational temperature derived
from other species which trace the same component and for which we were able to obtain a reliable RD. 
As described in the next sections, we assumed: 
(i) $T_{\rm rot}$ = 20 -- 35 K
for the envelope (from c-C$_{3}$H$_{2}$),
(ii) $T_{\rm rot}$ = 50 -- 70 K for the outflows
(from SO and H$_{2}$CO), and (iii)
$T_{\rm rot}$ = 60 -- 150 K for the L1551 
hot corino (from CH$_{3}$OH).

\subsubsection{Envelope Tracers}
\label{envelope}

Since c-C$_{3}$H$_{2}$ is detected in all the sources and shows a well-defined Gaussian profile with $FWHM$ $\sim$ 1.5 km s$^{-1}$, we used this species as a reference to define the temperature range for the envelope ($T_{\rm rot}$ = 20 -- 35 K).
Such temperatures have been adopted to estimate the column densities of species observed through one emission line only.
For lines (due to other species) broader than 1.5 km s$^{-1}$, 
we integrated the line intensity in the velocity range $\pm$1 km s$^{-1}$ around $V_{\rm sys}$ to disentangle the envelope component.
Tables \ref{L1489-CD}, \ref{B5-CD}, \ref{L1455-CD}, and \ref{L1551-CD} report the derived $T_{\rm rot}$ and $N_{\rm tot}$ values obtained for the envelope component. Source averaged column densities can be obtained from beam averaged column densities by applying the filling factors reported for c-C$_{3}$H$_{2}$ in Table \ref{LVGList}.
Finally, note that for L1551-IRS5, we had 
the possibility to apply the RD approach for both
ortho and para D$_{2}$CO species separately. 
Figure \ref{RD-L1551} shows that the present data do 
not allow us to reveal the expected o/p statistical values (3:1), 
likely due to poor statistics as well as low S/N value of the spectra.

\subsubsection{Outflow Tracers}
\label{outflow}

We detected line wings up to $\sim \pm$10 km s$^{-1}$ with respect to the systemic velocity for a large number of
species beside the CO isotopologues: H$_{2}$CO, SO, SO$_{2}$, $^{13}$CS, DCN, H$_{2}$S, DCO$^{+}$, and H$_{2}$CS.
This high-velocity emission (see e.g. Fig. \ref{B5_Spectra}, \ref{L1455_Spectra} and \ref{L1551_Spectra}) is plausibly due to the outflow motions,
and is detected in all the sources. On the other hand we do not see high-velocity outflow motion in the line profiles of L1489-IRS.

In order to analyze the outflow contribution, we used the emission integrated at high velocities 
in the residual spectra after subtraction of the brighter emission at systemic velocity due to the envelope fit with a Gaussian (with $FWHM$ $\sim$  1.5 km s$^{-1}$ consistent with envelope emission).
Rotational temperatures and column densities derived 
for the high-velocity outflow components are reported in Tables \ref{B5-CD}, \ref{L1455-CD}, \ref{L1551-CD}. For the species where the rotational diagram analysis cannot be performed, because only one line is detected, we derived the column densities by assuming
$T_{rot}$ $\sim$ 50 -- 70 K, following what found using
H$_{2}$CO. 

\subsubsection{Hot Corino Tracers} \label{hot corino}

Only L1551-IRS5 shows emission lines with linewidths of $\sim$4 km s$^{-1}$ due
to several iCOMs: CH$_{3}$OH, CH$_{3}$CN, CH$_{3}$CHO, and
HCOOCH$_3$. This points to the presence of a hot corino,
in agreement with the very recent ALMA results by
\citet{Bianchi2020} who imaged a hot corino using CH$_{3}$OH, CH$_{2}$DOH, HCOOCH$_{3}$, and t-CH$_{3}$CH$_{2}$OH.
The chemical richness looks to be predominantly associated
with the Northern component of the L1551-IRS5 binary system,
peaking at +7.5 km s$^{-1}$. The L1551-IRS5 system,
including both a hot corino and an envelope will 
be discussed in Sect. \ref{L1551Case}. 
In light of these findings, we derived the 
integrated emissions due to the hot corino also for
OCS, O$^{13}$CS, and H$_{2}$S, which show
linewidths larger than 4 km s$^{-1}$, but with a well defined emission peak at +7.5 km s$^{-1}$. 
In addition, OCS and H$_2$S show a broad emission
(larger than 4 km s$^{-1}$) also in L1455-IRS1, where methanol lines are narrow ($FWHM$ = 1.6 km s$^{-1}$). To be coherent with the L1551-IRS5 analysis
we assumed a high temperature for the gas emitting OCS and H$_2$S in L1455-IRS1 as done for L1551-IRS5. The origin of the OCS and H$_2$S emission will be discussed 
in details in Sect. \ref{diversity}.

We integrated the line intensities 
$\pm$2 km s$^{-1}$ around the hot corino velocity.
The derived $T_{\rm rot}$ and column densities derived using
the RD approach are listed in Tables \ref{L1551-CD}.
Rotational temperatures lie between 42 K and 148 K,
while beam averaged column densities are $\sim$10$^{12}$ -- 10$^{14}$ cm$^{-2}$. For species for which only a single transition is detected, we assumed $T_{\rm rot}$ = 60 -- 150 K and we derived column densities. Source averaged column densities can be obtained from beam averaged column densities by applying the filling factors reported for CH$_{3}$OH and CH$_{3}$CN in Table \ref{LVGList}.

%%%%%%%%%%%%%%%%%%%%%%%%%%%%%%%%%
%% ROTATIONAL DIAGRAM ANALYSIS %%
%%%%%%%%%%%%%%%%%%%%%%%%%%%%%%%%%
\hspace*{-8cm}
\begin{table}[ht]
  \caption{Number of detected lines (S/N $\geq$ 3), range of upper level energies of the detected transitions, temperatures (K) and beam averaged column densities (cm$^{-2}$) derived from the rotational diagram analysis.}

 { \begin{tabular}{lcccc}
\hline
Species & $N_{lines}$ & $E_{u}$  & $T_{rot}$ & $N_{tot}$  \\
 & & (K) & (K) &(cm$^{-2}$)\\
 \hline
 \hline
 \multicolumn{5}{c}{L1489-IRS}\\
\hline
SO & 2 & $35-44$ & 6 (2) & $2(3)\times10^{14}$\\
p-H$_{2}$CO & 3 & $21-68$ & 23 (3) & $43(13)\times10^{11}$\\
c-C$_{3}$H$_{2}$ & 3 & $19-39$ & 19 (5) & $5(2)\times10^{11}$\\
\hline
\hline
\multicolumn{5}{c}{B5-IRS1}\\
\hline
SO & 2 & $35-44$ & 8 (2) & $1(2)\times10^{14}$\\
p-H$_{2}$CO & 3 & $21-68$ & 24 (3) & $6(2)\times10^{12}$\\
c-C$_{3}$H$_{2}$ & 4 & $19-6$1 & 18 (3) & $10(3)\times10^{11}$\\
\hline
\hline
\multicolumn{5}{c}{L1455-IRS1}\\
\hline
SO & 2 & $35-44$ &  10 (4) & $4(5)\times10^{13}$\\
OCS & 2 & $100-110$&$59(102)$ & $2(5)\times10^{13}$ \\
p-H$_{2}$CO & 3 & $21-68$ &  68(25) & $20(6)\times10^{12}$\\
HDCS & 2 & $42-51$ & 13 (9) & $2(5)\times10^{12}$ \\
c-C$_{3}$H$_{2}$ & 4 & $19-61$ & 16 (3) & $13(5)\times10^{11}$\\
\hline
\hline
\multicolumn{5}{c}{L1551-IRS5}\\
\hline
SO & 2 & $35-44$ & 40 (50) & $2(3)\times10^{13}$\\
SO$_{2}$ & 2 & $19-94$ & 44 (8) & $8(2)\times10^{12}$\\
OCS$^{b}$&  2 & $100-110$& 64 (121) & $2(5)\times10^{15}$\\
p-H$_{2}$CO & 3 & $21-68$ & 41 (9) & $4(13)\times10^{13}$\\
o-D$_{2}$CO & 2 & $28-50$ & 18 (4) & $3(2)\times10^{12}$\\
p-D$_{2}$CO & 2 & $32-50$ & 28 (14) & $9(7)\times10^{12}$\\
c-C$_{3}$H$_{2}$ & 4 & $19-61$ & 24 (4) & $11(3)\times10^{12}$\\
H$_{2}$CCO & 2 & $64-76$ & 21 (13) & $2 (8)\times10^{13}$\\
CH$_{3}$OH$^a$  &9 &$45-374$ &148 (30) &$16(4)\times10^{13}$ \\
CH$_{3}$CN  &5 &$68-133$ &133 (81) & $1(5)\times10^{12}$ \\
CH$_{3}$CCH & 3 & $75-103$ & 23 (11) & $6(13)\times10^{13}$\\
CH$_{3}$CHO$^a$ & 3&$26-182$ &42 (5) &$22(6)\times10^{12}$ \\
\hline

\end{tabular}}
\tablefoot{($^a$) Column densities refer to the sum of the A and E type of methanol and acetaldehyde.
($^b$) Column density is corrected for the opacity as mentioned in Sect. \ref{iso}.
}
\label{RD}
\end{table}

\begin{table*}[ht]
  \caption{Source-averaged molecular column density, $N_{\rm tot}$, gas physical conditions (temperature, $T_{\rm kin}$, and density, $n_{\rm H_{2}}$), source size inferred from the LVG analysis, and beam-filling factor ($\eta_{bf}$) of c-C$_{3}$H$_{2}$, CH$_{3}$OH, and CH$_{3}$CN lines towards the four protostars in our sample (L1489-IRS, B5-IRS1, L1455-IRS1 and L1551-IRS5). The range of values reported correspond to the LVG solutions which reproduce the observations within the 1$\sigma$ confidence level.}
\centering
 { \begin{tabular}{lccccccc}
\hline
Species & $N_{\rm lines}$ & $E_{\rm up}$  & $T_{\rm kin}$ & $N_{\rm tot}$ & $n_{\rm H_{2}}$ &Source size &  $\eta_{bf}$ \\
 & & (K) & (K) &(cm$^{-2}$) &(cm$^{-3}$) & ($\arcsec$) & \\
\hline
\hline
\multicolumn{8}{c}{L1489-IRS}\\
c-C$_{3}$H$_{2}$ & 4 & $19 - 39$ & $5 - 25$ & $1\times10^{12}$-- $3\times10^{15}$ & $8\times10^{4}$ -- $3\times10^{6}$ & 2 -- 5 & 0.03 -- 0.17\\
\hline
\multicolumn{8}{c}{B5-IRS1}\\
%\hline
c-C$_{3}$H$_{2}$ & 4 & $19 - 39$ & $15 - 18$ & $1 - 30 \times10^{14}$ & $1 - 30 \times10^{4}$ & 2 -- 3 & 0.03 --0.07\\
\hline
\multicolumn{8}{c}{L1455-IRS1}\\
%\hline
c-C$_{3}$H$_{2}$ & 4 & $19 - 39$ & $10 - 25$ & $1 - 50 \times10^{13}$ &$8\times10^{4}$ -- $5\times10^{6}$& 3 -- 10 & 0.07 -- 0.45\\
\hline
\multicolumn{8}{c}{L1551-IRS5}\\
%\hline
c-C$_{3}$H$_{2}$ & 4 & $19 - 39$ & $15 - 20$ & $3\times10^{13}$ -- $1 \times10^{15}$ &$1\times10^{4}$ -- $3 \times10^{6}$ & 4 -- 8 & 0.12 -- 0.35\\
CH$_{3}$OH & 7 & $45 - 190$ & $50 - 105$ & $1 - 6 \times10^{18}$  &$0.5 - 30 \times10^{6}$ & 0.14 -- 0.18 & $1.6 - 2.6 \times10^{-4}$ \\
CH$_{3}$CN & 4 & $68 - 133$ & $105 - 135$ &  
$\geq \times10^{17}$ & $\geq3\times10^{4}$ & 0.11 -- 0.20 & $1.0 - 3.3 \times10^{-4}$\\
\hline

\end{tabular}}
\label{LVGList}
\end{table*}

%%%%%%%%%%%%%%%%%
%%%% 
%%%% LVG ANALYSIS
%%%%
%%%%%%%%%%%%%%%%%

\subsection{Non-LTE Large Velocity Gradient (LVG) Analysis}
\label{LVG}
 In order to derive the physical parameters of the emitting gas and the molecule column densities, we compare the line intensity measured in c-C$_{3}$H$_{2}$ (L1489-IRS, B5-IRS1, L1455-IRS1, and L1551-IRS5),  CH$_{3}$OH and CH$_{3}$CN (L1551-IRS5) with ones predicted by the non-LTE LVG code grelvg \citep{Ceccarelli2003}. 
 We compute the line intensities for a grid of molecular column densities, $N_{\rm x}$, gas densities, $n_{\rm H_{2}}$, and temperatures, $T_{\rm kin}$. Then the predicted line intensities are compared with the observed ones to determine the best fit, i.e. the values of $N_{\rm x}$, $n_{\rm H_{2}}$, $T_{\rm kin}$, and the source size $\theta$ which gives the minimum ${\chi}^2$.

 We carried out the non-LTE analysis only for three species (c-C$_{3}$H$_{2}$, CH$_{3}$OH, and CH$_{3}$CN) based on the following criteria: (1) at least four transitions have been detected, 
 (2) the detected transitions cover a significant range of upper level energies, (3)  collisional coefficients are available in the literature. 
 For c-C$_{3}$H$_{2}$, we used the collisional coefficients with He, computed by \citet{Chandra2000} and provided by the BASECOL database \citep{Dubernet2013}; for CH$_{3}$OH, we used the collisional coefficients of both A and E transitions with para-H$_{2}$, computed by \citet{Rabli2010} and provided by the BASECOL database \citep{Dubernet2013}. For CH$_{3}$CN, we 
 used the H$_{2}$ collisional coefficients by \citet{Green1986}, provided by the LAMDA database \citep{Schoier2005}.  
 
 When computing the line intensities with grelvg we assumed: (1) spherical geometry to compute the line escape probability \citep{deJong1980}; (2) a A-CH$_{3}$OH/E-CH$_{3}$OH ratio equal to 1; (3) the ortho-H$_{2}$/para-H$_{2}$ ratio  equal to 3; (4) a linewidth of 1 km s$^{-1}$ for c-C$_{3}$H$_{2}$, 3 km s$^{-1}$ for CH$_{3}$OH, and 2.5 km s$^{-1}$ for CH$_{3}$CN (as derived from our observations). For CH$_{3}$OH and CH$_{3}$CN, we used the integrated line intensities for the hot corino component as described in Sect. \ref{hot corino}. The uncertainties on the line integrated intensities, due to the error on calibration and the error on the Gaussian fit, amount to 30$\%$ of the observed intensities. 

\begin{figure}
\includegraphics[width=9cm]{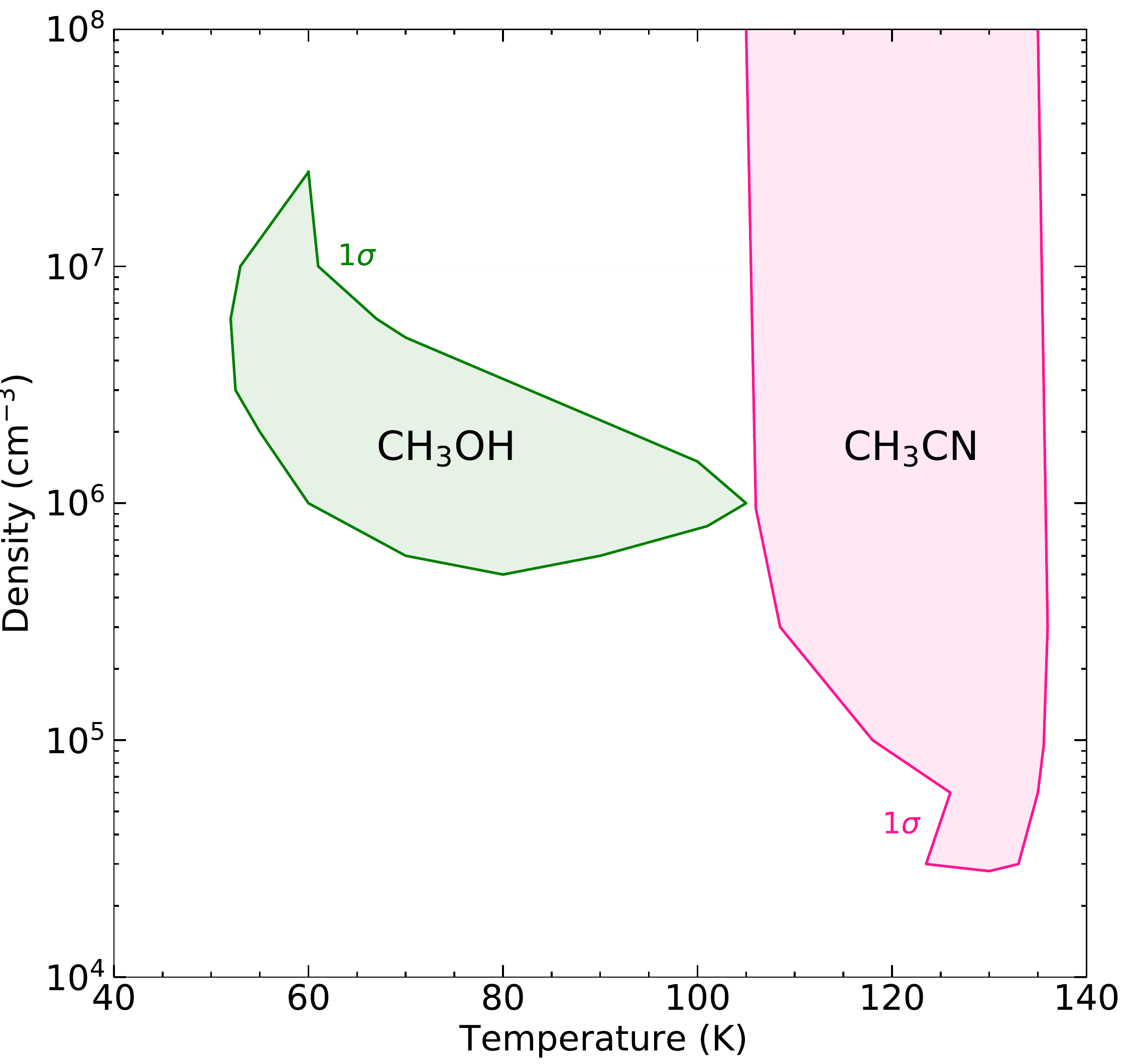}
\caption{Result of the LVG analysis for L1551-IRS5. The inferred range of gas density and temperature are summarized in Table \ref{LVGList}.
${\chi}^2$ contour plots as a function of the gas density and temperature. The green, and pink contours show the 1$\sigma$ confidence level for CH$_{3}$OH, and CH$_{3}$CN respectively for representative value of \textit{N(CH$_{3}$OH)} = 5 $\times$ 10$^{18}$ cm$^{-2}$ with source size of 0$\farcs$18 and \textit{N(CH$_{3}$CN)} $\geq$1 $\times$ 10$^{17}$ cm$^{-2}$ with source size of 0$\farcs$11.}
\label{LVG_densityandT}
\end{figure}

In the next sub-sections, we describe the results obtained by applying the LVG analysis for each species. Table \ref{LVGList} summarizes  the results of the LVG analysis by showing the number of lines, the range of upper level energy  used for the analysis, kinetic temperatures, total column densities, H$_{2}$ densities and the source size of the species which reproduces the observations within 1$\sigma$ confidence level. From the estimated source size we also derive the beam filling factor of the emitting species, $\eta_{bf}$, reported in the last column of Table \ref{LVGList}. Figure \ref{LVG_densityandT} shows the ranges of temperature, $T_{\rm kin},$ and H2 density, $n_{\rm H_{2}}$, obtained from the analysis of the CH$_3$OH and CH$_3$CN emission towards L1551-IRS5.

\begin{figure}
 \includegraphics[width=9cm]{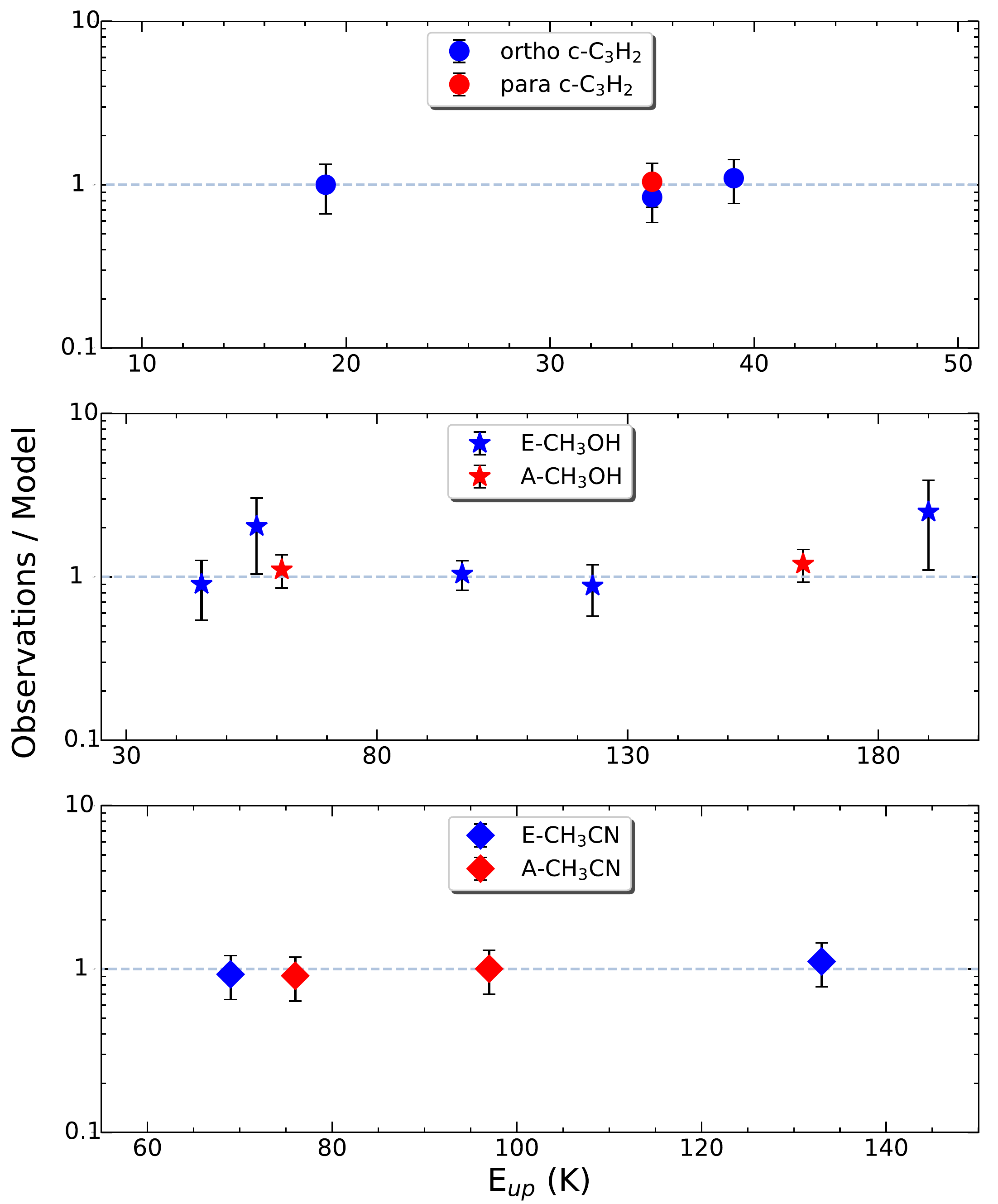}
 \caption{Ratios of the observed line intensities to
 those predicted by the best-fitting LVG models as a function of upper level energy of the considered transitions towards L1551-IRS5. Upper panel: Blue and red circles are for the ortho and para c-C$_{3}$H$_{2}$, respectively. Middle 
 panel: Blue and red stars show the E and A transitions of CH$_{3}$OH respectively. Bottom panel: Blue and red diamonds are
 for the E and A CH$_{3}$CN lines respectively.
 }
\label{LVG_M&O}
   \end{figure}

Finally, Fig. \ref{LVG_M&O} shows the ratio between the observed over predicted line intensities as a function of upper level energy of the transition.

\subsubsection{LVG analysis of c-C$_{3}$H$_{2}$ emission}
\label{c-c3h2}
We ran a grid of LVG models to obtain predictions of the detected lines, for a range of total (para- c-C$_{3}$H$_{2}$ and ortho- c-C$_{3}$H$_{2}$) column density \textit{N(c-C$_{3}$H$_{2}$)} from 10$^{11}$ cm$^{-2}$ to 10$^{17}$ cm$^{-2}$, a gas density $n_{\rm H_{2}}$ from 10$^{4}$ cm$^{-3}$ to 10$^{8}$ cm$^{-3}$ and a kinetic temperature $T_{\rm kin}$ in the 5 -- 80 K range. 

The 1$\sigma$ confidence level of the observed c-C$_{3}$H$_{2}$ lines give similar results for the four sources, i.e. the column density values (\textit{N(c-C$_{3}$H$_{2}$)}) are 1 $\times$ 10$^{12}$ -- 3 $\times$ 10$^{15}$ cm$^{-2}$, $n_{\rm H_{2}}$ of 1 $\times$ 10$^{4}$ -- 5 $\times$ 10$^{6}$ cm$^{-3}$, $T_{\rm kin}$ from 5 K to 25 K, and source sizes spanning between 2$\arcsec$ and 10$\arcsec$. Table \ref{LVGList} reports the results for each source.

The overall analysis is in agreement with a temperature of 25 K, a column density of $\sim$2 $\times$ 10$^{14}$ cm$^{-2}$ for the low mass star formation region L1527 \citep{Yoshida2015}.

These results are consistent with what was obtained from the LTE analysis. This is not surprising considering that the LVG analysis predicts that the detected transitions are optically thin ($\tau$ values are from 0.03 to 0.37).  
Moreover, the $n_{\rm cr}$ of the detected c-C$_{3}$H$_{2}$ lines \citep{Chandra2000} are between 4 $\times$ 10$^{5}$ cm$^{-3}$ and 3 $\times$ 10$^{6}$ cm$^{-3}$. Therefore the gas density $n_{\rm H_{2}}$ estimated through the LVG analysis towards the four protostars is larger than $n_{\rm cr}$ in many cases.

This implies that $n_{\rm H_{2}}$ towards L1551-IRS5 is higher than $n_{\rm cr}$. As a result, our LTE analysis is excellently consistent with LVG result. The $n_{\rm H_{2}}$ values towards 4 protostars between 1 $\times$ 10$^{4}$ cm$^{-3}$ and 5 $\times$ 10$^{6}$ cm$^{-3}$, are mostly higher than $n_{\rm cr}$. 

\subsubsection{LVG analysis of CH$_{3}$OH and CH$_{3}$CN emission} \label{LVG of L1551}

We ran the LVG code for  CH$_{3}$OH and CH$_{3}$CN in L1551-IRS5. The LVG code computes the predicted line intensities for a large grid of models ($\geq$ 10000),
assuming methanol column densities between 1 $\times$ 10$^{16}$  cm$^{-2}$ and 1 $\times$ 10$^{20}$ cm$^{-2}$, $n_{\rm H_{2}}$ in the 10$^{5}$ -- 10$^{7}$ cm$^{-3}$ range, and a gas kinetic temperature 
up to 150 K. The best fit, with the ${\chi}^2$ of 1.0,
gives $T_{\rm kin}$, $n_{\rm H_{2}}$ and \textit{N(CH$_{3}$OH)} equal to 80 K, 1 $\times$ 10$^{6}$ cm$^{-3}$, and 5 $\times$ 10$^{18}$ cm$^{-2}$, respectively. The size $\theta$ is 0$\farcs$18 (25 au). 
The line opacities ranges from 0.3 to 12, indicating that
line emissions are moderately thick. Figure \ref{LVG_densityandT} reports the $n_{\rm H_{2}}$ -- $T_{\rm kin}$ plot for the column density, N, and size, $\theta$, which minimizes the ${\chi}^2$, while Fig. \ref{LVG_M&O} report the ratio between the observed and predicted intensities for the best fit solution. Note that the critical densities $n_{\rm cr}$ of the transitions \citep{Rabli2010} lie from 4 $\times$ 10$^{4}$ to 2 $\times$ 10$^{5}$ cm$^{-3}$.

Since $n_{\rm H_{2}}$ is at least one order of magnitude higher than $n_{\rm cr}$, LTE is a reasonable assumptions for the CH$_{3}$OH emission. 
As a result, after correcting for the line opacities and beam filling factor,
the column densities estimated from the LTE analysis are in agreement with those inferred with the LVG code. Note that the CH$_{3}$OH 4$_{-2,3}$ -- 3$_{-1,2}$E (45 K) is known to be a 
Class I-type methanol maser \citep{Hunter2014, Chen2019}, observed also towards low mass star forming regions, mostly in Perseus \citep{Kalenskii2006,Kalenskii2012}. 
However, once considering the 50 -- 150 K temperature range, the
CH$_{3}$OH 4$_{-2,3}$ -- 3$_{-1,2}$E  line could be masing only very weakly: --$\tau$ less than 8$\times$10$^{-1}$ down to 8$\times$10$^{-5}$, which is consistent with $\tau$ $\sim$ --0.1 obtained for L1551-IRS5 from the LVG analysis.

For methyl cyanide, we considered column densities from 1 $\times$ 10$^{12}$ cm$^{-2}$ to 1 $\times$ 10$^{19}$ cm$^{-2}$, and $n_{\rm H_{2}}$  between 1 $\times$ 10$^{4}$ cm$^{-3}$ and 3 $\times$ 10$^{6}$ cm$^{-3}$. We used the same temperatures grid assumed for methanol (50 -- 150 K). The best fit with the lowest ${\chi}^2$ (0.3) is well-constrained for $T$ and $n_{\rm H_{2}}$ equal to 130 K, 1 $\times$ 10$^{5}$ cm$^{-3}$. On the other hand, the column density is not well-constrained:  \textit{N(CH$_{3}$CN)} $\geq$ 1 $\times$ 10$^{17}$ cm$^{-2}$. The size $\theta$ is equal to 0$\farcs$11 (16 au). The ratio between observed and modeled lines for CH$_{3}$CN is presented in the bottom panel in Fig. \ref{LVG_M&O}. Lines are extremely optically thick ($\tau$ values are between 245 and 367). 
The critical densities $n_{\rm cr}$  of  
the observed transitions \citep{Green1986} 
are $\sim$ 2 $\times$ 10$^{6}$ cm$^{-3}$. 
The 1$\sigma$ solutions for $n_{\rm H_{2}}$
indicated values higher than 3 $\times$ 10$^{4}$ cm$^{-3}$.
As a consequence, the LTE approximation could not be 
satisfied. However, as for methanol, the 
column densities derived using the LTE and LVG analysis are in agreement once correcting for the beam filling factor and opacities.

The analysis of the CH$_3$OH and CH$_3$CN emission 
indicates that the emission in the single-dish spectra is dominated by a hot corino.
This is consistent with what was recently found using ALMA emission with the LVG results of  $^{13}$CH$_{3}$OH on the same source  \citep{Bianchi2020}. They found $T$, $n_{\rm H_{2}}$, \textit{N(CH$_{3}$OH)} and $\theta$ equal to 100 K, 1 -- 1.5 $\times$ 10$^{8}$ cm$^{-3}$, 1 $\times$ 10$^{19}$ cm$^{-2}$, and 0$\farcs$15 respectively. The small difference between the two studies is due to the optical depth. They found that all CH$_{3}$OH lines are optically thick ($\tau$ $>$ 50), then they used the $^{13}$CH$_{3}$OH ($\tau$ $\sim$ 2) emission for their LVG analysis.
 
\begin{figure*}
  \centering
 \includegraphics[width=17cm]{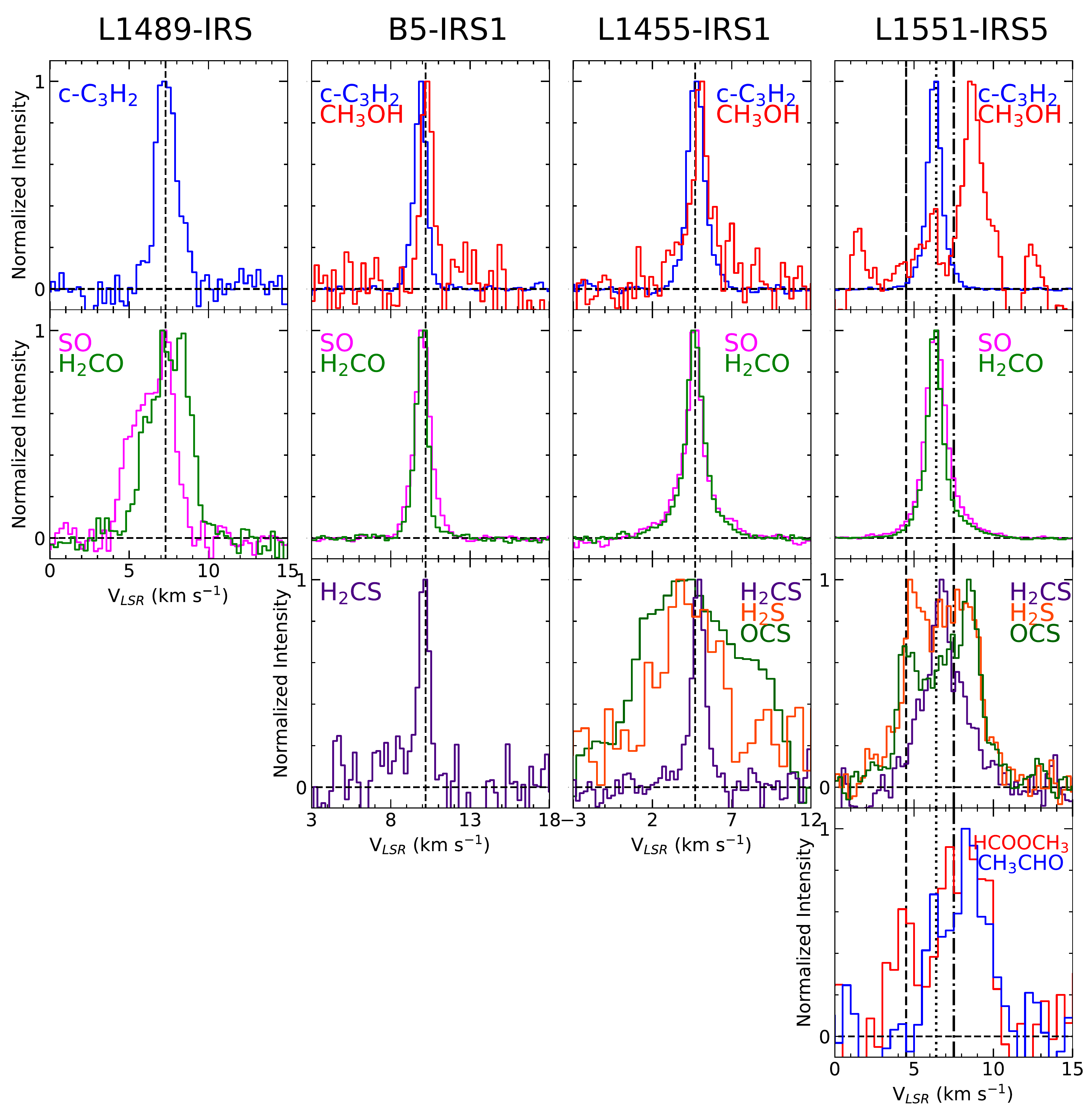}
\caption{Normalized intensity profiles of selected species towards L1489-IRS, B5-IRS1, L1455-IRS1, and L1551-IRS5. The profiles of SO, c-C$_{3}$H$_{2}$, H$_{2}$CO (in all sources), CH$_{3}$OH, CH$_{3}$CHO, HCOOCH$_{3}$ (in L1551-IRS5), and OCS (in L1455-IRS1) have been obtained by stacking the detected transitions for each species in order to maximize the S/N ratio. CH$_{3}$OH (4$_{-2,3}$ -- 3$_{-1,2}$E) in B5-IRS1 and L1455-IRS1 is the transition at 218440 MHz; H$_{2}$CS (7$_{1,7}$ -- 6$_{1,6}$) in B5-IRS1, L1455-IRS1, and L1551-IRS5 is the transition at 236727 MHz;  H$_{2}$S (2$_{2,0}$ -- 2$_{1,1}$) in L1455-IRS1 and L1551-IRS5 is the transition at 216710 MHz; OCS (19 -- 18) in L1551-IRS5 is the transition in 231060 MHz.} 
\label{diversityFIG}
 \end{figure*}

%%%%%%%%%%%%%%%%%
%%%% 
%%%% DISCUSSION
%%%%
%%%%%%%%%%%%%%%%%

\section{Discussion}
\label{discussion}
%%%%%%%%%%%%%%%%%
%%%% 
%%%% CHEMICAL DIVERSITY
%%%%
%%%%%%%%%%%%%%%%%
\subsection{Chemical Diversity}
\label{diversity}

The present dataset provides a chemical survey at 1.3mm of iCOMs, S-bearing, D-bearing, O-bearing, N-bearing and molecular ion species including their rare isotopologues towards Class I protostars as presented in the previous sections. Figure \ref{diversityFIG} shows the normalized intensity profiles of selected species towards the observed Class I sources.

\subsubsection{Narrow c-C$_{3}$H$_{2}$ and CH$_{3}$OH emission from the envelope}

The c-C$_{3}$H$_{2}$ lines are narrow ($\sim$1 km s$^{-1}$) in all the four protostars, and the non-LTE LVG analysis shows that the emission comes from the molecular envelope. More specifically, the c-C$_{3}$H$_{2}$ source sizes vary between 2$\arcsec$ -- 10$\arcsec$ ($\sim$450 -- 3000 au), depending on the source. Our results are in agreement with previous surveys showing that  c-C$_{3}$H$_{2}$ traces the molecular envelope. More specifically,
carbon chains such as c-C$_{3}$H$_{2}$ have been used to identify the
so called Warm Carbon Chain Chemistry (WCCC sources), characterized by a rich hydrocarbon on scales larger than 1000 au, 
instead of a hot corino iCOMs rich gas
\citep[e.g.,][and references therein]{Sakai2013}.
The prototype of this class kind of objects is L1527 \citep[e.g.,][]{Sakai2008,Sakai2010}. As a matter of fact, c-C$_{3}$H$_{2}$ traces the envelope including its inner portion (close to 100 au), as revealed by high-angular resolution interferometric observations \citep[e.g.,][]{Sakai2014a, Sakai2014b}. One of the proposed scenario on the origin of WCCC objects is illumination from Interstellar Radiation Field (ISRF) \citep{Spezzano2016}. Another one is related to the duration of the UV-shielded dense core prior to the protostellar heating of the gas. The longer this phase the higher the abundance of complex O-bearing species on dust mantles (e.g. methanol). On the opposite short starless phase are thought to favour the abundance of small hydrocarbons on dust surfaces \citep[e.g.,][]{Sakai2008, Sakai2013}. Once formed on dust, in order for the CH$_{4}$ to sublimate, temperature should be 20 -- 60 K \citep[see their Figure 1]{Collings2004}. This can happen in the protostar surrounding as well as when a cloud is UV illuminated, then being associated with a photodissociation region (PDR). \citet{Pety2005,Cuadrado2015} claimed that small carbon chains can be formed in the photodissociation regions (PDRs). We note that our targets are located in highly extincted regions ($A_{\rm V}$ $\geq$ 6 mag for Perseus \citep{Kirk2006} and $A_{\rm V}$ $\geq$ 4 mag for Taurus \citep{Pineda2010}) with dense environment ($n_{\rm H_{2}}$ $\geq$ 1 $\times$ 10$^{5}$). Therefore UV radiation would not easily penetrate into such dense regions, affecting only the external layers.

Interestingly, CH$_{3}$OH, as traced by the transition at 214.8 GHz ($E_{\rm up}$ = 45 K), shows in B5-IRS1, L1455-IRS1 and L1551-IRS5 a line profile 
peaking at  $V_{\rm sys}$ and with the same narrow profile exhibited by c-C$_{3}$H$_{2}$. This supports the idea that methanol emitting through
lines with $FWHM$ $\sim$ 1 km s$^{-1}$ coexists with c-C$_{3}$H$_{2}$, and
thus it is tracing extended emission and not the hot corino region.
The methanol column density (derived assuming the temperature of the envelope)
is 1 -- 2 $\times$ 10$^{13}$ cm$^{-2}$ (B5-IRS1), 
7 -- 8 $\times$ 10$^{13}$ cm$^{-2}$ (L1455-IRS1), and 5 -- 7 $\times$ 10$^{13}$ cm$^{-2}$ (L1551-IRS5).
These values are consistent with what is reported by \citet{Oberg2014} 
and \citet{Graninger2016} using
3mm IRAM-30m surveys: 1 $\times$ 10$^{13}$ cm$^{-2}$ (B5-IRS1),
and 7 $\times$ 10$^{13}$ cm$^{-2}$ (L1455-IRS1).
The detection of CH$_{3}$OH requires an energetic process 
to inject the dust mantle composition into the gas phase. 
As described before, a possibility is that the narrow lines of CH$_{3}$OH
and c-C$_{3}$H$_{2}$ trace the external skin of the high-density protostellar
envelope. Another mechanism to create UV radiation in low mass protostar can be seen along the extended outflow cavity walls \citep{vanKempen2009}. Particularly, these layers are rich in terms of warm H$_{2}$, [FeII], [Si II] since these ionized atoms are formed after dissociation process of molecular material. Indeed, B5-IRS1, L1455-IRS1, and L1551-IRS5
are associated with molecular outflows (see Sect. \ref{sample}).
These three sources (and indeed not L1489-IRS, where no narrow methanol lines are observed), are also associated,
according to the c2d Spitzer catalog, with  extended [FeII], [SiII], warm and hot H$_{2}$ knots \citep[see their Table 2]{Lahuis2010}. On the other hand \citet{White2006} argued that narrow CH$_{3}$OH line comes from a toroidal surface, due to a jet or x-ray source nearby the L1551-IRS5 which heat the dust grains in order to release methanol release into the gas phase.

\begin{figure}
 \includegraphics[width=9cm]{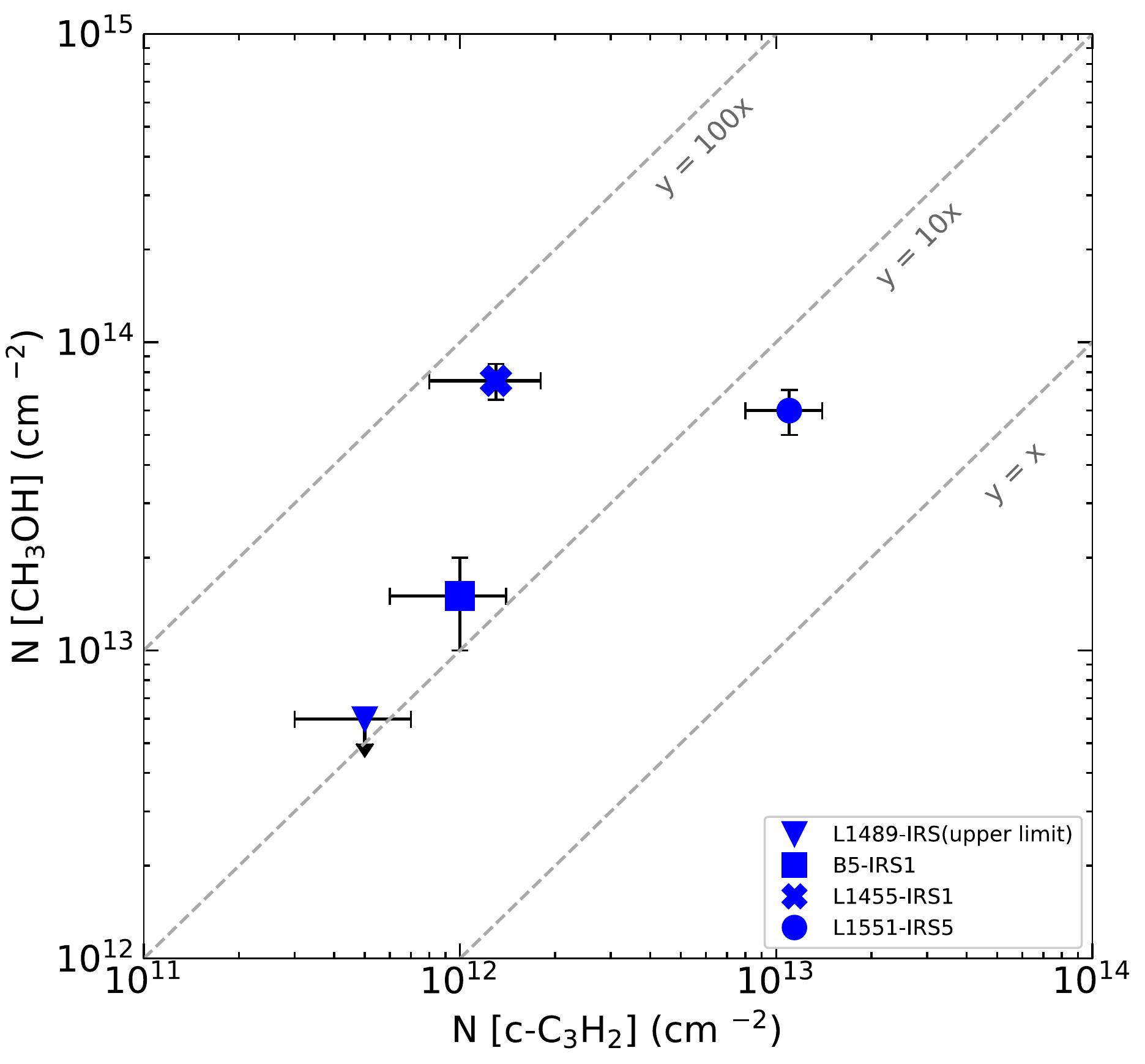}
\caption{Correlation of the column densities of CH$_{3}$OH and c-C$_{3}$H$_{2}$ towards the four protostars in our sample. The sources are labeled in the legend in the bottom right corner. Dashed lines indicate CH$_{3}$OH/c-C$_{3}$H$_{2}$ column density ratio of 100, 10, and 1.
}
\label{ch3oh_c3h2}
  \end{figure}

Figure \ref{ch3oh_c3h2} shows the correlation of the column densities of c-C$_{3}$H$_{2}$ and CH$_{3}$OH. Assuming that both molecules trace the same gas, the plot indicates that the abundance of methanol with respect to 
c-C$_{3}$H$_{2}$ is at least 10 times larger. The plot is well consistent with the correlation reported by \citet{Higuchi2018}, who observed, using single-dish telescopes, 36 Class 0/I protostars in Perseus (see their Fig. 8c) and by \citet{Bouvier2020}, who reported similar correlation between column densities of CCH and CH$_{3}$OH (see their figure 10).
The proposed scenario can be verified only using higher angular resolution observations
to image and compare the spatial distribution of the narrow CH$_{3}$OH
and c-C$_{3}$H$_{2}$ emission.
Finally, note that the results of the present survey indicates that L1551-IRS1 harbors 
an extended WCCC region with a size of $\sim$1200 au (the result of the LVG analysis on c-C$_{3}$H$_{2}$, see in Sect. \ref{c-c3h2}) and with detected c-C$_{3}$H and CH$_{3}$CCH emissions as well as a hot corino region. This source will be discussed in details in Sect. \ref{L1551Case}.

\subsubsection{iCOMs broad line emission from hot corino}

L1551-IRS5 is associated with a hot corino \citep{Bianchi2020}, and
shows broad ($\sim$3 km s$^{-1}$) line profiles of CH$_{3}$OH, CH$_3$CN,  CH$_{3}$CHO, and HCOOCH$_{3}$ (see the bottom right panel of Fig. \ref{diversityFIG}). 

In addition, the LVG solutions on CH$_{3}$OH and CH$_{3}$CN show compact source sizes (0$\farcs$18 and 0$\farcs$11, respectively) and 
kinetic temperatures larger than 50 K, up to 135 K. This is consistent with the chemistry expected for methanol, i.e. a formation uniquely on dust mantles  \citep[e.g.,][]{Tielens1982, Rimola2014}, and a thermal desorption due
to protostellar heating  \citep[e.g.,][]{Charnley1992,Maret2005}.
In principle CH$_{3}$CN can be formed both through the gas-phase and grain-surface reactions \citep[e.g.][]{Walsh2014}. CH$_{3}$CHO, and HCOOCH$_{3}$ are thought to form either on the grain surfaces before evaporating onto the gas phase or in the gas phase using simpler species from the dust mantle
\citep[e.g.][and references therein]{Garrod2006,Balucani2015,Vazart2020,Jorgensen2020}.

On the other hand, we did not detect any hot corino activity towards the other three protostars of our sample: L1489-IRS, B5-IRS1, and L1455-IRS1. 
We report 3$\sigma$ upper limits on the column densities of iCOMs in Table \ref{upperlimits} by assuming $T_{\rm rot}$ = 150 K and a $FWHM$ = 3 km s$^{-1}$. The column densities are lower than $\sim$10$^{13}$ cm$^{-3}$ (CH$_3$OH, HCOOCH$_3$), and
$\sim$10$^{11}$ -- 10$^{12}$ cm$^{-3}$ (CH$_3$CN, CH$_3$CHO).
Recently, \citet{Yang2021} mapped with ALMA several O-bearing and N-bearing iCOMs at high $E_{\rm up}$ (up to $\sim$850 K) towards Class 0 and Class I protostars in Perseus, L1455-IRS1 and B5-IRS1 among them.
\citet{Yang2021} imaged a hot corino towards these sources, while no extended
emission is present. No column density is reported for L1455-IRS1, while 
for B5-IRS1 they report a $N_{\rm CH_3OH}$ = 5 $\times$ 10$^{15}$ cm$^{-2}$
measured in the inner 0$\farcs$5. Once diluted in the IRAM-30m beam,
the methanol column density of the hot corino would be $\sim$10$^{13}$ cm$^{-2}$. This value is consistent with the upper limits 
on the iCOMs column densities reported in Table \ref{upperlimits}.
Conversely, the present IRAM-30m detection of the extended narrow CH$_3$OH emission in B5-IRS1 (as well as those reported by \citet{Oberg2014}) could have been filtered out by the ALMA images by \citet{Yang2021}.

\begin{table}
  \caption{Upper limits (3$\sigma$) on the column densities of the iCOMs detected in L1551-IRS5 towards L1489-IRS, B5-IRS1, and L1455-IRS1.}
\centering
\begin{tabular}{lccc}
\hline
\hline
Species & \multicolumn{3}{c}{$N_{\rm tot}$$^a$}  \\
 & \multicolumn{3}{c}{(cm$^{-2}$)} \\
\hline \noalign {\medskip}
 & L1489-IRS & B5-IRS1 & L1455-IRS1 \\
\hline 
\\
CH$_{3}$OH &$\leq$ $2\times10^{13}$ &$\leq$ $1\times10^{13}$& $\leq$ $2\times10^{13}$\\
% {\medskip}
CH$_{3}$CN  &$\leq$ $2\times10^{11}$ & $\leq$ $2\times10^{11}$&$\leq$ $2\times10^{11}$ \\
%{\medskip}
CH$_{3}$CHO  &$\leq$ $8\times10^{11}$ & $\leq$ $5\times10^{11}$&$\leq$ $6\times10^{11}$ \\
%{\medskip}
HCOOCH$_{3}$ &$\leq$ $7\times10^{12}$&$\leq$ $5\times10^{12}$ & $\leq$ $6\times10^{12}$ \\
\hline
\end{tabular}

\tablefoot{($^a$) Column densities are beam averaged. The values are derived assuming $T_{\rm rot}$ equal to 150 K, i.e. assuming a hot corino region, as found in L1551-IRS5.}
\label{upperlimits}
\end{table}

\subsubsection{H$_2$S, H$_2$CS, and OCS: Hot corinos and circumbinary disks}
\label{S-bearing}

The abundance of sulphur in our Solar System
is S/H = 1.8 $\times$ 10$^{-5}$ \citep{Anders1989}.
However, it is well known that sulphur in star forming regions is
depleted by at least two orders of magnitude with respect to
such value \citep[e.g.][]{Tieftrunk1994,Wakelam2004,Laas2019,vanthoff2020}. 
In addition, the main S-carrier species on dust grains, has not been yet firmly identified \citep{Boogert2015}.
Historically the most plausible suspect was H$_2$S, which is expected
to be formed by surface reactions in the dust mantle due to high hydrogenation and mobility of H. An alternative solution is OCS
\citep[e.g.][]{Wakelam2004,Codella2005,Podio2014,Holdship2016,Taquet2020},
which is potentially formed on grain surfaces by an addition of oxigen to the CS molecule. However, neither H$_2$S nor OCS have been observed
in interstellar ices \citep{Boogert2015}. Also the possible role 
of H$_2$CS has been discussed in light of observations of protostellar shocks, where the dust composition is injected into the gas phase \citep{Codella2005}, but H$_{2}$CS is now thought to be mainly formed by different neutral-neutral and gas phase reactions of CH$_{3}^{+}$ and S \citep{Yamamoto2017}. 
Interestingly, H$_2$S and H$_2$CS have been recently detected and imaged using ALMA towards a number of protoplanetary disks \citep{Phuong2018,LeGal2019,Loomis2020,Codella2020}. 
This makes instructive the observation of H$_2$S, H$_2$CS, and OCS  towards Class I objects in order to investigate how
the molecular complexity of the early star formation stages changes with time. 

Narrow ($\leq$1 km s$^{-1}$) H$_{2}$CS emission has been observed towards both B5-IRS1 and L1455-IRS1, suggesting an emission from the molecular
gas of the envelope. L1551-IRS5 shows a $\sim$3 km s$^{-1}$ wide H$_{2}$CS line
centered at the systemic velocity which is consistent with the envelope origin.
On the contrary, H$_2$S and OCS detected
towards L1455-IRS5 and L1551-IRS5 are completely different with respect to H$_2$CS, both showing broad ($\geq$ 5 km s$^{-1}$) emission.
Rotational diagrams of OCS ($E_{\rm up}$ $\geq$ 100 K) in both objects (see Fig. \ref{RD-L1455} and \ref{RD-L1551}) suggest that it traces warm gas. 
This is in agreement with the idea that OCS and H$_{2}$S are among the expected S-bearing species released from ice mantles of dust grains in hot corinos. 
Interestingly, L1455-IRS1 and L1551-IRS5 are binary systems
resolved by interferometric images \citep[e.g.][]{Tobin2018, Takakuwa2020, Yang2021} associated with circumbinary disks. More specifically, \citet{Takakuwa2020} showed that OCS
traces not only the two protostars but also and mainly the circumbinary disk around the two L1551-IRS5 objects. The OCS  as traced by ALMA
emits in the same velocity range here observed with the IRAM-30m.
These findings, coupled with the similarity of OCS and H$_{2}$S spectra in L1551-IRS5 and L1455-IRS1 suggest that also H$_2$S is associated with
the circumbinary disks of the sources. 
To confirm this hypothesis, it is needed to observe the structures with high resolution images. The complexity of L1551-IRS5 spectra (produced by emission due to two hot corinos and an envelope) will be discussed in details in Sect. \ref{L1551Case}. 

%%%%%%%%%%%%%%%%%
%%%% 
%%%% THE CASE OF L1551
%%%%
%%%%%%%%%%%%%%%%%
\subsection{The L1551-IRS5 system}
\label{L1551Case}

The L1551-IRS5 is a binary system composed by 
two protostars: the so called northern and southern components \citep[e.g.,][and references therein]{Cruz2019, Bianchi2020}.
L1551-IRS5 is a unique protostar among our sample since it presents both hydrocarbons and a hot corino activity with iCOMs emission as discussed in Sect. \ref{LVG of L1551}, \ref{hot corino}, and \ref{diversity}.

The systemic velocities of the northern 
and southern components are +7.5 km s$^{-1}$ and +4.5 km s$^{-1}$, respectively \citep{Bianchi2020}.
The velocity of
the circumbinary envelope is  +6.4 km s$^{-1}$, 
derived using c-C$_2$H$_3$ and C$^{18}$O.
This measurement is in agreement with
the C$^{18}$O images obtained with ALMA
by \citet{Takakuwa2020}. 
In light of the information on the systemic velocity of
the L1551-IRS5 system, the IRAM-30m spectra of the region
can be analyzed in details. 

\begin{figure}
    \centering
 \includegraphics[width=9cm]{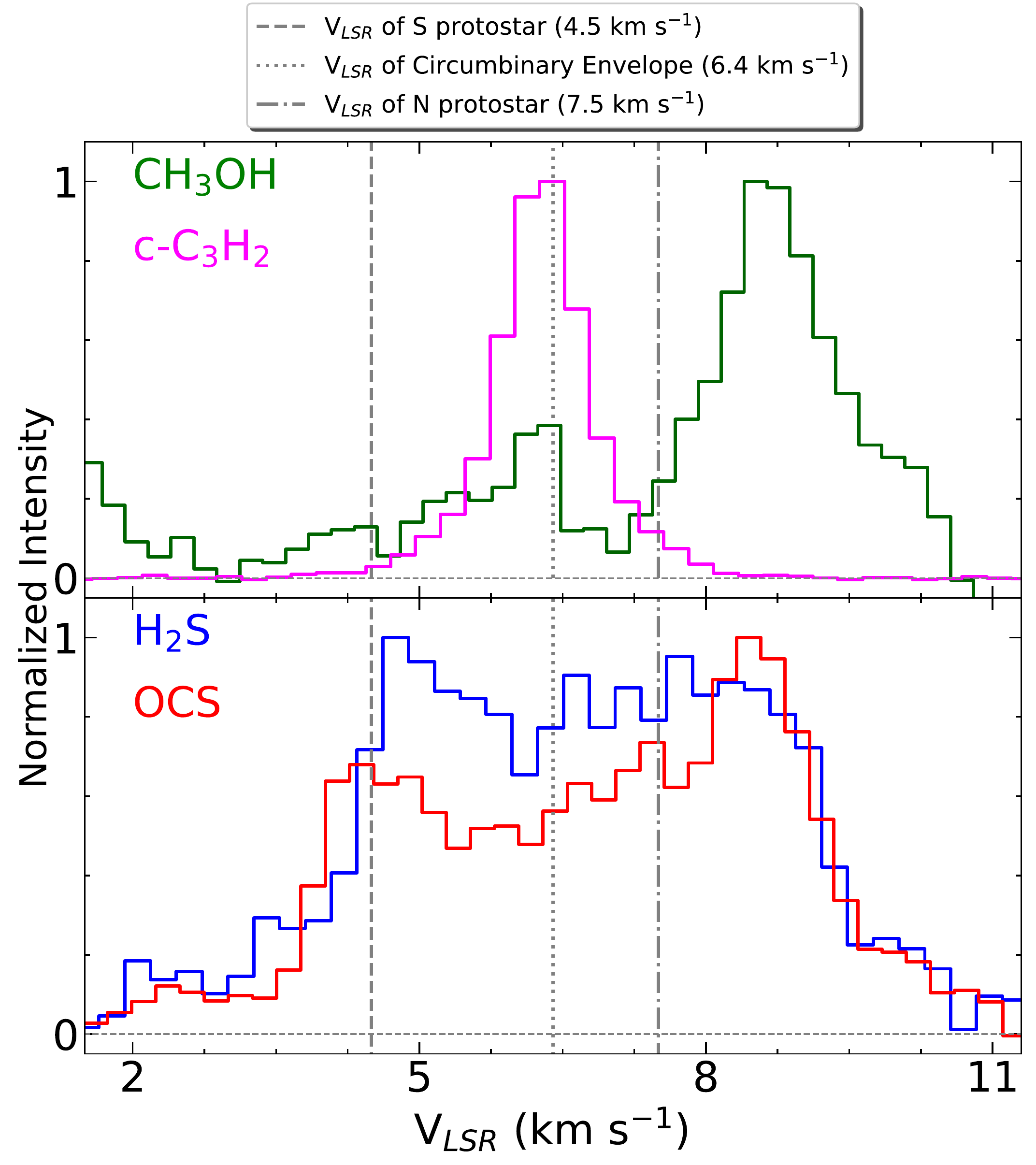}
\caption{Normalized intensity profiles of 4 species towards L1551-IRS5. The spectra of CH$_{3}$OH in green and c-C$_{3}$H$_{2}$ in pink in the upper panel are obtained after stacking the detected transitions (see Fig. \ref{diversityFIG}). Spectra of H$_{2}$S in blue and OCS in red in the lower panel are transitions 2$_{2,0}$ -- 2$_{1,1}$ and 19 -- 18 respectively (see Fig. \ref{L1551_Spectra}). The dashed and dash-dotted lines represent the systemic velocities of the southern protostar (4.5 km s$^{-1}$) and the northern protostar (7.5 km s$^{-1}$) respectively \citep{Bianchi2020} while the dotted line represents the systemic velocity of the circumbinary envelope (6.4 km s$^{-1}$, measured using the C$^{18}$O 2 -- 1 line (this work)).}
\label{L1551-3peaks}
  \end{figure}

Figure \ref{L1551-3peaks} summarizes the L1551-IRS5 spectra
as observed using different molecules. 
Most of detected species peak at the envelope velocity 
+6.4 km s$^{-1}$ such as c-C$_{3}$H$_{2}$.
CH$_{3}$OH as well as other detected iCOMs (CH$_{3}$CN, CH$_{3}$CHO, and HCOOCH$_{3}$) 
peaks at the velocity associated with the Northern protostars, which is red shifted by
$\sim$1.5 km s$^{-1}$ with respect to +7.5 km s$^{-1}$.
This finding is well in agreement with the ALMA images of
iCOMs emission by \citet{Bianchi2020}, who showed
that the northern protostar is associated with a rotating
hot corino, being the NW red shifted emission brighter than the SE blue shifted one. As shown in Sect. \ref{LVG},
the LVG results indicate that  CH$_{3}$OH and CH$_{3}$CN have T$_{kin}$ $\sim$ 100 K and $\theta$ $\sim$ 0$\farcs$15, tracing the hot corino region around the northern protostar. 

Finally, H$_{2}$S (blue line in Fig. \ref{L1551-3peaks}) and OCS show profiles which differ from that of the envelope and of the northern hot corino.
There are two peaks, one again red shifted with respect to the velocity of the northern component, but, in addition, a second emission peak occurs at 4.5 km s$^{-1}$, i.e. the velocity of the southern components of the L1551-IRS5
binary system.
Our H$_2$S and OCS spectra can be interpreted in light of the C$^{18}$O and OCS ALMA observations at 10 au spatial scale reported by \citet{Takakuwa2020}, who reported the presence of two circumstellar disks associated with the two companions of the binary system, plus emission from the circumbinary disk. 

To conclude, the present OCS profile
observed with the IRAM-30m is due to the sum of 
emission close to the protostars as well as emission due
to the circumbinary disk. 
Given the similarity of the OCS and H$_2$S profiles we expect that, if observed with a 10 au resolution,
also H$_2$S emission would be image 
a combination of emission from the hot corinos and the circumbinary disk.

\subsection{Chemical Evolution Along the Star Formation Process}  
%%%%%%%%%%%%%%%%%
%%%% 
%%%% DEUTERATION and isotopologues
%%%%
%%%%%%%%%%%%%%%%%
\subsubsection{Deuterium Fractionation} \label{Deuteration}
As we stressed in Sect. \ref{Intro}, 
the molecular deuteration is
very important to understand the chemical evolution from prestellar cores to comets. 
Up to today, a collection of deuterated species have been detected from the prestellar phase to small bodies of the Solar System \citep[e.g.][]{Caselli2012,Ceccarelli2014}. However, to our best knowledge, there is a limited number of studies on deuteration in Class I protostars.

By studying the prototypical Class I protostars, SVS13-A, \citet{Bianchi2017, Bianchi2-2019} reported a decrease of deuteration of methanol, thioformaldehyde, and formaldehyde with respect to what is found towards Class 0 objects. With the results obtained for the four Class I sources in our sample, we can verify what is found in SVS13-A. Table \ref{DRatios} reports the deuteration measurements towards our sample.

For double deuterated formaldehyde in L1489-IRS, B5-IRS1, L1455-IRS1, and L1551-IRS5, we measured the abundance ratios [D$_{2}$CO]/[H$_{2}$CO] = 
N$_{\rm D_{2}CO}$/N$_{\rm H_{2}CO}$. From the abundance ratios, we derive the deuterium fractionation as D/H $=\sqrt{[D_2CO]/[H_2CO]}$ \citep{Manigand2019}. We found D/H of $9\% - 13\%$, $\sim 24\%-55\%$, $\sim28\%-53\%$, and $\sim45\%-84\%$, respectively.

\begin{figure}
    \centering
 \includegraphics[width=9cm]{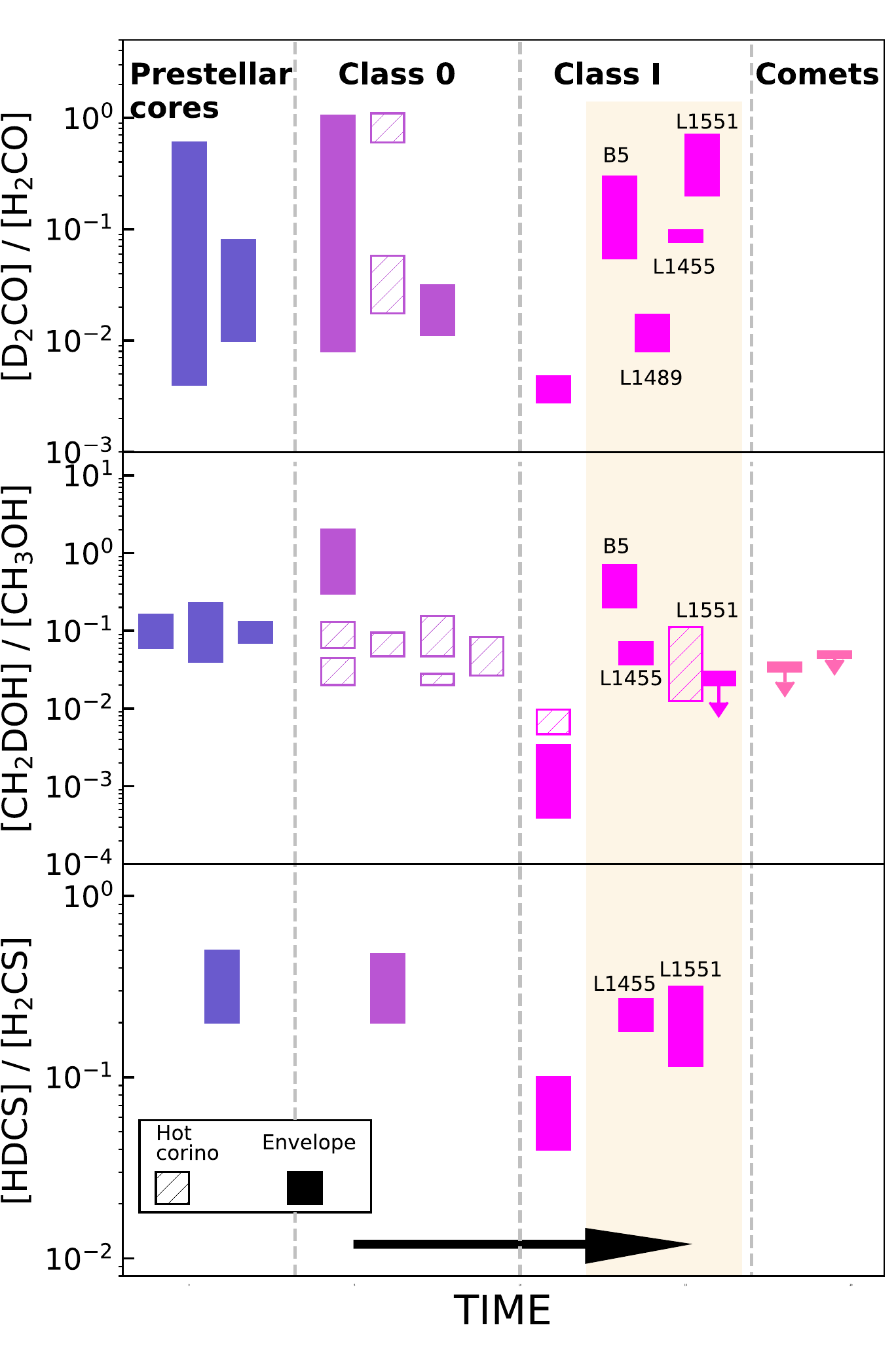}
\caption{[XD]/[XH] abundance ratios measured for organic species in sources at different evolutionary stages: prestellar cores (D$_{\rm 2}$CO \citep{Bacmann2003, Tanarro2019}, CH$_{\rm 2}$DOH \citep{Bizzocchi2014, Tanarro2019,Ambrose2021}, and HDCS \citep{Marcelino2005,Vastel2018}), Class 0 protostars (D$_{\rm 2}$CO \citep{Parise2006, Watanabe2012, Persson2018, Manigand2020}, CH$_{\rm 2}$DOH \citep{Parise2006, Bianchi2-2017, Jor2018, Jacobsen2019, vanGelder2020, Manigand2020}, and HDCS \citep{Drozdovskaya2018}, Class I protostar (SVS13-A; D$_{\rm 2}$CO, CH$_{\rm 2}$DOH \citep{Bianchi2017}, and HDCS \citep{Bianchi2-2019}), comets (Hale-Bopp; CH$_{\rm 2}$DOH \citep{Crovisier2004} 67P/C-G;  CH$_{\rm 2}$DOH \citep{Drozdovskaya2021}). The [XD]/[XH] values inferred for L1489-IRS, B5-IRS1, L1455-IRS1, L1551-IRS5 are in magenta (from this work). The dashed bars represent values inferred for the hot corinos ($r$ $\leq$ 100 au, T $\geq$ 100 K) while the filled bars indicate the measurements derived for extended envelope.
Dra}

\label{D/H}
  \end{figure}

Figure \ref{D/H} compares our results with what has been obtained for SVS13-A as well as with the deuteration levels measured, using single-dish telescopes, in Class 0 sources, and
prestellar cores \citep{Bacmann2003,Parise2006,Watanabe2012,Bianchi2017,Tanarro2019}. We also added interferometric measurements towards Class 0 sources \citep{Persson2018, Manigand2020}.
Figure \ref{D/H}  shows that, once increased the number of [D$_2$CO]/[H$_2$CO]
measurements towards Class I, there is no significant
change moving from the prestellar core to the Class I stage.

In the case of L1551-IRS5, we have emission of methanol and single deuterated methanol, CH$_2$DOH, from different regions, which are kinematically distinct in the line spectra (see Fig. \ref{L1551-3peaks}), such as the envelope and the hot corino. Thus, we estimate the abundance ratios, [CH$_2$DOH]/[CH$_3$OH], for the hot corino and the envelope separately and we indicate these different measurements in the Fig. \ref{D/H} by using different symbols. 

The corresponding D/H ratios are derived as D/H = 1/3 [CH$_2$DOH]/[CH$_3$OH] \citep{Manigand2019} and are 6\% -- 23\% (B5-IRS1), 1\% -- 2\% (L1455-IRS1), and  $\leq$ 1\% (L1551-IRS5 envelope). In addition, for the L1551-IRS5 hot corino, [CH$_2$DOH]/[CH$_3$OH] is $\sim2\%- 10\%$, which gives a D/H ratio of 0.7\% -- 3\%.
Figure \ref{D/H} reports, as done for [D$_2$CO]/[H$_2$CO], the comparison of our measurements with what is presented in literature, from prestellar cores to Class I protostars \citep[e.g.,][]{Bianchi2-2017,Jor2018, Jacobsen2019, vanGelder2020, Manigand2020}. Hot corinos are highlighted by dashed areas.
Overall, the measurements are consistent ([CH$_2$DOH]/[CH$_3$OH] $\sim 0.1$ within an order of magnitude) moving along the different evolutionary stages. An exception is provided by the SVS13-A envelope, which shows a lower D/H ratio ($\sim 10^{-3}$). 
Interestingly, upper limits on the [CH$_2$DOH]/[CH$_3$OH] ratio obtained for comets \citep[Hale-Bopp and 67P/C-G;][]{Crovisier2004, Drozdovskaya2021} suggest that the methanol deuteration is not significantly different from the previous stages (from prestellar core to Class I protostar). This could suggest that methanol formed onto dust grains can be inherited from earlier phases \citep{Drozdovskaya2021}.
A better statistics, provided by an increase of the measurements for comets, could help to test the inheritance scenario suggested by Fig. \ref{D/H}.

%%%%%%%%%%%%%%%%%%%%

\begin{table*}[ht]
\begin{centering}
\caption{Abundance ratios of the deuterated molecule over the non deuterated one and inferred D/H ratios in the observed Class I sources. }     
\label{DRatios}
{\begin{tabular}{lcccccccc}
        \hline
        \hline
             & \multicolumn{2}{c}{L1489-IRS} & \multicolumn{2}{c}{B5-IRS1} &  \multicolumn{2}{c}{L1455-IRS1} & \multicolumn{2}{c}{L1551-IRS5} \\
            
 & Abundance & D/H &  Abundance & D/H &  Abundance & D/H &  Abundance & D/H \\
 & (\%)      & (\%)&  (\%)      & (\%)&  (\%)      & (\%)&  (\%)      & (\%)\\
\hline
{\medskip}
$\rm $D$_{2}$CO/H$_{\rm 2}$CO & 0.8 -- 1.8 $^{\rm E}$ & 9 --13 & 6 -- 30 $^{\rm E}$ & 24 -- 55 & 8 -- 28 $^{\rm E}$ & 28 -- 53 & 20 -- 70 $^{\rm E}$ & 45 -- 84\\
\hline
{\medskip}
$\rm $CH$_{2}$DOH/CH$_{\rm 3}$OH & -- &  --  & 20 -- 70 $^{\rm E}$ & 6 -- 23 & 4 -- 7 $^{\rm E}$ & 1 -- 2 &   $\leq$ 3 $^{\rm E}$ & $\leq$ 1 \\
{\medskip}
 &  &   &   &  &  &  & 2 -- 10 $^{\rm H}$ & 0.7 -- 3 \\
\hline
{\medskip}
HDCS/H$_{\rm 2}$CS & --  & --  & --  & -- & 18 -- 27 $^{\rm E}$  & 9 -- 14  & 10 -- 30 $^{\rm E}$ & 5 -- 15 \\ 
\hline
\end{tabular}}
\begin{tablenotes}
\item[] E: Envelope. H: Hot corino. 
\end{tablenotes}
\end{centering}
\end{table*}
    
Finally, Fig. \ref{D/H} reports also the
[HDCS]/[H$_2$CS] abundance ratio in L1455-IRS1 and L1551-IRS5.
The D/H ratios are derived as D/H = 1/2 [HDCS]/[H$_2$CS] \citep{Manigand2019} and are 9\%--14\% (L1455-IRS1), and 5\%--15\% (L1551-IRS5).
These two measurements together with the limited
number of previous measurements so far reported by \citet{Marcelino2005, Vastel2018, Drozdovskaya2018, Bianchi2019}. It suggests 
that also for H$_2$CS the level of deuterium fractionation is
constant from the prestellar core to the Class I stage. 
A comparison between the deuteration of S-bearing molecules with that of complex organics could be instructive for astrochemical models of star forming regions \citep[][]{Taquet2019}. Indeed, thioformaldehyde is
one of the few S-bearing species imaged
in protoplanetary disks \citep{LeGal2019,Codella2020}.
\citet{Marcelino2005} modelled the Barnard 1 dark cloud using a steady-state gas-phase chemical network in environments with different level of Sulfur depletion. According to their results, [HDCS]/[H$_{2}$CS] ratio is $\sim$ 0.2, which is consistent with the ratios that we estimated towards L1455-IRS1 and L1551-IRS5.

In general, the early stages of low-mass star formation are characterized by high deuteration \citep[see e.g.][]{Parise2004,Parise2006,Aikawa2013,Caselli2012, Ceccarelli2014}.
In high-density and low-temperature prestellar cores, where CO is highly depleted, the D/H abundance in the gas phase is dramatically enhanced. In addition, the deuteration observed
at the protostellar stages, when the dust mantles
are injected back to the gas phase, can be used as a 
fossils to retrace the history of interstellar ices.
The deuteration of a molecule can
indirectly tell us which was the density (and
temperature) when this species
froze out \citep{Aikawa2013,Taquet2012,Taquet2014,Taquet2019}.
More specifically, CH$_2$DOH and D$_2$CO increase their abundance on dust grain surfaces once several mantle layers have been deposited due to freeze-out. As a consequence, the inner protostellar regions, where all the ice mantle is evaporated, is expected to be less deuterated with respect to the outer envelope region where only
the external layers is injected into the gas phase.
Figure \ref{D/H} shows that, with the exception of SVS13-A, the measurements of [CH$_2$DOH]/[CH$_3$OH] for envelopes and hot corinos is $\sim0.01-0.1$. Within one order of magnitude, these numbers are consistent with the ranges provided by \citet{Taquet2014,Taquet2019}.
Fig. \ref{D/H} shows that the 
observed [D$_2$CO]/[H$_2$CO] ratios are spread over two orders of
magnitude ($\sim$0.01 -- 1). The theoretical models by \citet{Taquet2014} predict abundance ratios from 10$^{-4}$ to 10$^{-1}$, 
a range which is too large for a proper comparison with observations.
The observations are also not fit by the gas–grain reaction network applied to a one-dimensional radiative hydrodynamic model by
\citet{Aikawa2013}, predicting a deuteration lower by about three orders of magnitude.

In conclusion, when the large sample of Class I sources is considered, the CH$_3$OH and (doubly) H$_2$CO 
deuteration observed from the prestellar core to Class I protostellar phases is roughly constant and similar to that measured in comets. This suggests that the D/H value set at the prestellar stage is inherited by the later phases of the star forming process. However recent Rosetta measurements of D/H in the organic refractory component of cometary dust particles in 67P/C-G \citep{Paquette2021} showed that D/H is 1.6 $\pm$ 0.5 $\times$ 10$^{-3}$, i.e. an order of magnitude lower than deuteration in the organics measured in gas phase towards, prestellar cores, protostars, and the cometary coma in Fig. \ref{D/H}. This questions the inheritance of D/H from the protostellar phase. Therefore, more observations of D/H in the organic refractory molecules towards comets are needed to test the chemical inheritance hypothesis.
  
\subsubsection{Chemical Richness} \label{evolution}

As reported in Sect. \ref{Intro}, Class I sources are expected to play a key role in understanding if and how
the chemical composition is transmitted from the earliest stages of a Sun-like forming star, i.e. from the parental cloud (prestellar cores) to the small-bodies in the Solar System (such as comets). 
To address this ambitious goal, we compare the abundance ratios of iCOMs, as well as the abundance ratios of simpler molecules ([CH$_{3}$OH]/[H$_{2}$CO] and [H$_{2}$CS]/[H$_{2}$CO]) obtained for the Class I protostars in our sample with the values obtained for prestellar cores, hot corinos and envelopes around Class 0 protostars, protoplanetary disks, and comets.

 \begin{figure*}
   \centering
 \includegraphics[width=17cm]{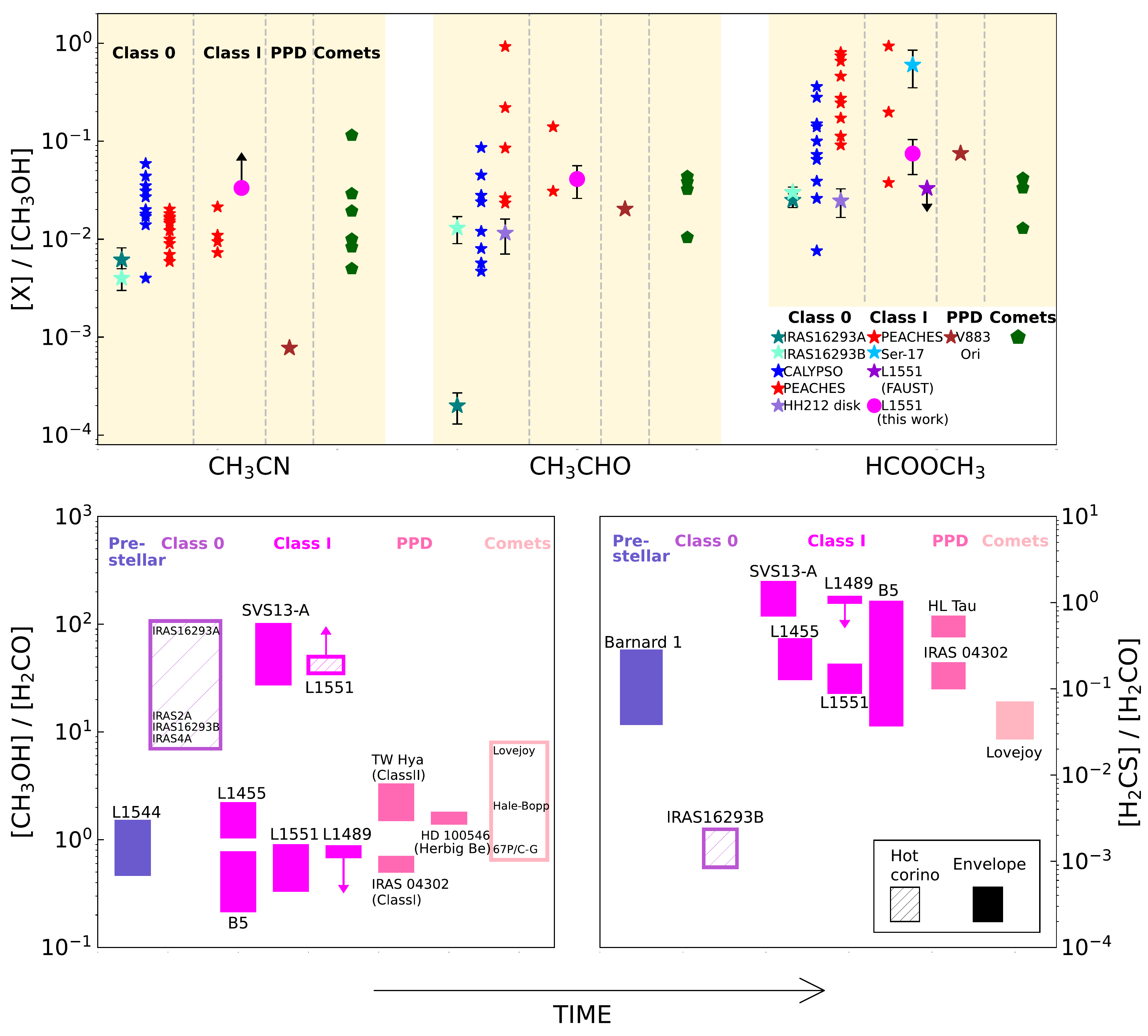}
\caption{{\it Upper panel:} Abundance ratios of iCOMs with respect to methanol as measured at different evolutionary stages. The three panels refer to CH$_3$CN, CH$_3$CHO, and HCOOCH$_3$, respectively, and each one is divided into four parts with gray dashed lines indicating Class 0 sources, Class I sources, protoplanetary disks (PPD), and comets. The abundance ratios estimated for L1551-IRS5 in this work are indicated by a magenta circle. The abundance ratios for the other sources are taken from the literature (references in the text) and labelled in the legend in the bottom right corner. {\it Bottom panels:} Abundance ratios of [CH$_{3}$OH]/[H$_{2}$CO] (Left) and [H$_{2}$CS]/[H$_{2}$CO] (Right) for our Class I sources (magenta bars) are compared with the values of  prestellar cores, Class 0 sources, PPDs, and comets from the literature, as labeled. 
The dashed bar in the left panel indicates hot corino measurements, while the dashed bar in the right panel indicates the measurement obtained in the inner 60 au towards the IRAS16293B protostar. The references of the measurements from literature are reported in Sect. \ref{evolution}.}
\label{icomsAbundance}
 \end{figure*}

In the first panel of Fig. \ref{icomsAbundance}, we show the ratios of the iCOMs CH$_{3}$CN (as a lower limit), CH$_{3}$CHO, and HCOOCH$_{3}$ with respect to CH$_{3}$OH for the L1551-IRS5 hot corino (magenta circle).
To estimate these abundance ratios, we use the column densities of CH$_{3}$CN and CH$_{3}$OH obtained from the LVG analysis which takes into account the source size. For CH$_{3}$CHO and HCOOCH$_{3}$, we assume a source size of 0$\farcs$18 (i.e. the source size of CH$_{3}$OH) and correct for the beam filling factor.
The [CH$_{3}$CN]/[CH$_{3}$OH], [CH$_{3}$CHO]/[CH$_{3}$OH], and [HCOOCH$_{3}$]/[CH$_{3}$OH] abundance ratios are $\geq$0.02, 0.04 $\pm$ 0.015, and 0.07 $\pm$ 0.01, respectively. 
Our measurements are compared 
with those derived towards chemically rich hot corinos in the literature.
We consider only interferometric measurements in order to minimize effects due to beam dilution.
The Class 0 sources are: IRAS16293-2422A and B \citep[PILS ALMA survey;][]{Jorgensen2016, Jor2018, Calcutt2018, Manigand2020}, HH212 \citep{Lee2019}, the CALYPSO IRAM PdBI sources \citep{Belloche2020}, and the PEACHES ALMA sources \citep{Yang2021}. For the Class I sources we report: Ser-17 \citep{Bergner2019}, L1551-IRS5 \citep[FAUST ALMA][]{Bianchi2020}, and the PEACHES ALMA sources \citep{Yang2021}.
In addition, we plot the values obtained for the V883 Ori
protoplanetary disk \citep{LeeJeong2019}, as well as what is measured towards several comets, namely: Hale-Bopp, Hyakutake, LINEAR, Lovejoy, 73P/SW3/B, 73P/SW3/C, 67P summer hemisphere, and 67P winter hemisphere \citep{LeRoy2015}. 

Figure \ref{icomsAbundance} shows that the abundance ratios between CH$_3$CN, CH$_{3}$CHO and HCOOCH$_{3}$ with respect to CH$_3$OH in L1551-IRS5 
are in agreement (within an order of magnitude) with those measured in 
Class 0 and other Class I hot corinos. In addition, Fig. \ref{icomsAbundance} shows that similar values are derived for the V883 Ori protoplanetary disk and for comets. 
The only exception is given by the [CH$_{3}$CN]/[CH$_{3}$OH] ratio in V883 Ori, which is 
roughly one order of magnitude lower than what is shown by the rest of the sample.
To conclude, the overall suggestion provided by Fig. \ref{icomsAbundance} is that the 
molecular complexity characterizing an evolutionary stage of low-mass star formation could 
be inherited from the previous, earlier phase. Clearly, in order to verify this picture, the statistics needs to be improved, in particular, that about the more evolved Class II disks.

In addition to iCOMs, Fig. \ref{icomsAbundance} also shows the [CH$_{3}$OH]/[H$_{2}$CO] and [H$_{2}$CS]/[H$_{2}$CO] abundance ratios derived for our Class I sources by considering the emission from both the hot corinos and the envelopes.
These abundance ratios are compared with those obtained for: (i)  the prestellar cores L1544 \citep{Tanarro2019} and Barnard 1 \citep{Marcelino2005}; (ii) the Class 0 IRAS16293-2422A and B \citep{Jor2018,Persson2018, Manigand2020, Drozdovskaya2018}, NGC1333-IRAS 4A and IRAS 2A \citep{Taquet2015}; (iii) the Class I SVS13-A \citep{Bianchi2017, Bianchi2019}; (iv) the protoplanetary disks IRAS 04302+2247, HL Tau, TW Hya, and HD 100546 \citep{Walsh2016,Carney2019,Codella2020,Podio2020, Garufi2021,Booth2021}; and (v) the comets Lovejoy, Hale-Bopp and 67P/C-G \citep{Biver2015,Rubin2019}. 
The range of the [CH$_{3}$OH]/[H$_{2}$CO] abundance ratios towards L1489-IRS, B5-IRS1, L1455-IRS1 and L1551-IRS5 are 0.4 -- 0.9, 0.2 -- 0.7, 1.0 -- 2.2, 0.3 -- 0.9 respectively. 
These ratios (derived towards extended envelope) are consistent with what measured in protoplanetary disks and comets as well as in prestellar cores (Fig. \ref{icomsAbundance}, Bottom-Left) with the exception of the envelope of SVS13-A.

On the other hand, the abundance ratios for the four hot corinos derived from interferometric
observations of Class 0 protostars (dashed box) is up to 2 orders of magnitude higher. 
From our survey we derive for the L1551-IRS5 hot corino a lower limit on
the [CH$_{3}$OH]/[H$_{2}$CO] ratio which is in agreement with the measures
derived for Class 0 hot corinos. Thus, [CH$_{3}$OH]/[H$_{2}$CO] in the L1551-IRS5 envelope (0.3 -- 0.9)
is lower than that in the L1551-IRS5 hot corino ($\geq$30).
This effect plausibly reflects the methanol release from icy dust into the gas phase due to thermal evaporation, richer in methanol with respect to the envelope and/or prestellar objects where CH$_3$OH is produced due to non-thermal processes. 

Finally, Fig. \ref{icomsAbundance} (Bottom-Right) shows the H$_{2}$CS over H$_{2}$CO ratio. The reason of measuring this ratio is that if both species are formed in the gas-phase by reactions of atomic O or S with the methyl group CH$_3$ (as in the molecular layer of disks \citep[e.g.][]{Fedele2020}), then their abundance ratio can be used to estimate the elemental S/O abundance ratio. 
Figure \ref{icomsAbundance} shows that the prestellar core
Barnard 1 and the envelopes around Class I objects are associated with a quite similar [H$_{2}$CS]/[H$_{2}$CO] ratio, between few 10$^{-2}$ and $\sim$ 1.
Also the ratios for protoplanetary disks and comets fall in the same range.
Conversely, the [H$_{2}$CS]/[H$_{2}$CO] as measured in the inner 60 au towards the
Class 0 IRAS16293B is definitely lower (see dashed box; $\sim$ 10$^{-3}$, \citealt{Persson2018,Drozdovskaya2018}). 
Only further measurements on Solar System scales
will benchmark if this difference reflects a real chemical difference with respect
to what is measured in the more extended envelope.

%%%%%%%%%%%%%%%%%
%%%% 
%%%% CONCLUSIONS
%%%%
%%%%%%%%%%%%%%%%%
\section{Summary}
\label{summary}

We have presented a chemical survey of four Class I protostars (L1489-IRS, B5-IRS1, L1455-IRS1,
and L1551-IRS5) using IRAM-30m single-dish observations at 1.3 mm.  The main conclusions are summarized as follows.

\begin{itemize}
 
\item We  detected 157 lines due to 27 species from simple diatomic and triatomic molecules to iCOMs. Namely:
C-chains (c-C$_{3}$H, c-C$_{3}$H$_{2}$, and CH$_{3}$CCH), iCOMs (CH$_{3}$OH, CH$_{3}$CN, CH$_{3}$CHO, and HCOOCH$_{3}$), N-bearing species ($^{13}$CN, C$^{15}$N, and HNCO), S-bearing molecules (SO, $^{34}$SO, SO$_{2}$, $^{13}$CS, OCS, O$^{13}$CS, CCS, H$_{2}$S, H$_{2}$CS, and H$_{2}$C$^{33}$S), SiO, and deuterated molecules  (DCO$^{+}$, N$_{2}$D$^{+}$, CCD, DCN, HDCS, D$_{2}$CO, and CH$_{2}$DOH). More specifically we detected: 17 transitions due to 10 species in L1489-IRS, 29 transitions due to 15 species in B5-IRS1, and 36 transitions due to 21 species in L1455-IRS1. L1551-IRS5 is associated with the highest number of molecules with 75 transitions from 27 species.

\item
We used the standard rotational diagram approach in order to determine the rotational temperatures and to derive the column densities. When applicable, we corrected the obtained column densities
from the opacity derived using the emission of rarer isotopologues.
The rotational temperature for diatomic and triatomic species in L1489-IRS, B5-IRS1, and L1455-IRS1 is low, being  $\leq$20 K suggesting an origin in the envelopes of these sources.
Conversely, L1551-IRS5 shows iCOMs emission which traces the inner protostellar region with $T_{\rm rot}$, up to $\sim$150 K. The beam averaged column densities are between 10$^{11}$ cm$^{-2}$ and 10$^{13}$ cm$^{-2}$
for all the species but SO and CH$_{3}$OH for which we find $N_{\rm tot}$ of a few 10$^{14}$ cm$^{-2}$.

 \item 
 We observed different line profiles: (i) narrow ($\sim$1 km s$^{-1}$)
 Gaussians centered at systemic velocities, (ii) broader (3 -- 4 km s$^{-1}$)
 lines also at systemic velocities, and (iii) line wings extending up to $\sim$30 km s$^{-1}$ with respect to systemic velocity. The latter have been observed in CO, SO, and H$_2$CO and signatures of the molecular protostellar outflows.
 Narrow profiles are expected to trace extended emission, plausibly the 
 molecular envelope. The LVG analysis of the narrow
 c-C$_2$H$_3$ line emission indicates temperatures in the 5 -- 25 K range, densities larger than 10$^3$ cm$^{-3}$, and sizes between 2$\arcsec$ and 10$\arcsec$. On the other
 hand, all iCOMs show broad emission in L1551-IRS5: the LVG analysis of CH$_3$OH and CH$_3$CN emission points to
 kinetic temperatures larger than 50 K up to 135 K,
 densities larger than 3 $\times$ 10$^5$ cm$^{-3}$,
 and a size of 0$\farcs$11 -- 0$\farcs$18 ($\sim$ 16 -- 25 au), confirming
 the presence of a hot corino.
 
\item 
The CH$_3$OH 4$_{\rm 2,3}$--3$_{\rm 1,2}$E line ($E_{\rm up}$ = 45 K) shows narrow ($\sim$1 km s$^{-1}$) line emission, and has the same profile
of the envelope tracer c-C$_{3}$H$_{2}$. The emission is expected to trace the external layers of the envelope or the outflow cavities hit by UV radiations.

\item We suggest that H$_{2}$S and OCS trace hot corinos as well as circumstellar disks in previously confirmed binary systems, L1455-IRS1 and L1551-IRS5. To confirm this hypothesis, it is needed to observe these S-bearing species with high resolution images
in several binaries.

\item We derived the elemental deuterium fractionation through the following
abundance ratios: D$_{2}$CO/H$_{2}$CO, HDCS/H$_{2}$CS, and CH$_{2}$DOH/CH$_{3}$OH. The derived D/H ratios in the envelope are:
10\%--70\% (D$_{2}$CO/H$_{2}$CO), 5\%--15\% (HDCS/H$_{2}$CS), and 1\%--23\% (CH$_{2}$DOH/CH$_{3}$OH). For the L1551-IRS5 hot corino we
derive for deuterated methanol D/H $\sim$ 0.7\%--3\%.
These measurements are similar to what reported 
for earlier star forming stages, supporting the inheritance scenario
from prestellar cores to Class I sources.

\item We computed the column density ratios between three iCOMs (CH$_{3}$CN, CH$_{3}$CHO and HCOOCH$_{3}$) and  CH$_{3}$OH in the L1551-IRS5 hot corino, which are 
between 0.04 and 0.07. The comparison of these values with what is measured in
Class 0, other Class I sources, protoplanetary disks, and comets suggests that the iCOMs abundance ratios
are constant during the formation process of a Sun-like star.
This supports the hypothesis that the chemical complexity at the planet forming stage is inherited from
the earlier star formation stages.

\item
The [CH$_3$OH]/[H$_2$CO] abundance ratio in L1551-IRS5 is higher in the hot corino than in the envelope, in agreement with 
an increase of the methanol abundance due to thermal evaporation of the dust mantles
in the hot corino.

\end{itemize}

The results obtained with our IRAM-30m 1.3 mm observations provide crucial informations on the chemical richness of Class I sources and on the chemical evolutionary path along the low-mass star forming process. More observations of the different evolutionary stages are needed to better understand the evolution of the abundance ratios. In particular, higher spatial resolution observations and images are needed to disentangle the complexity of each components around protostars. In this context, NOEMA Large Program SOLIS \citep[Seeds Of Life In Space,][]{Ceccarelli2017},
the ALMA Large Program FAUST   \footnote{\label{note1}\url{http://stars.riken.jp/faust/fausthome.html}}\citep[Fifty AU Study of the Chemistry in the Disk/Envelope System of Solar-like Protostars,][]{Bianchi2020},
as well as the ALMA project PEACHES 
\citep[The Perseus ALMA Chemistry Survey,][]{Yang2021} are expected to provide much more information on the molecular complexity for larger samples of Solar-like protostars.

\begin{acknowledgements}
We thank the IRAM-30m staff for its
valuable help during the observations. This project has received funding from the European Union’s Horizon 2020 research and innovation
programme under the Marie Skłodowska-Curie grant agreement No 811312 for the project "Astro-Chemical Origins” (ACO) and from the European Research Council (ERC) under the European
Union's Horizon 2020 research and innovation programme, for the Project “The Dawn of Organic
Chemistry” (DOC), grant agreement No 741002. We also thank Paola Caselli and Luca Dore for fruitful discussion regarding the hyperfine components.
\end{acknowledgements}

\bibliographystyle{aa} % style aa.bst
%\bibliography{./References.bib} 
\bibliography{30m_ClassI_Mercimek.bib}

%%%%%%%%%%%%%%%%%%%%%
%%% Appendix      %%
%%%%%%%%%%%%%%%%%%%%%
\appendix

\section{Detected lines and observed spectra}

\begin{figure*}
  \centering
 \includegraphics[width=18cm]{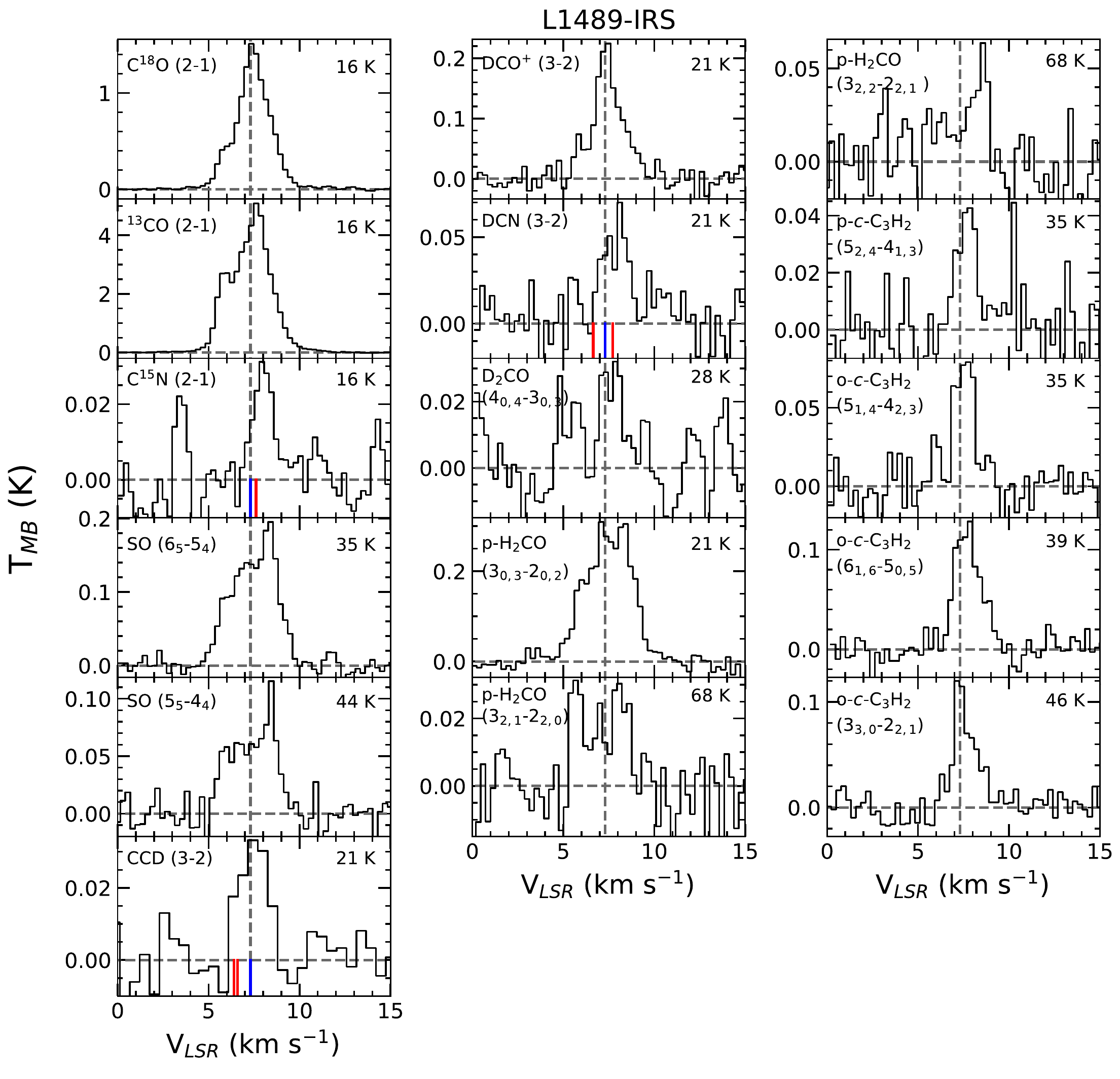}

\caption{Spectra observed towards L1489-IRS (see Table \ref{L1489-list} in $T_{\rm MB}$
scale). Species, transition, and upper level energy are reported in each panel. The vertical dashed line stands for the source systemic velocity as measured using the C$^{18}$O(2--1) line (+7.3 km s$^{-1}$). In the CCD(3--2) spectrum, the velocity scale is centered on the brightest (3 D$^{2}$) hyperfine component (vertical blue line) at 216372.8 MHz; the vertical red lines indicate the offset in velocity of the fainter components (1.9 D$^{2}$ at 216373.3 MHz; 1.4 D$^{2}$ at 216373.2 MHz. This spectrum is smoothed to 0.40 MHz. In C$^{15}$N(2--1) spectrum (in the lower panel) the velocity scale is centered on the brightest (6 D$^{2}$) hyperfine component (vertical blue line) at 219934.8 MHz; the vertical red line indicates the offset in velocity of the fainter components (4 D$^{2}$ at 219934.0 MHz). In DCN(3--2) spectrum the velocity scale is centered on the brightest (35 D$^{2}$) hyperfine component (vertical blue line) at 217238.3 MHz; the vertical red lines indicate the offset in velocity of the fainter components (20 D$^{2}$ at 217238.6 MHz; 16 D$^{2}$ at 217238.3 MHz). 
} 
\label{L1489_Spectra}
 \end{figure*}

In this section, we present the detected molecules with the spectra for each protostar in Figure \ref{L1489_Spectra} -- \ref{L1551_Spectra}.
In Figure \ref{CO1} and \ref{CO2}, we present the observed spectra of CO and its isotopologues with their isotopic ratios. We also report, for each source and for each detected line,  the spectral line parameters as well as the results of the spectral fit, namely: frequency (MHz), telescope $HPBW$ $(\arcsec)$, upper level energies $E_{\rm up}$ (K),  the S$\mu$ $^{2}$ product (D$^{2}$),  rms (mK),  peak temperature (mK), spectral resolution $\delta$v (km s$^{-1}$), peak velocity (km s$^{-1}$),  $FWHM$ (km s$^{-1}$), velocity integrated line intensity $I_{int}$ (mK km s$^{-1}$) from Table \ref{L1489-list} to \ref{L1551-list}.

%%%%%%%%%%%%%%%%%%%%%%%%%%%%
%% L1489-IRS SPECTRA %%%%%%%
%%%%%%%%%%%%%%%%%%%%%%%%%%%%
\newpage
 
%%%%%%%%%%%%%%%%%%%%%%%%%%%%
%% B5-IRS1 SPECTRA   %%%%%%%
%%%%%%%%%%%%%%%%%%%%%%%%%%%%

 \begin{figure*}
  \centering
 \includegraphics[width=18cm]{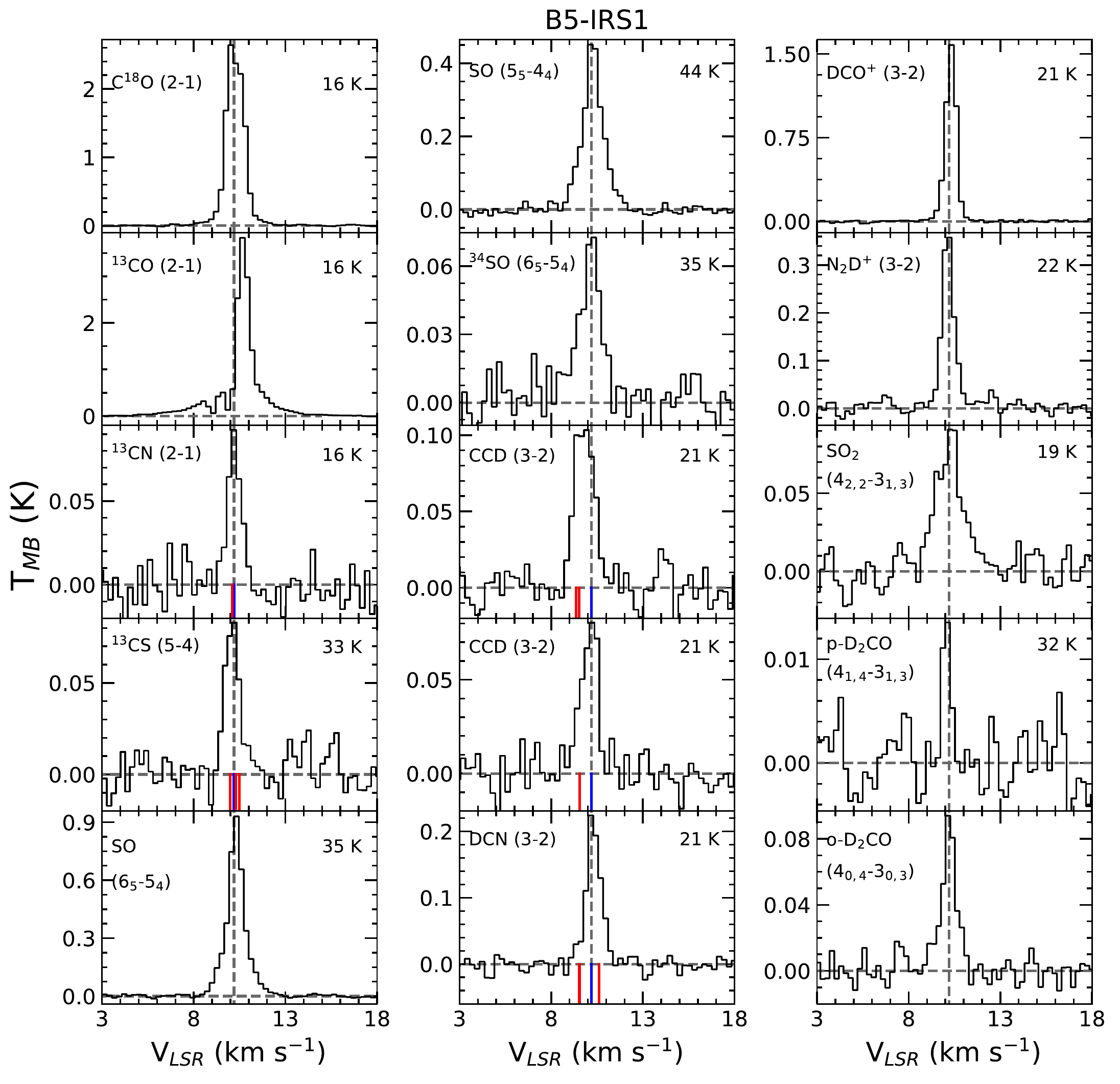}

\caption{Spectra observed towards B5-IRS1 (see Table \ref{B5-list} in $T_{\rm MB}$
scale). Species, transition, and upper level energy are reported in each panel. The vertical dashed line stands for the source systemic velocity as measured using the C$^{18}$O(2--1) line (+10.2 km s$^{-1}$). In the $^{13}$CN(2--1) spectrum, the velocity scale is centered on the brightest (8 D$^{2}$) hyperfine component (vertical blue line) at 217467.1 MHz; the vertical red line indicates the offset in velocity of the fainter component (5 D$^{2}$ at 217467.1 MHz). $^{13}$CS(5--4) spectrum, the velocity scale is centered on the brightest (38 D$^{2}$) hyperfine component (vertical blue line) at 231220.7 MHz; the vertical red lines indicate the offset in velocity of the fainter components (21 D$^{2}$ at 231220.7 MHz; 17 D$^{2}$ at 231220.6 MHz; 1 D$^{2}$ at 231220.5 MHz). In the CCD(3--2) spectrum (the middle panel of the middle column), the velocity scale is centered on the brightest (3 D$^{2}$) hyperfine component (vertical blue line) at 216372.8 MHz; the vertical red lines indicate the offset in velocity of the fainter components (1.9 D$^{2}$ at 216373.3 MHz; 1.4 D$^{2}$ at 216373.2 MHz). In lower panel, CCD(3--2) spectrum the velocity scale is centered on the brightest (2 D$^{2}$) hyperfine component (vertical blue line) at 216428.3 MHz; the vertical red line indicate the offset in velocity of the fainter components (0.7 D$^{2}$ at 216428.8 MHz). In DCN(3--2) spectrum the velocity scale is centered on the brightest (35 D$^{2}$) hyperfine component (vertical blue line) at 217238.3 MHz; the vertical red lines indicates the offset in velocity of the fainter components (20 D$^{2}$ at 217238.6 MHz; 16 D$^{2}$ at 217238.3 MHz).
} 
\label{B5_Spectra}
 \end{figure*}
  \addtocounter{figure}{-1}
\begin{figure*}
  \centering
 \includegraphics[width=18cm]{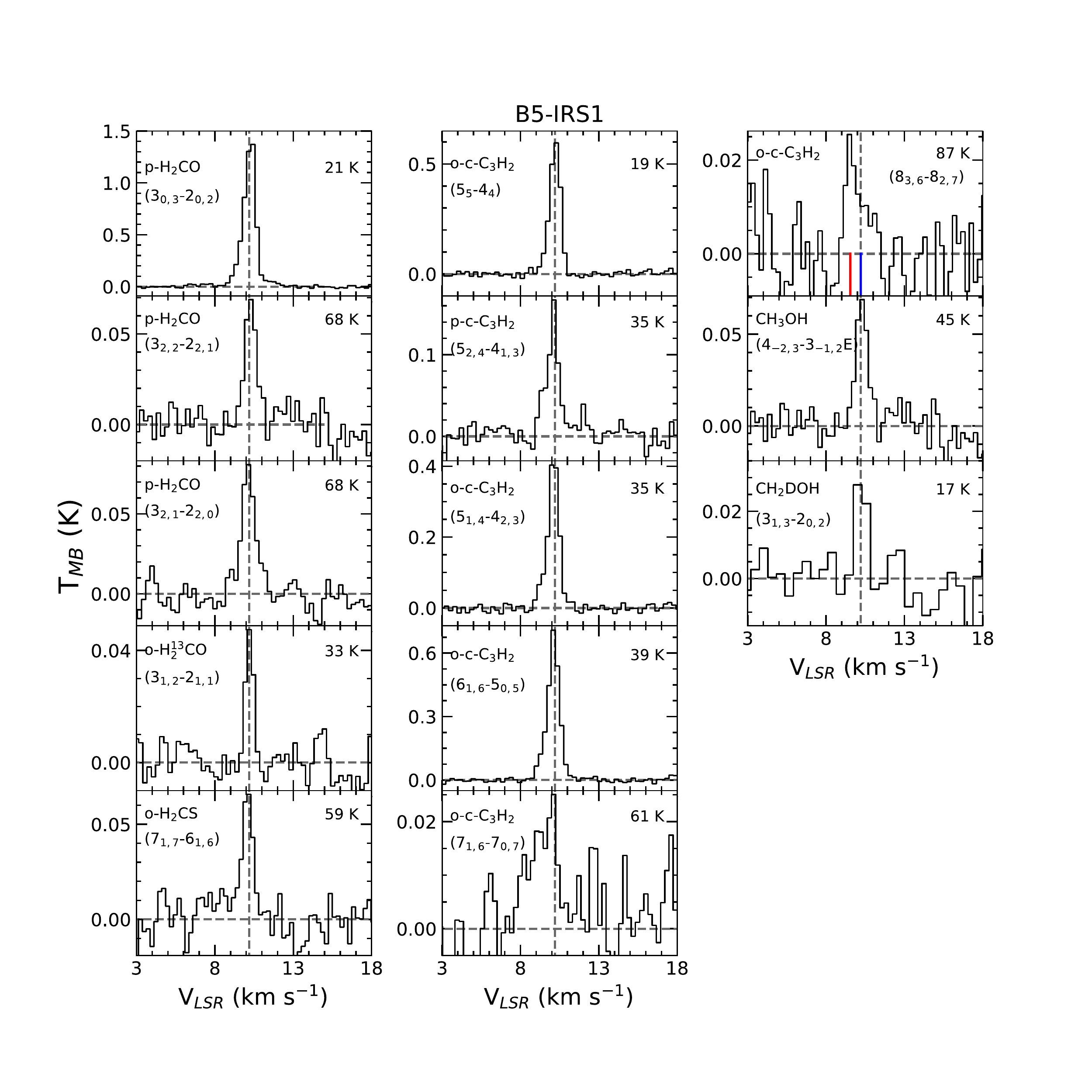}

\caption{Continued: Spectra observed towards B5-IRS1 (see Table \ref{B5-list} in $T_{\rm MB}$
scale). In o- c-C$_{3}$H$_{2}$ spectrum, the velocity scale is centered on the brightest (63 D$^{2}$) hyperfine component (vertical blue line) at 218449.4 MHz; the vertical red line indicates the offset in velocity of the fainter components (21 D$^{2}$ at 216448.8 MHz). CH$_{2}$DOH spectrum is smoothed to 0.40 MHz.
}
 \end{figure*}
 %%%%%%%%%%%%%%%%%%%%%%%%%%%%
%% L1455-IRS1 SPECTRA %%%%%%%
%%%%%%%%%%%%%%%%%%%%%%%%%%%%
 \begin{figure*}
  \centering
 \includegraphics[width=18cm]{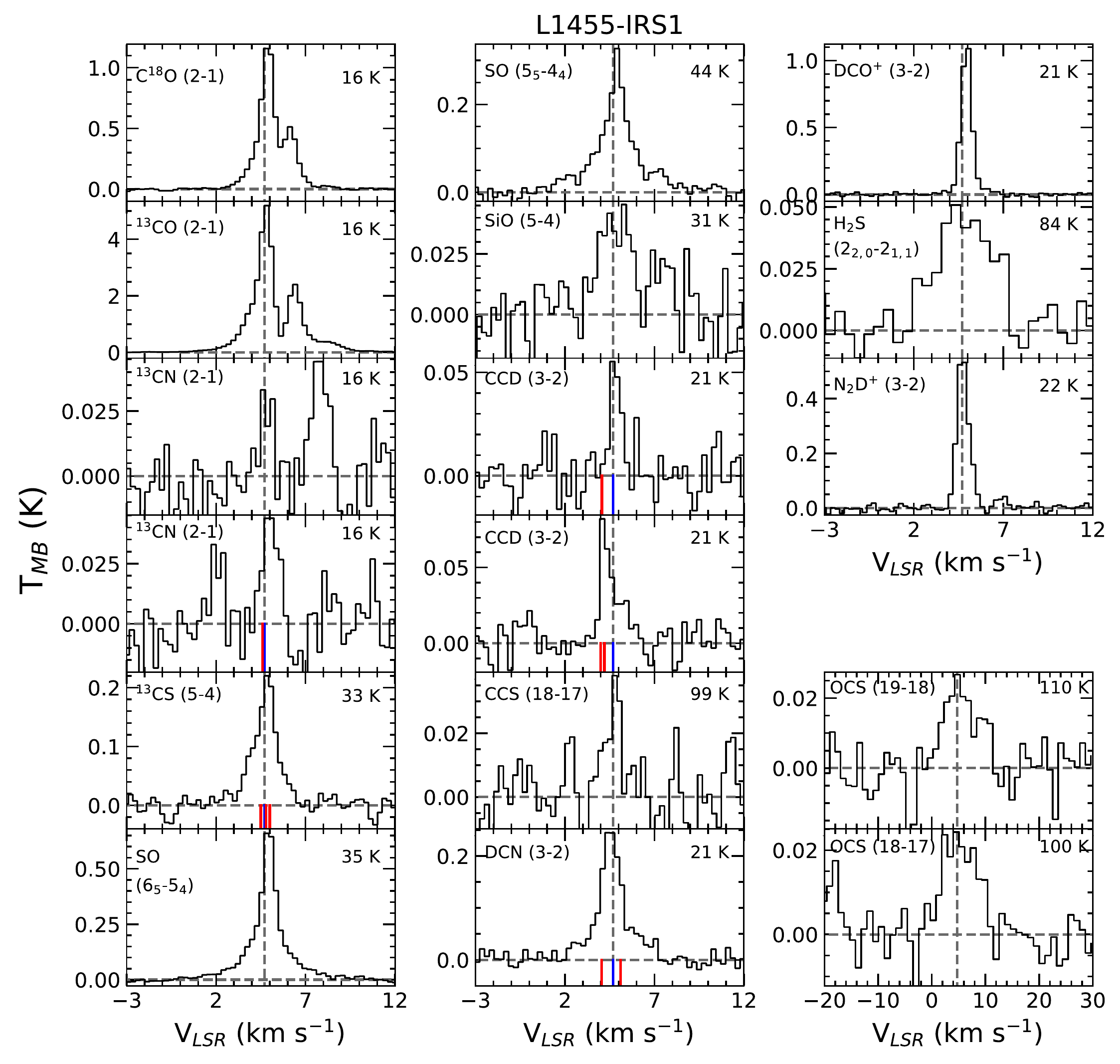}

\caption{Spectra observed towards L1455-IRS1 (see Table \ref{L1455-list} in $T_{\rm MB}$
scale). Species, transition, and upper level energy are reported in each panel. The vertical dashed line stands for the source systemic velocity as measured using the C$^{18}$O(2--1) line (+4.7 km s$^{-1}$). In the $^{13}$CN(2--1) spectrum, the velocity scale is centered on the brightest (8 D$^{2}$) hyperfine component (vertical blue line) at 217467.1 MHz; the vertical red line indicates the offset in velocity of the fainter component (5 D$^{2}$ at 217467.1 MHz). $^{13}$CS(5--4) spectrum, the velocity scale is centered on the brightest (38 D$^{2}$) hyperfine component (vertical blue line) at 231220.7 MHz; the vertical red lines indicate the offset in velocity of the fainter components (21 D$^{2}$ at 231220.7 MHz; 17 D$^{2}$ 231220.6 MHz; 1 D$^{2}$ at 231220.5 MHz). In the CCD(3--2) spectrum (4th row) the velocity scale is centered on the brightest (2 D$^{2}$) hyperfine component (vertical blue line) at 216428.3 MHz; the vertical red line indicates the offset in velocity of the fainter components (0.7 D$^{2}$ at 216428.8 MHz). In lower panel, CCD(3--2) spectrum the velocity scale is centered on the brightest
(3 D$^{2}$) hyperfine component (vertical blue line) at 216372.8 MHz; the vertical red lines indicate the offset in velocity of the fainter components (1.9 D$^{2}$ at 216373.3 MHz; 1.4 D$^{2}$ at 216373.2 MHz). In DCN(3--2) spectrum the velocity scale is centered on the brightest (35 D$^{2}$) hyperfine component (vertical blue line) at 217238.3 MHz; the vertical red lines indicate the offset in velocity of the fainter components (20 D$^{2}$ at 217238.6 MHz; 16 D$^{2}$ at 217238.3 MHz).
} 
\label{L1455_Spectra}
 \end{figure*}
  \addtocounter{figure}{-1}
  \begin{figure*}
  %\centering
 \includegraphics[width=18cm]{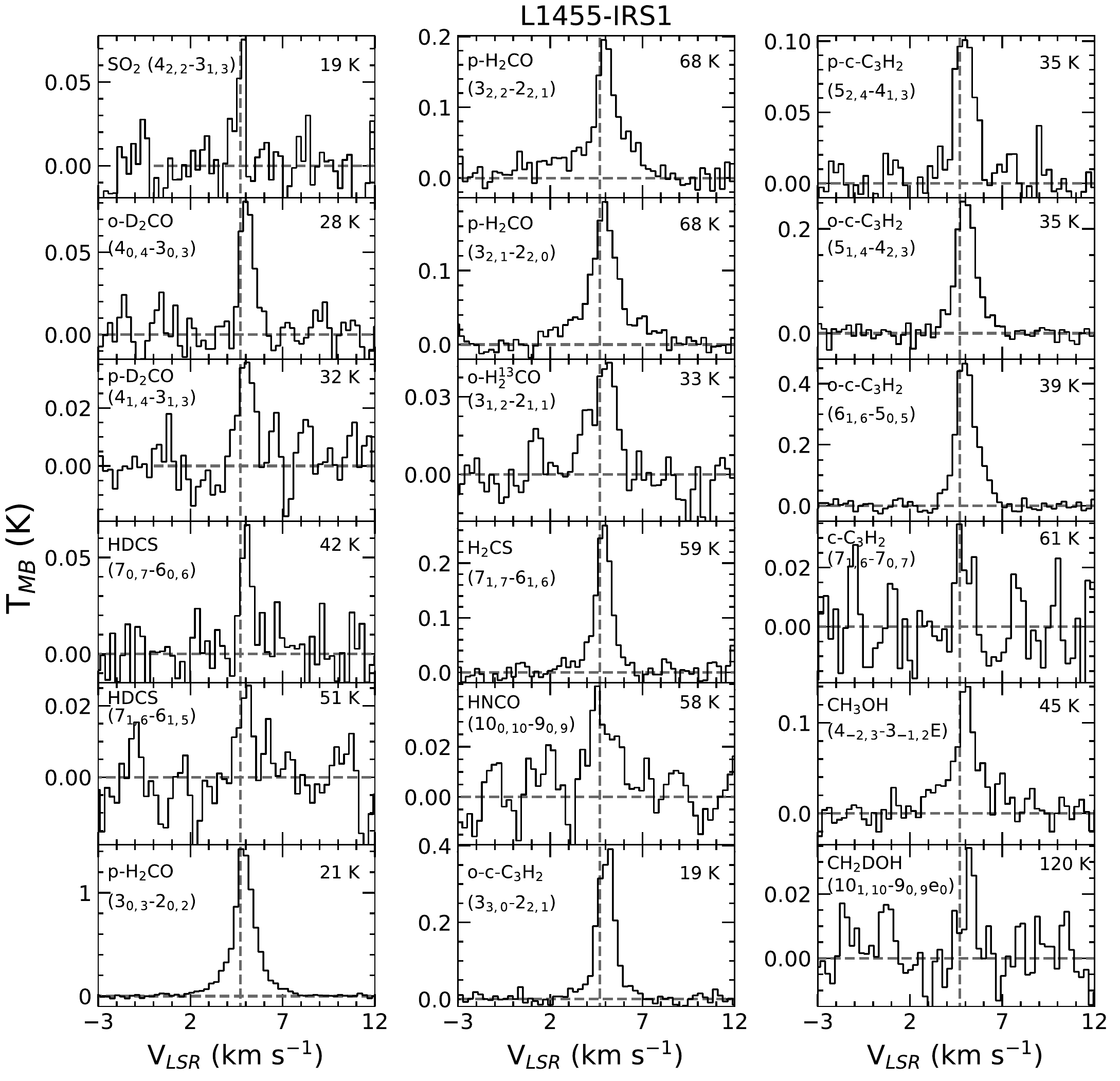}

\caption{Continued: Spectra observed towards L1455-IRS1 (see Table \ref{L1455-list} in $T_{\rm MB}$
scale).
} 
 \end{figure*}
 %%%%%%%%%%%%%%%%%%%%%%%%%%%%
%% L1551-IRS5 SPECTRA %%%%%%%
%%%%%%%%%%%%%%%%%%%%%%%%%%%%
   \begin{figure*}
  \centering
 \includegraphics[width=18cm]{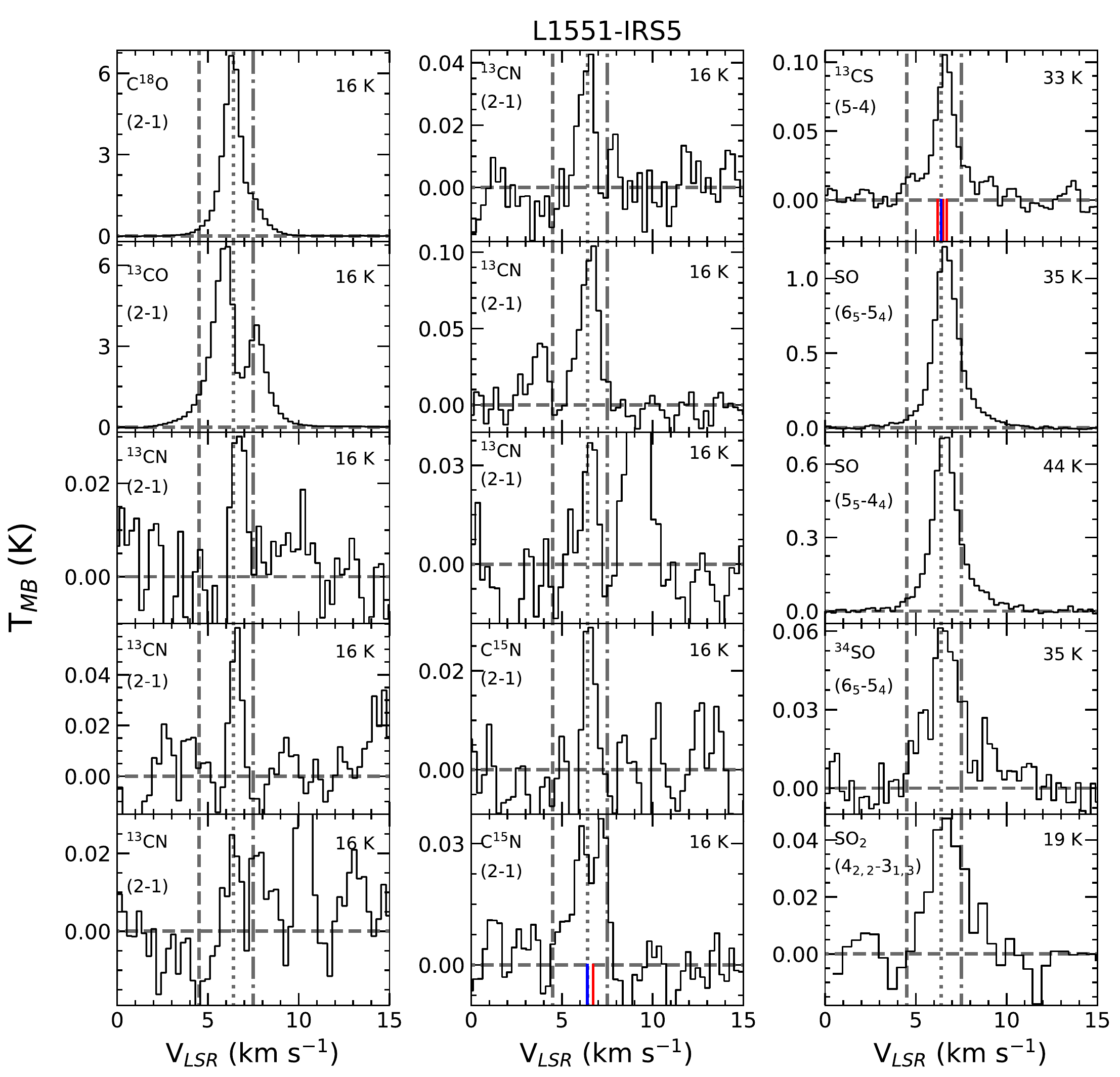}

\caption{Spectra observed towards L1551-IRS5 (see Table \ref{L1551-list} in $T_{\rm MB}$
scale). Species, transition, and upper level energy are reported in each panel. The vertical dotted line stands for the source systemic velocity as measured using the C$^{18}$O(2--1) line (+6.4 km s$^{-1}$). The vertical dashed line and the vertical dashed-dotted line stand for the systemic velocities of the binary system associated with the southern source (+4.5 km s$^{-1}$) and northern source (+7.5 km s$^{-1}$) respectively. $^{13}$CS(5--4) spectrum, the velocity scale is centered on the brightest (38 D$^{2}$) hyperfine component (vertical blue line) at 231220.7 MHz; the vertical red lines indicate the offset in velocity of the fainter components (21 D$^{2}$ at 231220.7 MHz; 17 D$^{2}$ at 231220.6 MHz; 1 D$^{2}$ at 231220.5 MHz). In C$^{15}$N(2--1) spectrum (in the lower panel) the velocity scale is centered on the brightest (6 D$^{2}$) hyperfine component (vertical blue line) at 219934.8 MHz; the vertical red line indicate the offset in velocity of the fainter components (4 D$^{2}$ at 219934.0 MHz).
 } 
\label{L1551_Spectra}
 \end{figure*}
  \addtocounter{figure}{-1}
    \begin{figure*}
  \centering
 \includegraphics[width=18cm]{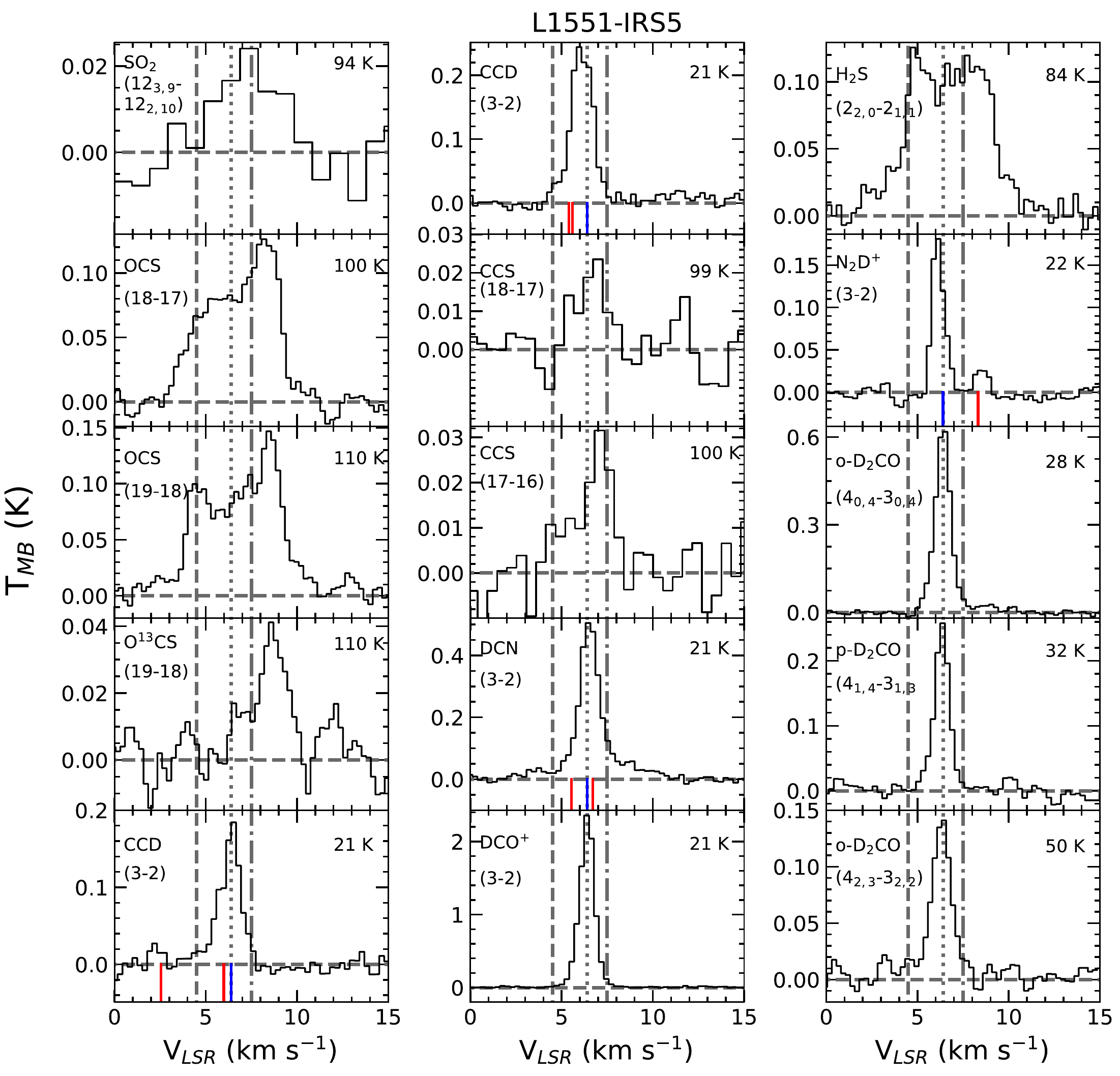}
\caption{Continued: Spectra observed towards L1551-IRS5 (see Table \ref{L1551-list} in $T_{\rm MB}$
scale). In the lower left panel of the first column, CCD(3--2) spectrum the velocity scale is centered on the brightest (2 D$^{2}$) hyperfine component (vertical blue line) at 216428.3 MHz; the vertical red lines indicate the offset in velocity of the fainter components (0.3 D$^{2}$ at 216430.3 MHz; 0.7  D$^{2}$ at 216428.8 MHz). In the upper middle panel, CCD(3--2) spectrum the velocity scale is centered on the brightest (3 D$^{2}$) hyperfine component (vertical blue line) at 216372.8 MHz; the vertical red lines indicate the offset in velocity of the fainter components (1.9 D$^{2}$ at 216373.3 MHz; 1.4 D$^{2}$ at 216373.2 MHz). In DCN(3--2) spectrum the velocity scale is centered on the brightest (35 D$^{2}$) hyperfine component (vertical blue line) at in 217238.3 MHz; the vertical red lines indicate the offset in velocity of the fainter components (20 D$^{2}$ at 217238.6 MHz; 16 D$^{2}$ at 217238.3 MHz). In N$_{2}$D$^{+}$ spectrum, the velocity scale is centered on the brightest (312 D$^{2}$) hyperfine component (vertical blue line) at 231319.9 MHz; the vertical red line indicates the offset in velocity of the brightest transition of 40 fainter components (5 D$^{2}$ at 231324.5 MHz). } 

 \end{figure*}
  \addtocounter{figure}{-1}
   \begin{figure*}
  \centering
 \includegraphics[width=18cm]{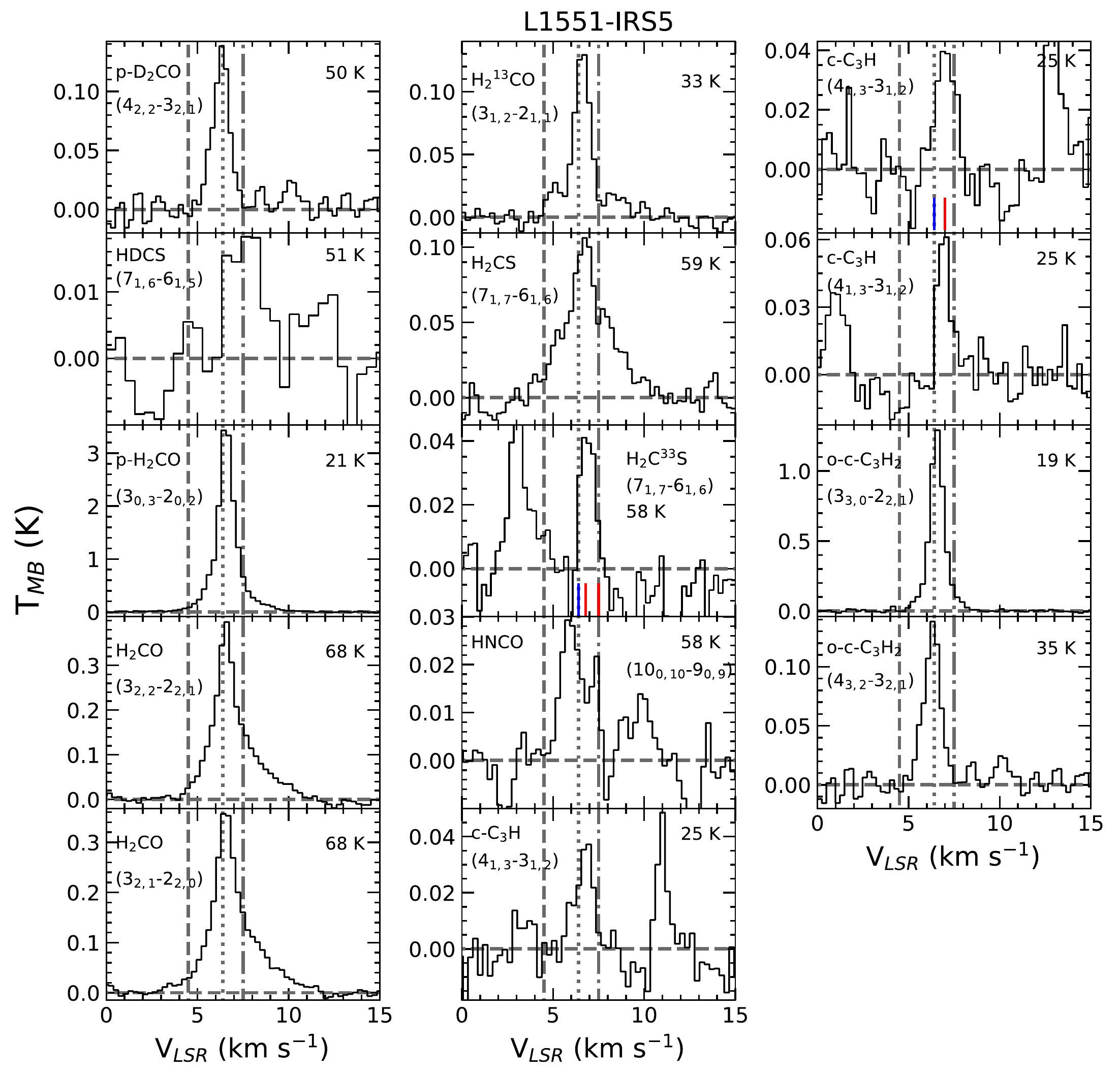}

\caption{Continued: Observed spectra observed towards L1551-IRS5 (see Table \ref{L1551-list} in $T_{\rm MB}$
scale). In c-C$_{3}$H(4$_{1,3}$--3$_{1,2}$) spectrum, the velocity scale is centered on the c-C$_{3}$H$_{2}$ emission (vertical blue line) at 216492.6 MHz; the vertical red line indicates possible contamination due to D$_{2}$CO(8$_{1,7}$--8$_{1,8}$) at 216492.4 MHz. H$_{2}$C$^{33}$S(7$_{1,7}$--6$_{1,6}$) spectrum, the velocity scale is centered on the brightest (67 D$^{2}$) hyperfine component (vertical blue line) at 234678.8 MHz; the vertical red lines indicate the offset in velocity of the fainter components (58 D$^{2}$ at 234678.7 MHz; 2 D$^{2}$ 234676.9 MHz).
} 

 \end{figure*}
 \addtocounter{figure}{-1}
 \begin{figure*}
  \centering
 \includegraphics[width=18cm]{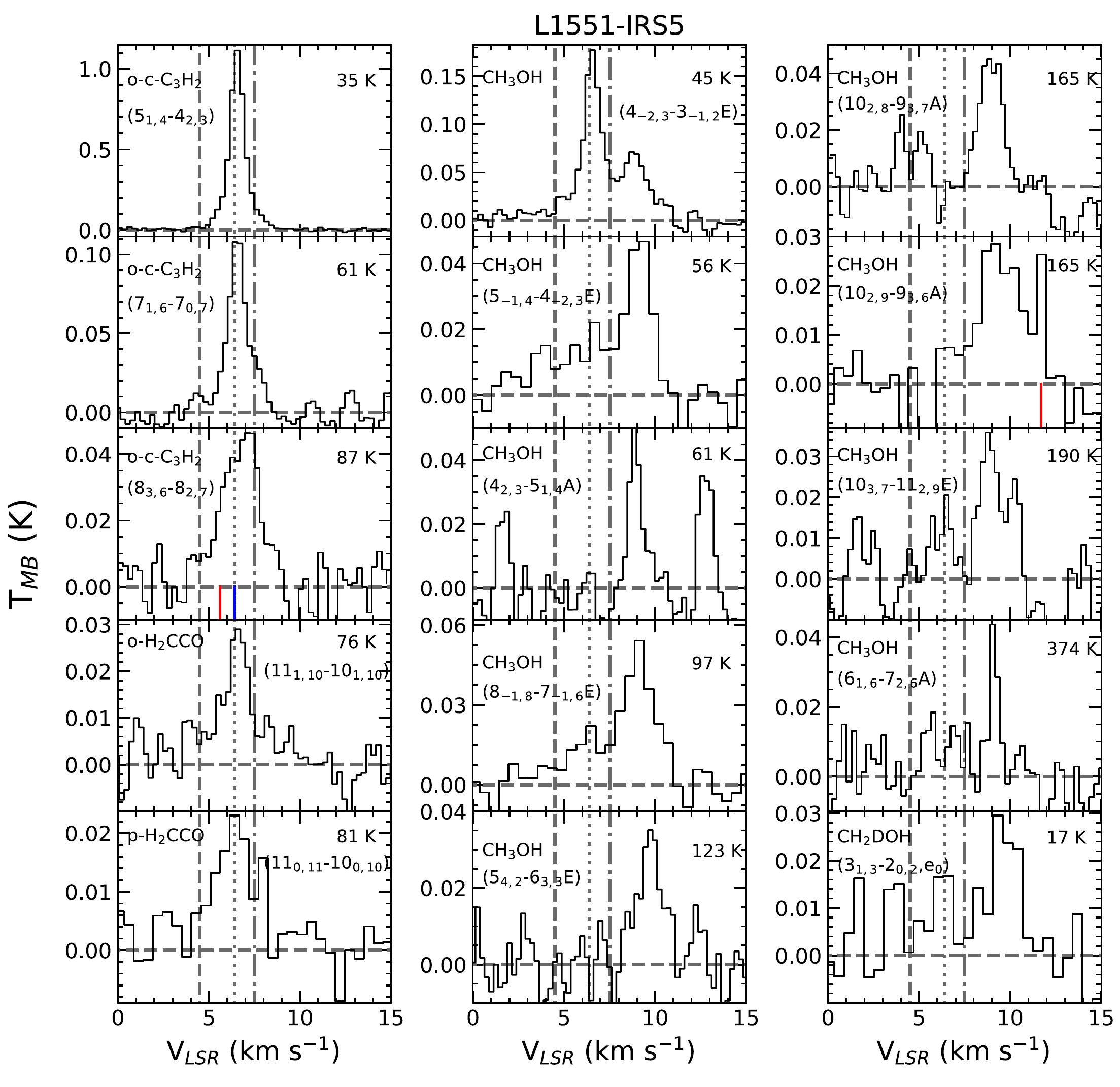}

\caption{Continued: Observed spectra of the molecules, towards L1551-IRS5, listed in Table \ref{L1551-list} (in $T_{\rm MB}$
scale). In o-c-C$_{3}$H$_{2}$ spectrum, the velocity scale is centered on the brightest (63 D$^{2}$) hyperfine component (vertical blue line) at 218449.4 MHz; the vertical red line indicate the offset in velocity of the fainter components (21 D$^{2}$ at 216448.8 MHz). In CH$_{3}$OH spectrum, the velocity scale is centered on the CH$_{3}$OH emission at 231281.1 MHz; the vertical red line indicates possible contamination due to CH$_{3}$CHO at 231278.5 MHz.
} 

 \end{figure*}
 
 \addtocounter{figure}{-1}
 \begin{figure*}
  \centering
 \includegraphics[width=18cm]{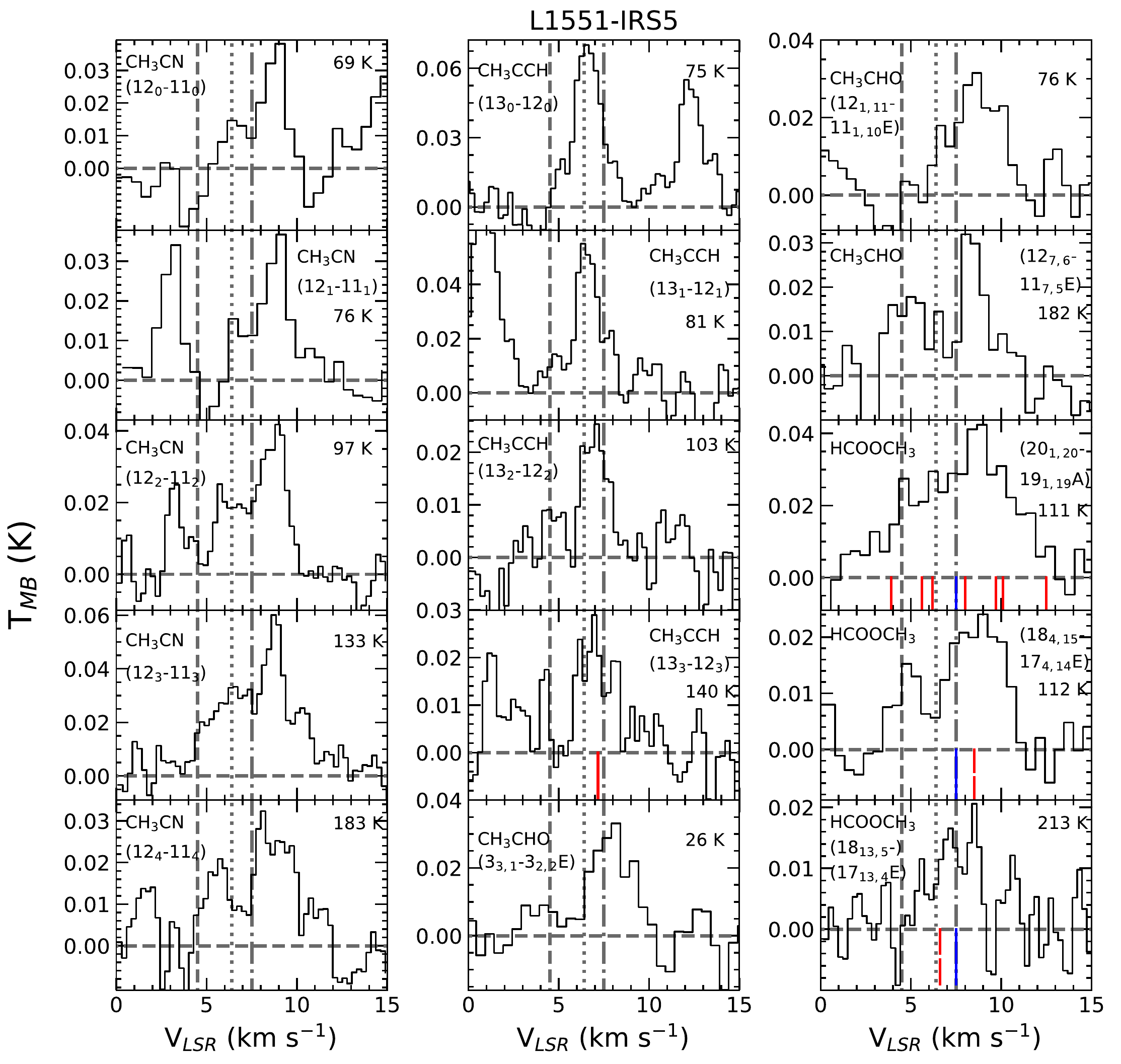}

\caption{Continued: Observed spectra of the molecules, towards L1551-IRS5, listed in Table \ref{L1551-list} (in $T_{\rm MB}$
scale). In CH$_{3}$CCH spectrum, the velocity scale is centered on the CH$_{3}$CCH emission at 222128.8 MHz; the vertical red line indicates possible contamination due to HCOOCH$_{3}$ at 222128.2 MHz. In three HCOOCH$_{3}$ spectra, the spectra are centered on the brightest component (vertical blue line) and the vertical red lines indicate the offset in velocity and strength of the fainter components reported in Table \ref{L1551-list}.
} 

 \end{figure*}
%%%%%%%%%%%%%%%%%%%%%%%%%%%%%%%%%
%% CO spectra in L1489 and B5 %%%
%%%%%%%%%%%%%%%%%%%%%%%%%%%%%%%%
 \begin{figure*}
 \centering
 \includegraphics[width=8cm]{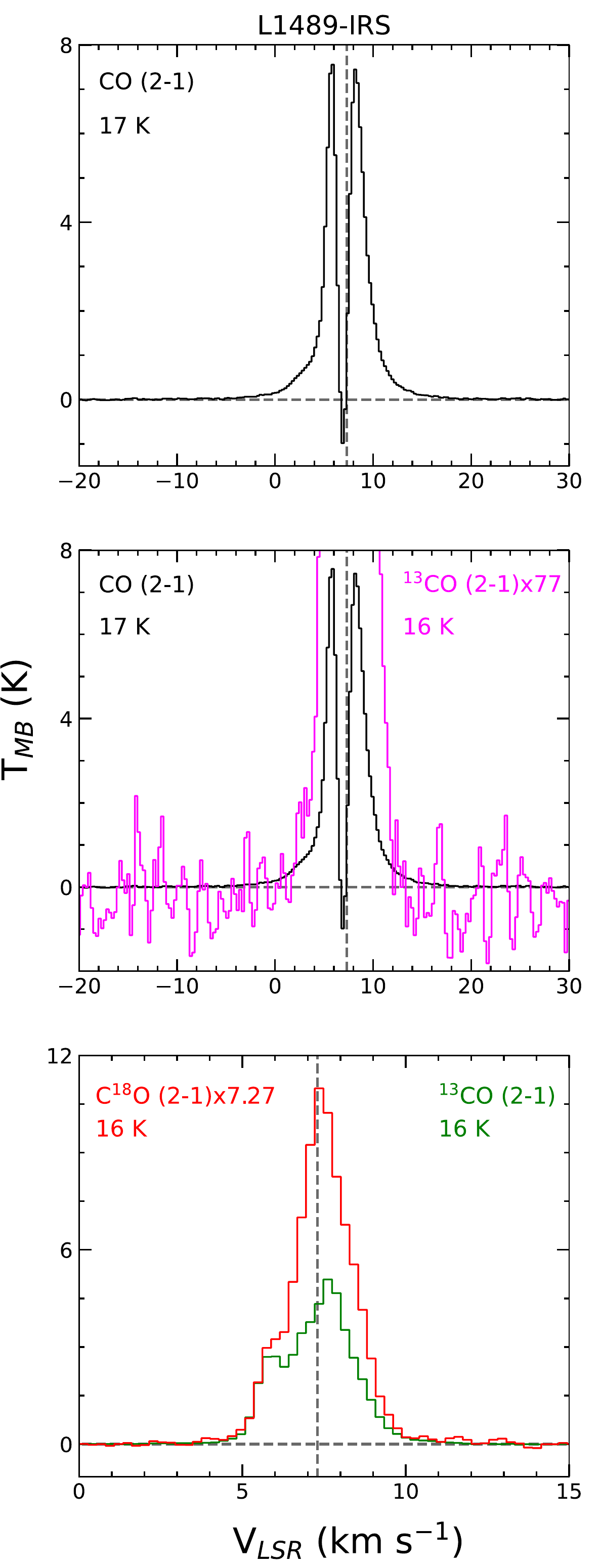}
 \includegraphics[width=8cm]{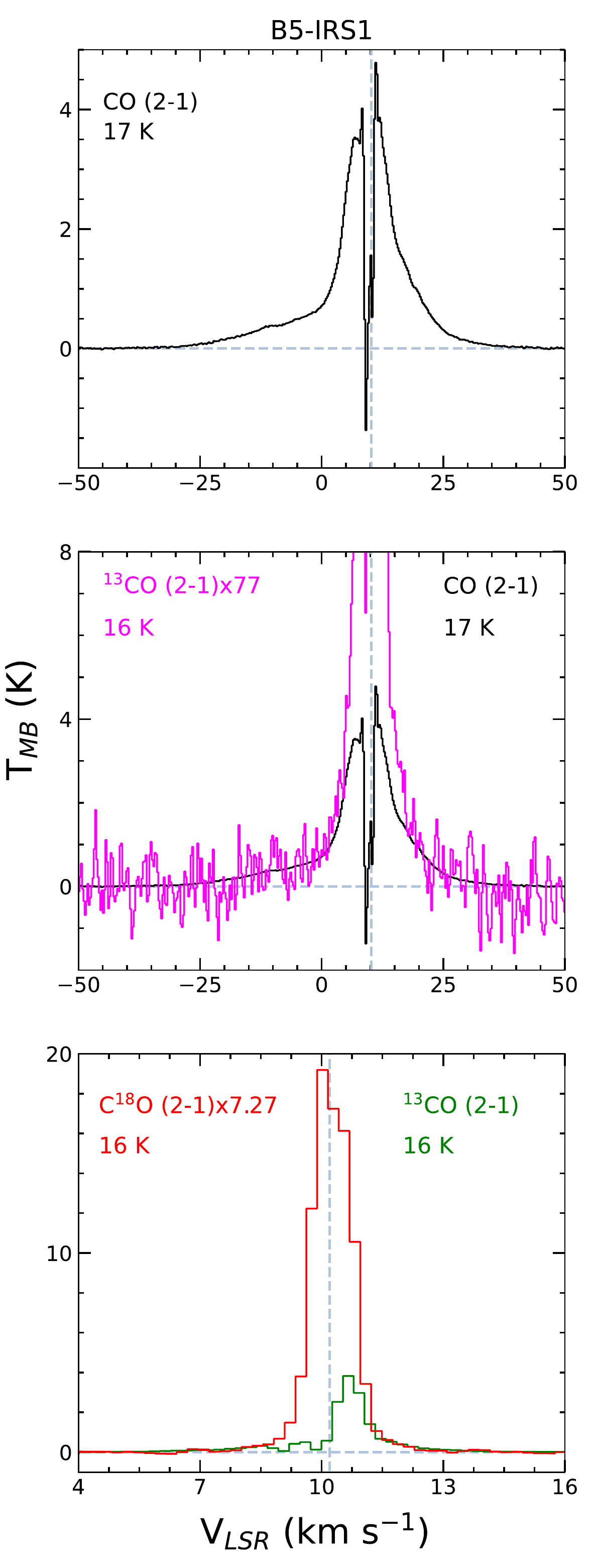}
\caption{Observed spectra of CO isotopologues (in T$_{MB}$ scale) towards L1489-IRS (left column) and B5-IRS1 (right column). Species, transition and upper level energy are reported in each panel. For panels in the left column, the vertical dashed lines stand for the source systemic velocity as measured using C$^{18}$O (2-1) emission (+7.3 km s$^{-1}$). For panels in the right column, the vertical dashed lines stand for the source systemic velocity as measured using C$^{18}$O (2-1) emission (+10.2 km s$^{-1}$).{\it Top panels:} the CO (2-1) transition. {\it Middle panels:} zoom of the CO (2-1) line wings (in black) and the $^{13}$CO (2-1) line wings (in magenta), scaled by a factor 77 in the T$_{MB}$ range from -2 K to 2.5 K. {\it Bottom panels:} $^{13}$CO (in green) and C$^{18}$O (in red) scaled by a factor 7.27 assuming isotopic ratios of $^{12}$C/$^{13}$C = 77 and $^{16}$O/$^{18}$O = 560 \citep{Milam2005}.
} 
\label{CO1}
 \end{figure*}
%%%%%%%%%%%%%%%%%%%%%%%%%%%%%%%%%%%
%% CO spectra in L1455 and L1551 %%
%%%%%%%%%%%%%%%%%%%%%%%%%%%%%%%%%%%
 \begin{figure*}
 \centering
 \includegraphics[width=8.1cm]{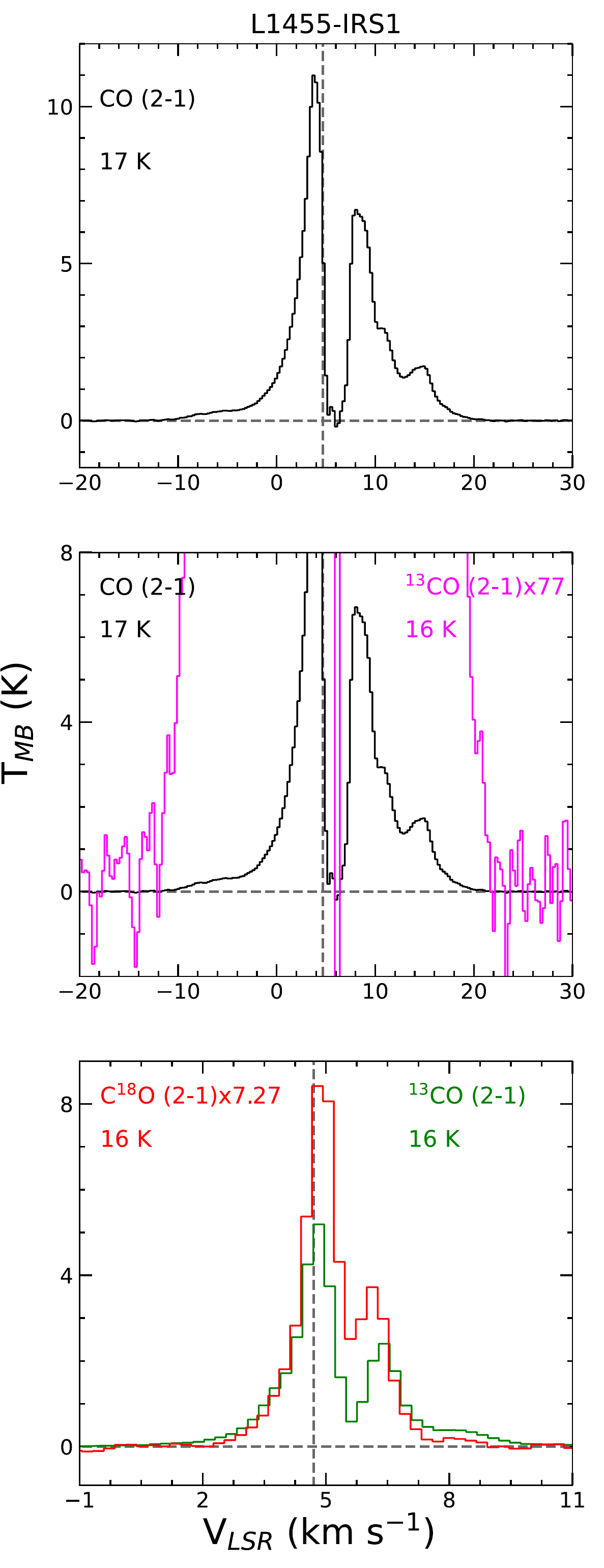}
 \includegraphics[width=8cm]{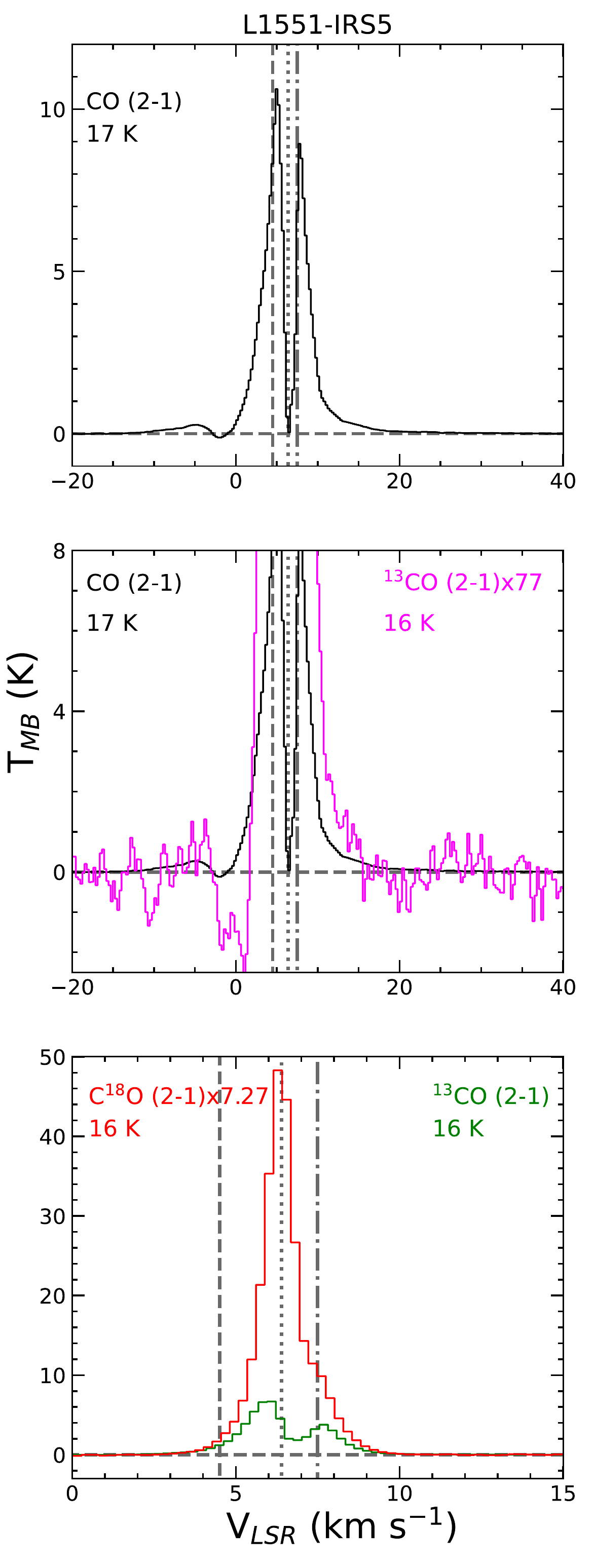}
\caption{Observed spectra of CO isotopologues (in T$_{MB}$ scale) towards L1455-IRS1 (left column) and L1551-IRS5 (right column). Species, transition and upper level energy are reported in each panel. For panels in the left column, the vertical dashed lines stand for the source systemic velocity as measured using C$^{18}$O (2-1) emission (+4.7 km s$^{-1}$). For panels in the right column, the vertical dotted line stands for the source systemic velocity as measured using the C$^{18}$O(2--1) line (+6.4 km s$^{-1}$). The vertical dashed line and the vertical dashed-dotted line stand for the systemic velocities of the binary system associated with the southern source (+4.5 km s$^{-1}$) and northern source (+7.5 km s$^{-1}$) respectively.{\it Top panels:} the CO (2-1) transition. {\it Middle panels:} zoom of the CO (2-1) line wings (in black) and the $^{13}$CO (2-1) line wings (in magenta), scaled by a factor 77 in the T$_{MB}$ range from -2 K to 2.5 K. {\it Bottom panels:} $^{13}$CO (in green) and C$^{18}$O (in red) scaled by a factor 7.27 assuming isotopic ratios of $^{12}$C/$^{13}$C = 77 and $^{16}$O/$^{18}$O = 560 \citep{Milam2005}.
} 
\label{CO2}
 \end{figure*}
 
 %%%%%%%%%%%%%%%%%%%%%%%
%% LISTS OF SPECIES %%
%%%%%%%%%%%%%%%%%%%%%%%
%\input{L1489-Table}

\newpage
%%%%%%%%%%%%%%%%%%%%%%%%
%%%%    L1489 List  %%%%
%%%%%%%%%%%%%%%%%%%%%%%%
\begin{table*}
\centering
%\tiny
%\scriptsize
%\normalsize
\caption{List of transitions and observed line properties detected towards L1489-IRS}
%{
\label{L1489-list}  
\resizebox{19cm}{!}{%
\begin{threeparttable}
 
\begin{tabular}{llccccccccccc}
\hline \hline \noalign {\medskip}

Species & Transition $^{a}$& $\nu$ $^{a}$ & $HPBW$ & $E_{\rm up}$ $^{a}$ & $S\mu^2$ $^{a}$ & rms $^{b}$ & ${\delta} V$ $ ^{b}$ & $T_{peak}$ $^{b}$ & $V_{\rm peak}$ $^{b}$ & $FWHM$ $^{b}$ & $I_{int}$ $^{b}$ & Database \\ 
& & (MHz) &$(\arcsec$)&(K) & (D$^{2}$) & (mK) &(km s$^{-1}$)& (K) & (km s$^{-1}$) & (km s$^{-1}$) &(K km s$^{-1}$) \\
\hline \noalign {\medskip}
%\hline
%\hline
{\medskip}
CO $ ^{COng}$ &   2 -- 1&  230538.0  &  11  &  17    &  0.02   &  11  &  0.25  &  -   &  -   &  -  &   $\geq${53}     &  CDMS   \\
{\medskip}
$^{13}$CO $^{COng}$   &   2 -- 1&  220398.7&  11&  16 &  0.05 &  11 &  0.25&  4.37 (0.37) &  +7.19 (0.01) &  2.7 (0.1) & 9.25 (0.01) &  CDMS \\
{\medskip}
C$^{18}$O        &  2 -- 1   &  219560.3  &  11  &  16    &  0.02   &  12   &  0.27  &  1.35 (0.07)   &  +7.32 (0.01)   &  2.2 (0.1)   &   3.09 (0.01)    &  CDMS  \\
{\medskip}
C$^{15}$N $ ^{h1}$ & 2 -- 1, J=5/2-3/2, &  219934.8  &  11  &  16  &  6 & 8  & 0.27  & 0.03 (0.01)  & +7.88 (0.11)  & 1.3 (0.3) &  0.04 (0.01) &  CDMS \\
&  F=3-2 & & & & & & & & \\
{\medskip}
SO &  5$_{5}$ -- 4$_{4}$   &  215220.6   &  11  &  44   &  12   &  13   &  0.27  & 0.12 (0.02)   &  +7.38 (0.11)   &  3.0 (0.2)  &   0.12 (0.15)  &  CDMS    \\
{\medskip}
SO   &   6$_{5}$ -- 5$_{4}$  &  219949.4  &  11  &  35   &  14    &  9   &  0.27  &  0.14 (0.02)   &  +7.45 (0.03)   &  2.9 (0.1)   &   0.52 (0.01)   &  CDMS   \\
{\medskip}
CCD $ ^{h2}$ & 3 -- 2, J=7/2-5/2,   &  216372.8  &  11  &  21   &  3   &  10   &  0.27  &  0.04 (0.01)    &  +7.24 (0.14)   &  1.2 (0.2)  &  0.06 (0.01)  &  CDMS   \\
 &   F=9/2-7/2  & & & & & & & & \\
{\medskip}
DCN $ ^{h3}$ &  3 -- 2, F=4-3   &  217238.3  &  11  &  21    &  35   &  10   &  0.27  &  0.05 (0.01)   &  +7.78 (0.13)   &  1.9 (0.5)  &  0.11 (0.06)  &  CDMS   \\
{\medskip}
DCO$^{+}$ &  3 -- 2   &  216112.6  &  11  &  21    &  142   &  22   &  0.27  &  0.17 (0.05)   &  +7.24 (0.05)   &  1.1 (0.1)   &  0.21 (0.01)  &  CDMS   \\
{\medskip}
o-D$_{2}$CO &   4$_{0,4}$ -- 3$_{0,3}$  &  231410.3  &  11  &  28   &  43   &  12   &  0.25  &   0.03 (0.01)   &  +7.25 (0.18)   &  1.4 (0.5)   &  0.01 (0.01)   &  JPL   \\
{\medskip}
p-H$_{2}$CO    &  3$_{0,3}$ -- 2$_{0,2}$   &  218222.1  &  11  &  21   &  16   &  15   &  0.27  &  0.30 (0.03)   &  +7.41 (0.03)  &  3.0 (0.1)    &   0.98 (0.02)  &  CDMS  \\ {\medskip}
p-H$_{2}$CO       &  3$_{2,2}$ -- 2$_{2,1}$  &  218475.6  &  11   &  68   &  9   &  15   &  0.27  &  0.06 (0.01)   &  +8.23 (0.14)   &   1.2 (0.3)   &   0.06 (0.01)  &  CDMS  \\
{\medskip}
p-H$_{2}$CO       &  3$_{2,1}$ -- 2$_{2,0}$  &  218760.0  &  11  &  68   &  9   &  10   &  0.27  &  0.03 (0.01)   &  +7.10 (0.28)   &  3.2 (0.5)  &  0.03 (0.01)  &  CDMS   \\
{\medskip}
o-\textit{c}-C$_{3}$H$_{2}$  &  3$_{3,0}$ -- 2$_{2,1}$   &  216278.8  &  11   &  19   &  46   &  12   &  0.27  &  0.10 (0.02)   &  +7.37 (0.06)   &  1.6 (0.4)  &  0.17 (0.01)  &  CDMS   \\
{\medskip}
o-\textit{c}-C$_{3}$H$_{2}$ &  6$_{1,6}$ -- 5$_{0,5}$   &  217822.1  &  11  &  39    &  175   &  12   &  0.27  &  0.10 (0.02)   &  +7.37 (0.06)   &   1.6 (0.4)  &  0.22 (0.01)  &  CDMS   \\ 
{\medskip}
o-\textit{c}-C$_{3}$H$_{2}$  &  5$_{1,4}$ -- 4$_{2,3}$   &  217940.0  &  11  &  35   &  110    &  15   &  0.27  &  0.08 (0.01)   &  +7.41 (0.07)   &  1.3 (0.2)  &  0.11 (0.01)   &  CDMS  \\ 
{\medskip}
p-\textit{c}-C$_{3}$H$_{2}$    &  5$_{2,4}$ -- 4$_{1,3}$   &  218160.5  &  11  &  35   &  37   &  9   &  0.27  &  0.04 (0.01)   &  +7.62 (0.12)   &  1.2 (0.2)   &   0.05 (0.01)  &  CDMS   \\

\end{tabular}
\begin{tablenotes}
\item[a] Frequencies and spectroscopic parameters have been extracted from the Jet Propulsion Laboratory (JPL) molecular database \citep{Pickett1998} and the Cologne Database for Molecular Spectroscopy \citep{Muller2005}. 
{\smallskip}
\item[b] Gaussian fit. 
{\smallskip}
\item[COng] Non-Gaussian profile. The line intensity is obtained by integrating the Zero Power Full Width ($ZPFW$) range. In particular $-$5, +20 km s$^{-1}$ for CO (2 -- 1) see in Figure \ref{CO1}. The value has to be considered a lower limit given the presence of absorption features.
{\smallskip}
 
{\smallskip}
\item[ng] Non-Gaussian profile. The line intensity is obtained by integrating the $ZPFW$ range.
{\smallskip}
\item[h1] The detected 2 -- 1, J=5/2--3/2, F=3--2 line consists of 2 hyperfine components with $S\mu^2$ $\leq$ 6 D$^2$ \citep{Muller2005} in a 0.8 MHz frequency interval. The line with the highest $S\mu^2$ line is reported (see e.g. Figures \ref{L1489_Spectra}).
{\smallskip}

\item[h2] The detected 3 -- 2, J=7/2--5/2, F=9/2--7/2 line consists of 3 hyperfine components with $S\mu^2$ $\leq$ 3 D$^2$ \citep{Muller2005} in a 0.5 MHz frequency interval. The line with the highest $S\mu^2$ line is reported (see e.g. Figures \ref{L1489_Spectra}).
{\smallskip}
\item[h3] The detected 3 -- 2, F=4--3 line consists of 3 hyperfine components with $S\mu^2$ $\leq$ 35 D$^2$ \citep{Muller2005} in a 0.3 MHz frequency interval. The line with the highest $S\mu^2$ line is reported (see e.g. Figures \ref{L1489_Spectra}).
\end{tablenotes} 
\end{threeparttable} }
\end{table*}

%%%%%%%%%%%%%%%%%%%%%%%%
%%%%    B5 List  %%%%%%
%%%%%%%%%%%%%%%%%%%%%%%%
\newpage

\begin{table*}
\centering
\caption{List of transitions and observed line properties detected towards B5-IRS1}
\label{B5-list}
\resizebox{19cm}{!}{%
 \begin{threeparttable}
\begin{tabular}{llccccccccccc}
\hline \hline \noalign {\medskip}

Species & Transition $^{a}$& $\nu$ $^{a}$ & $HPBW$ & $E_{\rm up}$ $^{a}$ & $S\mu^2$ $^{a}$ & rms $^{b}$ & ${\delta} V$ $ ^{b}$ & $T_{peak}$ $^{b}$ & $V_{\rm peak}$ $^{b}$ & $FWHM$ $^{b}$ & $I_{int}$ $^{b}$ & Database \\  
& & (MHz) &$(\arcsec$)&(K) & (D$^{2}$) & (mK) &(km s$^{-1}$)& (K) & (km s$^{-1}$) & (km s$^{-1}$) &(K km s$^{-1}$) \\
\hline \noalign {\medskip}

{\medskip}

CO $ ^{COng}$ &   2 -- 1&  230538.0  &  11  &  17    &  0.02   &  70  &  0.25  &  -   &  -   &  -  &   $\geq${53}     &  CDMS   \\
{\medskip}
$^{13}$CO $^{COng}$   &   2 -- 1&  220398.7&  11&  16 &  0.05 &  130 &  0.27&  3.77 (0.23) &  +10.58 (0.01) &  0.8 (0.1) &  4.53 (0.53) &  CDMS \\
{\medskip}
C$^{18}$O        &  2 -- 1   &  219560.3  &  11  &  16    &  0.02   &  9   &  0.27  &  2.69 (0.14)   &  +10.19 (0.01)   &  1.1 (0.1)   &   2.34 (0.01)    &  CDMS  \\
{\medskip}
$^{13}$CN $ ^{h1}$ & 2 -- 1, J=5/2-3/2,   &  217467.1  &  11  &  16    &  8   &  8   &   0.27  &  0.09 (0.05)   &  +10.18 (0.02)   &   0.9 (0.1)  &  0.06 (0.01)   &  CDMS  \\
&   F=4-3, F$_{1}$=3-2,  & & & & & & & & \\
{\medskip}
$^{13}$CS $ ^{h2}$&  5 -- 4   &  231220.7  &  11  &  33    &  38   &  6   &  0.25  &  0.08 (0.02)   &  +9.96 (0.03)   &  0.9 (0.1)  &  0.06 (0.01)  &  CDMS    \\
{\medskip}
SO &  5$_{5}$ -- 4$_{4}$   &  215220.6   &  11  &  44   &  12   &  9   &  0.27  &  0.42 (0.03)   &  +10.14 (0.01)   &  1.3 (0.1)  &   0.42 (0.01)  &  CDMS    \\
{\medskip}
SO   &   6$_{5}$ -- 5$_{4}$  &  219949.4  &  11  &  35   &  14    &  8   &  0.27  &  0.86 (0.06)   &  +10.17 (0.01)   &  1.1 (0.1)   &   0.71 (0.01)   &  CDMS   \\
{\medskip}
$^{34}$SO &  6$_{5}$ -- 5$_{4}$  &  215839.9   &  11  &  35    &  14   &  9   &  0.27  &  0.07 (0.01)   &  +10.01 (0.06)   &  1.4 (0.2)  &  0.07 (0.01)   &  CDMS    \\
{\medskip}
CCD $ ^{h3}$ & 3 -- 2, J=7/2-5/2,   &  216372.8  &  11  &  21   &  3   &  8   &  0.27  &  0.11 (0.01)    &  +9.66 (0.04)   &  1.2 (0.1)  &  0.11 (0.01)  &  CDMS   \\
 &   F=9/2-7/2  & & & & & & & & \\
{\medskip}
CCD $ ^{h4}$&  3 -- 2, J=5/2-3/2,   &  216428.3  &  11  &  21    &  2   &  8   &  0.27  &  0.08 (0.01)   &  +9.96 (0.05)   &  1.1 (0.1)   &   0.07 (0.01)  &  CDMS  \\
&  F=7/2-5/2  & & & & & & & & \\
{\medskip}
DCN $ ^{h5}$ &  3 -- 2, F=4-3   &  217238.3  &  11  &  21    &  35   &  7   &  0.27  &  0.22 (0.14)   &  +10.08 (0.02)   &  0.9 (0.1)  &  0.16 (0.01)  &  CDMS   \\
{\medskip}
DCO$^{+}$   &  3 -- 2   &  216112.6  &  11  &  21    &  142   &  5   &  0.27  &  1.59 (0.05)   &  +10.18 (0.01)   &  0.7 (0.1)   &  0.85 (0.01)  &  CDMS   \\
{\medskip}
N$_{2}$D$^{+}$ $ ^{h6}$ &   3 -- 2  &  231321.7  &  11  &  22   &  312   &  7   &  0.25  &  0.37 (0.02)   &  +10.01 (0.01)   &  0.8 (0.1)  &  0.23 (0.01)   &  JPL   \\
{\medskip}
SO$_{2}$   &   4$_{2,2}$ -- 3$_{1,3}$  &  235151.7  &  10  &  19   &  5   &  10   &  0.25  &   0.08 (0.01)   &  +10.07 (0.06)   &   1.6 (0.1)   &  0.07 (0.01)  &  CDMS  \\
{\medskip}
 p-D$_{2}$CO &  4$_{1,4}$ -- 3$_{1,3}$   &  221191.7   &  11   &  32   &  20   &  9   &  0.26  &  0.05(0.03)   &  +9.93(0.04)   &  0.5(0.1)  &  0.03(0.05)   &  JPL    \\
{\medskip}
o-D$_{2}$CO &   4$_{0,4}$ -- 3$_{0,3}$  &  231410.3  &  11  &  28   &  43   &  8   &  0.25  &   0.09 (0.02)   &  +10.09 (0.04)   &   0.9 (0.1)   &  0.07 (0.01)   &  JPL   \\

{\medskip}
p-H$_{2}$CO    &  3$_{0,3}$ -- 2$_{0,2}$   &  218222.1  &  11  &  21   &  16   &  8   &  0.27  &  1.40 (0.09)   &  +10.10 (0.01)   &  0.9 (0.1)    &   0.95 (0.09)  &  CDMS  \\ 
{\medskip}
p-H$_{2}$CO       &  3$_{2,2}$ -- 2$_{2,1}$  &  218475.6  &  11   &  68   &  9   &  9   &  0.27  &  0.09 (0.01)   &  +9.94 (0.04)   &   1.2 (0.1)   &   0.08 (0.01)  &  CDMS  \\
{\medskip}
p-H$_{2}$CO       &  3$_{2,1}$ -- 2$_{2,0}$  &  218760.0  &  11  &  68   &  9   &  7   &  0.27  &  0.08 (0.01)   &  +10.01 (0.03)   &  1.1 (0.1)  &  0.07 (0.01)  &  CDMS   \\
{\medskip}
o-H$_{2}^{13}$CO &   3$_{1,2}$ -- 2$_{1,1}$  &  219908.5  &  11  &  33    &  44   &  4   &  0.27  &  0.05 (0.02)   &  +10.08 (0.03)   &  0.6 (0.1)   &  0.02 (0.01)   &  CDMS  \\
{\medskip}
o-H$_{2}$CS &   7$_{1,7}$ -- 6$_{1,6}$  &  236727.0  &  10  &  59   &  56   &  6   &  0.25  &   0.07 (0.01)   &  +9.95 (0.04)   &  0.8 (0.1)   &  0.05 (0.01)  &  CDMS   \\
{\medskip}
o-\textit{c}-C$_{3}$H$_{2}$  &  3$_{3,0}$ -- 2$_{2,1}$   &  216278.8  &  11   &  19   &  46   &  7   &  0.27  &  0.61 (0.02)   &  +10.08 (0.01)   &  0.8 (0.1)  &  0.38 (0.01)  &  CDMS   \\
{\medskip}
o-\textit{c}-C$_{3}$H$_{2}$ &  6$_{1,6}$ -- 5$_{0,5}$   &  217822.1  &  11  &  39    &  175   &  8   &  0.27  &  0.67 (0.03)   &  +9.98 (0.01)   &   0.9 (0.1)  &  0.64 (0.01)  &  CDMS   \\ 
{\medskip}
o-\textit{c}-C$_{3}$H$_{2}$  &  5$_{1,4}$ -- 4$_{2,3}$   &  217940.0  &  11  &  35   &  110    &  8   &  0.27  &  1.42 (0.24)   &  +9.98 (0.01)   &  0.8 (0.1)  &  0.27 (0.01)   &  CDMS  \\ 
{\medskip}
p-\textit{c}-C$_{3}$H$_{2}$    &  5$_{2,4}$ -- 4$_{1,3}$   &  218160.5  &  11  &  35   &  37   &  9   &  0.27  &  0.14 (0.02)   &  +9.94 (0.01)   &  0.9 (0.1)   &   0.16 (0.04)  &  CDMS   \\
{\medskip}
p-\textit{c}-C$_{3}$H$_{2}$ &  8$_{2,6}$-8$_{1,7}$  & 218448.8  &  11 & 87  & 21  & - & -  & -  & -  &  -  &  -  &  CDMS  \\
{\medskip}
o-\textit{c}-C$_{3}$H$_{2}$   $^{c}$ &   8$_{3,6}$ -- 8$_{2,7}$   &  218449.4   &   11  &  87   &  63   &  5  &  0.27   &  0.02(0.01)   &  +9.69(0.19)   &   1.4(0.1)   &   0.03(0.01)   &   CDMS   \\
{\medskip}
o-\textit{c}-C$_{3}$H$_{2}$ $ ^{ng}$ &   7$_{1,6}$ -- 7$_{0,7}$   &  218732.7   &   11  &  61   &  33   &  5   &  0.27  &  0.02 (0.01)   &  +9.42 (0.41)   &   3.0 (0.1)   &   0.06 (0.01)   &   CDMS   \\
{\medskip}
CH$_{3}$OH      &  4$_{-2,3}$ -- 3$_{-1,2}$E   &  218440.1  &  11  &  45   &  14   &  7   &  0.27  &  0.07 (0.01)   &  +10.14 (0.04)   &  0.9 (0.1)   &   0.06 (0.01)  &  CDMS   \\
{\medskip}
CH$_{2}$DOH $ ^{ng}$ &   3$_{1,3}$ -- 2$_{0,2}$e$_{0}$   &   214701.7   &   11   &   17   &   2   &   7   &   0.54  &   0.03 (0.07)   &   +9.99 (0.04)  &  0.6 (0.1)   &   0.03 (0.07)   &  JPL  \\

 \bottomrule
\end{tabular}
\begin{tablenotes}
\item[a] Frequencies and spectroscopic parameters have been extracted from the Jet Propulsion Laboratory (JPL) molecular database \citep{Pickett1998} and the Cologne Database for Molecular Spectroscopy \citep{Muller2005}. 
{\smallskip}
\item[b] Gaussian fit. 
{\smallskip}
\item[COng] Non-Gaussian profile.  The line intensity is obtained by integrating the Zero Power Full Width ($ZPFW$) range. In particular $-$40, +40 km s$^{-1}$ for CO (2--1), +8, +12 km s$^{-1}$ for $^{13}$CO (2 -- 1) see in Figure \ref{CO1}. The value has to be considered a lower limit given the presence of absorption features.
{\smallskip}
\item[ng] Non-Gaussian profile.  The line intensity is obtained by integrating the $ZPFW$ range.
{\smallskip}
% 13CN
\item[h1] The detected 2 -- 1, J=5/2--3/2, F=4--3, F$_{1}$=3--2 line consists of 2 hyperfine components with $S\mu^2$ $\leq$ 8 D$^2$ \citep{Muller2005} in a 0.05 MHz frequency interval. The line with the highest $S\mu^2$ line is reported (see e.g. Figures \ref{B5_Spectra}).
% 13CS
\item[h2] The detected 5 -- 4 line consists of 4 hyperfine components with $S\mu^2$ $\leq$ 38 D$^2$ \citep{Muller2005} in a 0.0 MHz frequency interval. The line with the highest $S\mu^2$ line is reported (see e.g. Figures \ref{B5_Spectra}).
{\smallskip}
% CCD
\item[h3] 
The detected 3 -- 2, J=7/2--5/2, F=9/2--7/2 line consists of 3 hyperfine components with $S\mu^2$ $\leq$ 3 D$^2$ \citep{Muller2005} in a 0.5 MHz frequency interval. The line with the highest $S\mu^2$ line is reported (see e.g. Figures \ref{B5_Spectra}).
{\smallskip}
% CCD
\item[h4] The detected 3 -- 2, J=5/2--3/2, F=7/2--5/2 line consists of 2 hyperfine components with $S\mu^2$ $\leq$ 2 D$^2$ \citep{Muller2005} in a 0.4 MHz frequency interval. The line with the highest $S\mu^2$ line is reported (see e.g. Figures \ref{B5_Spectra}).
{\smallskip}
% DCN
\item[h5] The detected 3 -- 2, F=4--3 line consists of 3 hyperfine components with $S\mu^2$ $\leq$ 35 D$^2$ \citep{Muller2005} in a 0.7 MHz frequency interval. The line with the highest $S\mu^2$ line is reported (see e.g. Figures \ref{B5_Spectra}).
{\smallskip}
%N2D+
\item[h6] The detected 3 -- 2 line consists of 40 hyperfine components with $S\mu^2$ $\leq$ 312 D$^2$ in a 3 MHz frequency interval. The values are taken from L. Dore in private communication. The line with the highest $S\mu^2$ line is reported (see e.g. Figures \ref{B5_Spectra}).
{\smallskip}
\item[c] Line blended at the present spectral resolution (0.2 MHz). Possible contamination due to p-\textit{c}-C$_{3}$H$_{2}$ emission at 218448.8 MHz ($E_{\rm u}$ = 21 K, $S\mu^2$ = 87 D$^2$).

\end{tablenotes} 

     \end{threeparttable} }
\end{table*}
%%%%%%%%%%%%%%%%%%%%%%%%
%%%%    L1455 List  %%%%
%%%%%%%%%%%%%%%%%%%%%%%%
\begin{table*}
\centering
\caption{List of transitions and observed line properties detected towards L1455-IRS1}
\label{L1455-list}
\resizebox{19cm}{!}{%
 \begin{threeparttable}
\begin{tabular}{llccccccccccc}
\hline \hline \noalign {\medskip}

Species & Transition $^{a}$& $\nu$ $^{a}$ & $HPBW$ & $E_{\rm up}$ $^{a}$ & $S\mu^2$ $^{a}$ & rms $^{b}$ & ${\delta} V$ $ ^{b}$ & $T_{peak}$ $^{b}$ & $V_{\rm peak}$ $^{b}$ & $FWHM$ $^{b}$ & $I_{int}$ $^{b}$ & Database \\ 
& & (MHz) &$(\arcsec$)&(K) & (D$^{2}$) & (mK) &(km s$^{-1}$)& (K) & (km s$^{-1}$) & (km s$^{-1}$) &(K km s$^{-1}$) \\
\hline \noalign {\medskip}
{\medskip}
CO $ ^{COng}$ &   2 -- 1&  230538.0  &  11  &  17    &  0.02   &  11  &  0.25  &  -   &  -   &  -  &   $\geq${60}     &  CDMS   \\
{\medskip}
$^{13}$CO $^{COng}$   &   2 -- 1&  220398.7&  11&  16 &  0.05 &  80 &  0.27&  5.24 (0.18) &  +4.76 (0.02) &  2.4 (0.1) &  6.78 (0.18) &  CDMS \\
{\medskip}
C$^{18}$O $^{COng}$       &  2 -- 1   &  219560.3  &  11  &  16    &  0.02   & 17 & 0.27  &  1.15 (0.01)  &  +4.72 (0.01)  &  1.6 (0.1)  &  1.69 (0.01)  &  CDMS  \\
{\medskip}
$^{13}$CN $ ^{h1}$ & 2 -- 1, J=5/2-3/2,   &  217467.1  &  11  &  16    &  8   &  12   &   0.27  &   0.05 (0.04)  &  +4.90 (0.09)  &  1.0 (0.2)  &  0.05 (0.01)  &  CDMS  \\
&   F=4-3, F$_{1}$=3-2  & & & & & & & & \\
{\medskip}
$^{13}$CN  &  2 -- 1, J=5/2-3/2,  & 217469.2 & 11 & 16   & 4  &  13 & 0.27  &  0.03 (0.01)  &  +4.67 (0.14)  &  0.8 (0.3)  &  0.02 (0.01)  &  CDMS  \\
& F=2-1, F$_{1}$=3-2 & & & & & & & \\
{\medskip}
$^{13}$CS $ ^{h2}$&  5 -- 4   &  231220.7  &  11  &  33    &  38   &  12   &  0.25  &  0.19 (0.02)   &  +4.66 (0.03)   &  1.6 (0.1)  &  0.32 (0.01)  &  CDMS    \\
{\medskip}
SO &  5$_{5}$ -- 4$_{4}$   &  215220.6   &  11  &  44   &  12   &  16   &  0.27  &  0.24 (0.03)   &  +4.74 (0.03)   &  2.0 (0.1)  &   0.51 (0.02)  &  CDMS    \\
{\medskip}
SO   &   6$_{5}$ -- 5$_{4}$  &  219949.4  &  11  &  35   &  14    &  26   &  0.27  &  0.59 (0.08)   &  +4.80 (0.12)   &  1.0 (0.1)   &   0.65 (0.01)   &  CDMS   \\
{\medskip}
SiO $ ^{ng}$    &  5 -- 4  &  217105.0  &  11  &  31  &  48  &  12  & 0.27 &  0.05 (0.01)  &  +4.93 (0.29)  &  3.4 (0.9)  &  0.09 (0.01)  &  CDMS  \\
{\medskip}
CCD $ ^{h3}$ & 3 -- 2, J=7/2-5/2,   &  216372.8  &  11  &  21   &  3   &  11   &  0.27  &  0.07 (0.01)    &  +4.20 (0.09)   &  1.0 (0.3)  &  0.07 (0.01)  &  CDMS   \\
 &   F=9/2-7/2  & & & & & & & & \\
{\medskip}
CCD $ ^{h4}$&  3 -- 2, J=5/2-3/2,   &  216428.3  &  11  &  21    &  2   &  11   &  0.27  &  0.06 (0.01)   &  +4.68 (0.06)   &  0.7 (0.1)   &   0.04 (0.01)  &  CDMS  \\
&  F=7/2-5/2  & & & & & & & & \\
{\medskip}
CCS &  N=17-16, J=18-17   &  221071.1  &  11  &  99  & 149  &  8  & 0.26 &  0.03 (0.01)  &  +4.58 (0.11)  &  1.1 (0.3)  &  0.04 (0.01)  &  JPL  \\
{\medskip}
DCN $ ^{h5}$ &  3 -- 2, F=4-3   &  217238.3  &  11  &  21    &  35   &  13   &  0.27  &  0.24 (0.02)   &  +4.44 (0.02)   &  1.5 (0.1)  &  0.08 (0.01)  &  CDMS   \\
{\medskip}
DCO$^{+}$ &  3 -- 2   &  216112.6  &  11  &  21    &  142   &  12   &  0.27  &  1.14 (0.02)   &  +4.81 (0.01)   &  0.7 (0.1)   &  0.84 (0.01)  &  CDMS   \\
{\medskip}
H$_{2}$S $ ^{ng}$    &  2$_{2,0}$ -- 2$_{1,1}$  &  216710.4  &  11  &  84  &  2  &  12 & 0.54  &  0.05 (0.01)  &  +4.55 (0.17)  &  3.6 (0.3)  &  0.19 (0.01)  &  CDMS  \\
{\medskip}
N$_{2}$D$^{+}$ $ ^{h6}$ &   3 -- 2  &  231321.7  &  11  &  22   &  312   &  13   &  0.25  &  0.56 (0.03)   &  +4.60 (0.01)   &  0.8 (0.1)  &  0.47 (0.01)   &  JPL   \\
{\medskip}
OCS $^{ng}$ & 18 -- 17 & 218903.4 & 11 & 100 & 9 & 4  & 1.01 & 0.02 (0.01)  & +4.64 (0.47)  & 7.4 (0.9) &  0.19 (0.02) & CDMS   \\
{\medskip}
OCS $^{ng}$ &19 -- 18 & 231061.0 & 11 & 110 & 10 & 4  & 1.01 & 0.02 (0.01)  & +4.82 (0.50)  & 7.2 (0.9) &  0.18 (0.02) & CDMS   \\
{\medskip}
SO$_{2}$   &   4$_{2,2}$ -- 3$_{1,3}$  &  235151.7  &  10  &  19   &  5   &  10   &  0.25  &   0.08 (0.02)   &  +4.66 (0.03)   &   0.2 (0.1)   &  0.03 (0.01)  &  CDMS  \\
{\medskip}
p-D$_{2}$CO &  4$_{1,4}$ -- 3$_{1,3}$   &  221191.7   &  11   &  32   &  20   &  8   &  0.26  &  0.04 (0.01)   &  +4.89 (0.08)   &  1.0 (0.2)  &  0.04(0.01)   &  JPL    \\
{\medskip}
o-D$_{2}$CO &   4$_{0,4}$ -- 3$_{0,3}$  &  231410.3  &  11  &  28   &  43   &  9   &  0.25  &   0.08 (0.01)   &  +4.93 (0.04)   &   0.8 (0.1)   &  0.07 (0.01)   &  JPL   \\
{\medskip}
HDCS    &  7$_{0,7}$ -- 6$_{0,6}$  &  216662.4  &  11  &  42  &  19  &  12 & 0.27  &  0.07 (0.01)  &  +4.90 (0.05)  &  0.7 (0.1)  &  0.05 (0.01)  &  CDMS  \\
{\medskip}
HDCS &  7$_{1,6}$ -- 6$_{1,5}$  &  221177.1  &  11  &  51  &  19  &  8  & 0.26 &  0.02 (0.01)  &  +4.85 (0.11)  &  0.8 (0.3)  &  0.02 (0.01)  &  CDMS  \\
{\medskip}
p-H$_{2}$CO    &  3$_{0,3}$ -- 2$_{0,2}$   &  218222.1  &  11  &  21   &  16   &  32   &  0.27  &  1.27 (0.15)   &  +4.80 (0.01)   &  1.1 (0.1)    &   1.53 (0.01)  &  CDMS  \\ {\medskip}
p-H$_{2}$CO       &  3$_{2,2}$ -- 2$_{2,1}$  &  218475.6  &  11   &  68   &  9   &  15   &  0.27  &  0.16 (0.03)   &  +4.92 (0.05)   &   1.7 (0.2)   &   0.30 (0.02)  &  CDMS  \\
{\medskip}
p-H$_{2}$CO       &  3$_{2,1}$ -- 2$_{2,0}$  &  218760.0  &  11  &  68   &  9   &  13   &  0.27  &  0.16 (0.01)   &  +4.81 (0.03)   &  1.5 (0.1)  &  0.27 (0.01)  &  CDMS   \\
 {\medskip}
o-H$_{2}^{13}$CO &   3$_{1,2}$ -- 2$_{1,1}$  &  219908.5  &  11  &  33    &  44   &  9   &  0.27  &  0.04 (0.01)   &  +4.73 (0.12)   &  1.7 (0.3)   &  0.07 (0.01)   &  CDMS  \\
{\medskip}
o-H$_{2}$CS &   7$_{1,7}$ -- 6$_{1,6}$  &  236727.0  &  10  &  59   &  56   &  14   &  0.25  &   0.26 (0.01)   &  +4.78 (0.02)   &  1.0 (0.1)   &  0.29 (0.01)  &  CDMS   \\
{\medskip}
HNCO $ ^{ng}$ &  10$_{0,10}$ -- 9$_{0,9}$  &  219798.3  &  11  &  58  &  25  &  9  & 0.27 &  0.03 (0.01)  &  +4.69 (0.14)  &  1.7 (0.3)  &  0.06 (0.01)  &  CDMS  \\

\end{tabular}
\begin{tablenotes}
\item[a] Frequencies and spectroscopic parameters have been extracted from the Jet Propulsion Laboratory (JPL) molecular database \citep{Pickett1998} and the Cologne Database for Molecular Spectroscopy \citep{Muller2005}. 
{\smallskip}
\item[b] Gaussian fit. 
{\smallskip}
\item[COng] Non-Gaussian profile.  The line intensity is obtained by integrating the Zero Power Full Width ($ZPFW$) range. In particular $-$ 10, +20 km s$^{-1}$ for CO (2--1), +1, +10 km s$^{-1}$) for $^{13}$CO (2--1), and  +1, +9 km s$^{-1}$ for C$^{18}$O (2 -- 1) see in Figure \ref{CO2}. The value has to be considered a lower limit given the presence of absorption features.
{\smallskip}
\item[ng] Non-Gaussian profile.  The line intensity is obtained by integrating the $ZPFW$ range.
{\smallskip}
% 13CN
\item[h1] The detected 2 -- 1, J=5/2--3/2, F=4--3, F$_{1}$=3--2 line consists of 3 hyperfine components with $S\mu^2$ $\leq$ 8 D$^2$ \citep{Muller2005} in a 2.0 MHz frequency interval. The line with the highest $S\mu^2$ line is reported (see e.g. Figures \ref{L1455_Spectra}.
% 13CS
\item[h2] The detected 5 -- 4 line consists of 4 hyperfine components with $S\mu^2$ $\leq$ 38 D$^2$ \citep{Muller2005} in a 0.0 MHz frequency interval. The line with the highest $S\mu^2$ line is reported (see e.g. Figures \ref{L1455_Spectra}.
{\smallskip}
% CCD
\item[h3] 
The detected 3 -- 2, J=7/2--5/2, F=9/2--7/2 line consists of 3 hyperfine components with $S\mu^2$ $\leq$ 3 D$^2$ \citep{Muller2005} in a 0.5 MHz frequency interval. The line with the highest $S\mu^2$ line is reported (see e.g. Figures \ref{L1455_Spectra}.
{\smallskip}
% CCD
\item[h4] The detected 3 -- 2, J=5/2--3/2, F=7/2--5/2 line consists of 2 hyperfine components with $S\mu^2$ $\leq$ 2 D$^2$ \citep{Muller2005} in a 0.4 MHz frequency interval. The line with the highest $S\mu^2$ line is reported (see e.g. Figures \ref{L1455_Spectra}.
{\smallskip}
% DCN
\item[h5] The detected 3 -- 2, F=4--3 line consists of 3 hyperfine components with $S\mu^2$ $\leq$ 35 D$^2$ \citep{Muller2005} in a 0.7 MHz frequency interval. The line with the highest $S\mu^2$ line is reported (see e.g. Figures \ref{L1455_Spectra}.
{\smallskip}
%N2D+
\item[h6] The detected 3 -- 2 line consists of 40 hyperfine components with $S\mu^2$ $\leq$ 312 D$^2$ in a 3 MHz frequency interval. The values are taken from L. Dore in private communication. The line with the highest $S\mu^2$ line is reported (see e.g. Figures \ref{L1455_Spectra}).
\end{tablenotes} 
\end{threeparttable} }
\end{table*}
%%%%%%%%%%%%%%%%%%%%%%%%%%%%%%%
\addtocounter{table}{-1}
\begin{table*}
\centering
%\tiny
%\scriptsize
%\normalsize
\caption{Continued}
%{
\resizebox{19cm}{!}{%
\begin{threeparttable}
 
\begin{tabular}{llccccccccccc}
\hline \hline \noalign {\medskip}

Species & Transition $^{a}$& $\nu$ $^{a}$ & $HPBW$ & $E_{\rm up}$ $^{a}$ & $S\mu^2$ $^{a}$ & rms $^{b}$ & ${\delta} v$ $ ^{b}$ & $T_{peak}$ $^{b}$ & $V_{\rm peak}$ $^{b}$ & $FWHM$ $^{b}$ & $I_{int}$ $^{b}$ & Database \\  
& & (MHz) &$(\arcsec$)&(K) & (D$^{2}$) & (mK) &(km s$^{-1}$)& (K) & (km s$^{-1}$) & (km s$^{-1}$) &(K km s$^{-1}$) \\
\hline \noalign {\medskip}
 {\medskip}
o-\textit{c}-C$_{3}$H$_{2}$  &  3$_{3,0}$ -- 2$_{2,1}$   &  216278.8  &  11   &  19   &  46   &  14   &  0.27  &  0.39 (0.02)   &  +4.90 (0.01)   &  1.1 (0.1)  &  0.46 (0.01)  &  CDMS   \\
{\medskip}
o-\textit{c}-C$_{3}$H$_{2}$ &  6$_{1,6}$ -- 5$_{0,5}$   &  217822.1  &  11  &  39    &  175   &  13   &  0.27  &  0.46 (0.01)   &  +4.89 (0.01)   &   1.3 (0.1)  &  0.64 (0.01)  &  CDMS   \\ 
{\medskip}
o-\textit{c}-C$_{3}$H$_{2}$  &  5$_{1,4}$ -- 4$_{2,3}$   &  217940.0  &  11  &  35   &  110    &  13   &  0.27  &  0.26 (0.01)   &  +4.89 (0.02)   &  1.3 (0.1)  &  0.34 (0.01)   &  CDMS  \\ 
{\medskip}
p-\textit{c}-C$_{3}$H$_{2}$    &  5$_{2,4}$ -- 4$_{1,3}$   &  218160.5  &  11  &  35   &  37   &  13   &  0.27  &  0.11 (0.01)   &  +4.83 (0.03)   &  1.2 (0.1)   &   0.14 (0.01)  &  CDMS   \\
{\medskip}
o-\textit{c}-C$_{3}$H$_{2}$ $ ^{ng}$ &   7$_{1,6}$ -- 7$_{0,7}$   &  218732.7   &   11  &  61   &  33   &  7   &  0.53  &  0.04 (0.01)   &  +4.81 (0.15)  &   0.6 (3.2)   &   0.02 (0.01)   &   CDMS   \\
{\medskip}
CH$_{3}$OH      &  4$_{-2,3}$ -- 3$_{-1,2}$E   &  218440.1  &  11  &  45   &  14   &  14   &  0.27  &  0.12 (0.02))   &  +4.89 (0.06)   &  1.6 (0.2)   &   0.22 (0.02)  &  CDMS   \\
{\medskip}
CH$_{2}$DOH $ ^{ng}$ &   10$_{1,10}$ -- 9$_{0,9}$e$_{0}$   &   221391.8   &   11   &   120   &   5   &   8   &   0.26  &   0.03 (0.01)   &   +5.15 (0.06)  &  0.5 (0.2)   &   0.02 (0.01)   &  JPL  \\
%%%%%%%
\end{tabular}
    
\begin{tablenotes}
\item[a] Frequencies and spectroscopic parameters have been extracted from the Jet Propulsion Laboratory (JPL) molecular database \citep{Pickett1998} and the Cologne Database for Molecular Spectroscopy \citep{Muller2005}. 
{\smallskip}
\item[b] Gaussian fit. 
{\smallskip}
\item[ng] Non-Gaussian profile.  The line intensity is obtained by integrating the Zero Power Full Width ($ZPFW$) range.

\end{tablenotes} 

    \end{threeparttable} }
\end{table*}
%%%%%%%%%%%%%%%%%%%%%%%%
%%%%    L1551 List  %%%%
%%%%%%%%%%%%%%%%%%%%%%%%

\begin{table*}
\centering
%\tiny
%\scriptsize
%\normalsize
\caption{List of transitions and observed line properties detected towards L1551-IRS5}
%{
\label{L1551-list} 
\resizebox{19cm}{!}{%
\begin{threeparttable}
 
\begin{tabular}{llccccccccccc}
\hline \hline \noalign {\medskip}

Species & Transition $^{a}$& $\nu$ $^{a}$ & $HPBW$ & $E_{\rm up}$ $^{a}$ & $S\mu^2$ $^{a}$ & rms $^{b}$ & ${\delta} V$ $ ^{b}$ & $T_{peak}$ $^{b}$ & $V_{\rm peak}$ $^{b}$ & $FWHM$ $^{b}$ & $I_{int}$ $^{b}$ & Database \\ 
& & (MHz) &$(\arcsec$)&(K) & (D$^{2}$) & (mK) &(km s$^{-1}$)& (K) & (km s$^{-1}$) & (km s$^{-1}$) &(K km s$^{-1}$) \\
\hline \noalign {\medskip}
%\hline
%\hline
{\medskip}
CO $ ^{COng}$ &   2 -- 1&  230538.0  &  11  &  17    &  0.02   &  70  &  0.25  &  -   &  -   &  -  &   $\geq${40}     &  CDMS   \\
{\medskip}
$^{13}$CO $^{ng}$   &   2 -- 1&  220398.7&  11&  16 &  0.05 &  16 &  0.27&  6.72 (0.03) & +6.04 (0.02) & 2.9 (0.1) &15.13 (0.03)& CDMS\\
{\medskip}
C$^{18}$O        & 2 -- 1  & 219560.3 & 11 & 16   & 0.02  & 65  & 0.27 & 6.17 (0.50)  & +6.41 (0.01)  & 1.3 (0.1) &  8.39 (0.04) & CDMS   \\
{\medskip}
$^{13}$CN & 2 -- 1, J=3/2-1/2, & 217277.7 & 11 & 16   & 1  & 9 & 0.27  & 0.03 (0.01)  & +6.59 (0.09)  & 0.8 (0.2) &  0.03 (0.01) & CDMS   \\
&  F=1-1, F$_{1}$=1-0 & & & & & & & \\
{\medskip}
$^{13}$CN & 2 -- 1, J=3/2-1/2,  & 217303.2 & 11 & 16   & 6  & 16 & 0.27  & 0.02 (0.02)  & +6.44 (0.35)  & 1.0 (0.8) &  0.02 (0.01) & CDMS   \\
&  F=3-2, F$_{1}$=2-1 & & & & & & & \\
{\medskip}
$^{13}$CN &2 -- 1, J=5/2-3/2,  & 217305.9 & 11 & 16   & 0.13  & 15 & 0.27  & 0.02 (0.02)  & +6.30 (0.21)  & 0.6 (0.5) &  0.10 (0.01) & CDMS   \\
&   F=1-2, F$_{1}$=2-1 & & & & & & & \\
{\medskip}
$^{13}$CN & 2 -- 1, J=5/2-3/2,  & 217428.6 & 11 & 16   & 4   & 8 & 0.27  & 0.04 (0.01)  & +6.23 (0.07)  & 0.9 (0.1) &  0.04 (0.01) & CDMS   \\
&   F=3-2, F$_{1}$=2-1 & & & & & & & \\
 {\medskip}
$^{13}$CN$ ^{h1}$ & 2 -- 1, J=5/2-3/2,  & 217467.1 & 11 & 16   & 8  & 12 & 0.27  & 0.10 (0.01)  & +6.38 (0.05)  & 1.2 (0.1) &  0.13 (0.01) & CDMS   \\
&   F=4-3, F$_{1}$=3-2 & & & & & & & \\
 {\medskip}
$^{13}$CN & 2 -- 1,J=5/2-3/2,  & 217469.2 & 11 & 16   & 4  & 21 & 0.27  & 0.04 (0.01)  & +6.43 (0.17)  & 0.8 (0.4) &  0.03 (0.01) & CDMS   \\
&  F=2-1, F$_{1}$=3-2, & & & & & & & \\
{\medskip}
C$^{15}$N &   2 -- 1, 3/2-1/2,&  219722.5  &  11  &  16  &  4 & 7 & 0.27  & 0.03 (0.01)  & +6.38 (0.07)  & 0.6 (0.1) &  0.02 (0.01) & CDMS   \\
&  F=2-1 & & & & & & & \\
{\medskip}
C$^{15}$N $ ^{h2}$ &  2 -- 1, J=5/2-3/2, &  219934.8  &  11  &  16  &  6  & 7  & 0.27 & 0.38 (0.01)  & +6.36 (0.12)  & 2.0 (0.2) &  0.07 (0.01) & CDMS   \\
&  F=3-2 & & & & & & & \\
{\medskip}
$^{13}$CS $ ^{h3}$&  5 -- 4   &  231220.7  &  11  &  33    &  38   &  8   &  0.25  &  0.09 (0.01)   &  +6.51 (0.03)   &  1.4 (0.1)  &  0.14 (0.01)  &  CDMS    \\
{\medskip}
SO &  5$_{5}$ -- 4$_{4}$   &  215220.6   &  11  &  44   &  12   &  13   &  0.27  &  0.68 (0.04)   &  +6.48 (0.01)   &  1.7 (0.1)  &   1.23 (0.01)  &  CDMS    \\
{\medskip}
SO   &   6$_{5}$ -- 5$_{4}$  &  219949.4  &  11  &  35   &  14    &  19   &  0.27  &  1.15 (0.06)   &  +6.53 (0.01)   &  1.6 (0.1)   &   1.96 (0.02)   &  CDMS   \\
{\medskip}
$^{34}$SO & 6$_{5}$ -- 5$_{4}$ & 215839.9  & 11 & 35   & 14  & 8  & 0.27 & 0.05 (0.01)  & +6.46 (0.08)  & 2.3 (0.2) &  0.13 (0.01) & CDMS   \\
{\medskip}
OCS $^{ng}$ & 18 -- 17 & 218903.4 & 11 & 100 & 9 & 7  & 0.27 & 0.13 (0.01)  & +7.00 (0.06)  & 6.9 (0.1) &  0.48 (0.03) & CDMS   \\
{\medskip}
OCS $^{ng}$ &19 -- 18 & 231061.0 & 11 & 110 & 10 & 7  & 0.27 & 0.15 (0.01)  & +6.95 (0.15)  & 5.1 (0.1) &  0.48 (0.03) & CDMS   \\
{\medskip}
O$^{13}$CS $^{ng}$&  19 -- 18 & 230317.5 & 11 & 110 & 10 & 7  & 0.25 & 0.04 (0.01)  & +8.46 (0.11)  & 2.1 (0.3) &  0.04 (0.01) & CDMS   \\
{\medskip}
CCD $ ^{h4}$ & 3 -- 2, J=7/2-5/2,   &  216372.8  &  11  &  21   &  3   &  9   &  0.27  &  0.25 (0.01)    &  +5.98 (0.02)   &  1.4 (0.1)  &  0.37 (0.01)  &  CDMS   \\
 &   F=9/2-7/2  & & & & & & & & \\
{\medskip}
CCD $ ^{h5}$&  3 -- 2, J=5/2-3/2,   &  216428.3  &  11  &  21    &  2   &  35   &  0.27  &  0.27 (0.01)   &  +6.19 (0.43)   &  0.7 (0.8)   &   0.01 (0.01)  &  CDMS  \\
&  F=7/2-5/2  & & & & & & & & \\
{\medskip}
CCS &  17 -- 16, J=16-15   &  219142.7  &  11  &  100  & 134&  6  & 0.53 &  0.03 (0.01)  &  +6.51 (0.18)  &  1.8 (0.6)  &  0.07 (0.01)  &  JPL  \\
{\medskip}
CCS &  17 -- 16, J=18-17   &  221071.1  &  11  &  99  & 149  &  8  & 0.52 &  0.03 (0.01)  &  +6.41 (0.18)  &  2.9 (0.4)  &  0.05 (0.01)  &  JPL  \\
{\medskip}
DCN $ ^{h6}$ &  3 -- 2, F=4-3   &  217238.3  &  11  &  21    &  35   &  16   &  0.27  &  0.49 (0.03)   &  +6.40 (0.01)   &  1.4 (0.1)  &  0.71 (0.01)  &  CDMS   \\
{\medskip}
DCO$^{+}$  &  3 -- 2   &  216112.6  &  11  &  21    &  142   &  10   &  0.27  &  2.34 (0.06)   &  +6.26 (0.01)   &  0.9 (0.1)   &  2.37 (0.01)  &  CDMS   \\
{\medskip}
H$_{2}$S $ ^{ng}$    &  2$_{2,0}$ -- 2$_{1,1}$  &  216710.4  &  11  &  84  &  2  &  8 & 0.27  &  0.12 (0.02)  &  +6.58 (0.05)  &  5.1 (0.1)  &  0.65 (0.01)  &  CDMS  \\

\end{tabular}
    
\begin{tablenotes}
\item[a] Frequencies and spectroscopic parameters have been extracted from the Jet Propulsion Laboratory (JPL) molecular database \citep{Pickett1998} and the Cologne Database for Molecular Spectroscopy \citep{Muller2005}. 
{\smallskip}
\item[b] Gaussian fit. 
{\smallskip}
\item[ng] Non-Gaussian profile. The line intensity is obtained by integrating the Zero Power Full Width ($ZPFW$) range. In particular $-$15, +30 km s$^{-1}$ for CO (2 -- 1), +3, +10 km s$^{-1}$) for $^{13}$CO (2--1), and  +3, +10 km s$^{-1}$ for C$^{18}$O (2-1) see in Figure \ref{CO2}. The value has to be considered a lower limit given the presence of absorption features.
{\smallskip}
\item[ng] Non-Gaussian profile. The line intensity is obtained by integrating the $ZPFW$ range.
{\smallskip}
% 13CN
\item[h1] The detected 2 -- 1, J=5/2--3/2, F=4--3, F$_{1}$=3--2 line consists of 6 hyperfine components with $S\mu^2$ $\leq$ 8 D$^2$ \citep{Muller2005} in a 261 MHz frequency interval. The line with the highest $S\mu^2$ line is reported (see e.g. Figures \ref{L1551_Spectra}.
{\smallskip}
% C15N
\item[h2] The detected 2 -- 1, J=5/2--3/2, F=3--2 line consists of 3 hyperfine components with $S\mu^2$ $\leq$ 6 D$^2$ \citep{Muller2005} in a 289 MHz frequency interval. The line with the highest $S\mu^2$ line is reported (see e.g. Figures \ref{L1551_Spectra}.
{\smallskip}
% 13CS
\item[h3] The detected 5 -- 4 line consists of 4 hyperfine components with $S\mu^2$ $\leq$ 38 D$^2$ \citep{Muller2005} in a 0.0 MHz frequency interval. The line with the highest $S\mu^2$ line is reported (see e.g. Figures \ref{L1551_Spectra}.
{\smallskip}
% CCD
\item[h4] 
The detected 3 -- 2, J=7/2--5/2, F=9/2--7/2 line consists of 3 hyperfine components with $S\mu^2$ $\leq$ 3 D$^2$ \citep{Muller2005} in a 0.5 MHz frequency interval. The line with the highest $S\mu^2$ line is reported (see e.g. Figures \ref{L1551_Spectra}.
{\smallskip}
% CCD
\item[h5] The detected 3 -- 2, J=5/2--3/2, F=7/2--5/2 line consists of 3 hyperfine components with $S\mu^2$ $\leq$ 2 D$^2$ \citep{Muller2005} in a 3.0 MHz frequency interval. The line with the highest $S\mu^2$ line is reported (see e.g. Figures \ref{L1551_Spectra}.
{\smallskip}
% DCN
\item[h6] The detected 3 -- 2, F=4--3 line consists of 3 hyperfine components with $S\mu^2$ $\leq$ 35 D$^2$ \citep{Muller2005} in a 0.7 MHz frequency interval. The line with the highest $S\mu^2$ line is reported (see e.g. Figures \ref{L1551_Spectra}.
{\smallskip}
%N2D+

\end{tablenotes} 

    \end{threeparttable} }
\end{table*}

\addtocounter{table}{-1}
\begin{table*}
\centering
%\tiny
%\scriptsize
%\normalsize
\caption{Continued}
%{
\resizebox{19cm}{!}{%
\begin{threeparttable}
 
\begin{tabular}{llccccccccccc}
\hline \hline \noalign {\medskip}

Species & Transition $^{a}$& $\nu$ $^{a}$ & $HPBW$ & $E_{\rm up}$ $^{a}$ & $S\mu^2$ $^{a}$ & rms $^{b}$ & ${\delta} v$ $ ^{b}$ & $T_{peak}$ $^{b}$ & $V_{\rm peak}$ $^{b}$ & $FWHM$ $^{b}$ & $I_{int}$ $^{b}$ & Database \\  
& & (MHz) &$(\arcsec$)&(K) & (D$^{2}$) & (mK) &(km s$^{-1}$)& (K) & (km s$^{-1}$) & (km s$^{-1}$) &(K km s$^{-1}$) \\
\hline \noalign {\medskip}
%\hline
%\hline
{\medskip}
N$_{2}$D$^{+}$ $ ^{h7}$ &   3 -- 2  &  231321.7  &  11  &  22   &  312   &  8   &  0.25  &  0.18 (0.01)   &  +6.00 (0.01)   &  0.8 (0.1)  &  0.15 (0.01)   &  JPL   \\
{\medskip}
SO$_{2}$   &   4$_{2,2}$ -- 3$_{1,3}$  &  235151.7  &  10  &  19   &  5   &  6   &  0.50  &   0.04 (0.01)   &  +6.49 (0.09)   &   2.2 (0.2)   &  0.11 (0.01)  &  CDMS  \\
{\medskip}
SO$_{2}$  $^{ng}$  &   12$_{3,9}$ -- 12$_{2,10}$  &  237068.8  &  10  &  94   &  18   &  9   &  0.99  &   0.02 (0.01)   &  +6.79 (0.28)   &   3.6 (0.6)   &  0.08 (0.01)  &  CDMS  \\
{\medskip}
p-D$_{2}$CO &  4$_{1,4}$ -- 3$_{1,3}$   &  221191.7   &  11   &  32   &  20   &  9   &  0.26  &  0.25 (0.01)   &  +6.21 (0.01)   &  1.0 (0.1)  &  0.26 (0.01)   &  JPL    \\
{\medskip}
o-D$_{2}$CO &   4$_{0,4}$ -- 3$_{0,3}$  &  231410.3  &  11  &  28   &  43   &  9   &  0.25  &   0.62 (0.02)   &  +6.27 (0.01)   &   1.0 (0.1)   &  0.67 (0.01)   &  JPL   \\
{\medskip}
p-D$_{2}$CO &  4$_{2,3}$ -- 3$_{2,2}$& 233650.7& 11& 50 & 33 &  9  & 0.25& 0.14 (0.01) & +6.20 (0.03) & 1.3 (0.1)&  0.19 (0.01)& JPL  \\
{\medskip}
o-D$_{2}$CO &  4$_{2,2}$ -- 3$_{2,1}$ & 236102.1 & 10 & 50  & 33  &  11   & 0.25 & 0.13 (0.01)  & +6.19 (0.03)  & 1.1 (0.1) &  0.15 (0.01) & JPL   \\
{\medskip}
HDCS $^{ng}$ &  7$_{1,6}$ -- 6$_{1,5}$  &  221177.1  &  11  &  51  &  19  &  6  & 0.53 &  0.02 (0.01)  &  +7.36 (0.3)  &  2.5 (0.9)  &  0.05 (0.01)  &  CDMS  \\
{\medskip}
p-H$_{2}$CO    &  3$_{0,3}$ -- 2$_{0,2}$   &  218222.1  &  11  &  21   &  16   &  42   &  0.53  &  3.32 (0.23)   &  +6.45 (0.01)   &  1.1 (0.1)    &   3.95 (0.04)  &  CDMS  \\ 
{\medskip}
p-H$_{2}$CO       &  3$_{2,2}$ -- 2$_{2,1}$  &  218475.6  &  11   &  68   &  9   &  12   &  0.27  &  0.32 (0.01)   &  +6.50 (0.02)   &   2.0 (0.1)   &   0.69 (0.01)  &  CDMS  \\
{\medskip}
p-H$_{2}$CO       &  3$_{2,1}$ -- 2$_{2,0}$  &  218760.0  &  11  &  68   &  9   &  10   &  0.27  &  0.30 (0.03)   &  +6.52 (0.02)   &  2.2 (0.1)  &  0.70 (0.01)  &  CDMS   \\
 {\medskip}
o-H$_{2}^{13}$CO &   3$_{1,2}$ -- 2$_{1,1}$  &  219908.5  &  11  &  33    &  44   &  7   &  0.27  &  0.13 (0.01)   &  +6.47 (0.02)   &  1.1 (0.1)   &  0.15 (0.01)   &  CDMS  \\
{\medskip}
o-H$_{2}$CS &   7$_{1,7}$ -- 6$_{1,6}$  &  236727.0  &  10  &  59   &  56   &  8   &  0.25  &   0.09 (0.01)   &  +6.68 (0.06)   &  2.9 (0.1)   &  0.27 (0.01)  &  CDMS   \\
{\medskip}
o-H$_{2}$C$^{33}$S$^{h8}$ &  7$_{1,7}$ -- 6$_{1,6}$ & 234678.8 & 10 & 58  & 67   & 13  & 0.25  & 0.05 (0.01)  & +6.78 (0.09)  & 0.8 (0.2) &  0.04 (0.01) & CDMS   \\
&  F=17/2-15/2 & & & & & & & \\
{\medskip}
HNCO $ ^{ng}$ &  10$_{0,10}$ -- 9$_{0,9}$  &  219798.3  &  11  &  58  &  25  &  7  & 0.27 &  0.03 (0.01)  &  +6.04 (0.13)  &  2.0 (0.3)  &  0.09 (0.01)  &  CDMS  \\

{\medskip}
\textit{c}-C$_{3}$H & 4$_{1,3}$ -- 3$_{1,2}$  & 216488.2 & 11 & 25   & 23  & 11 & 0.27  & 0.06 (0.09)  & +6.80 (0.06)  & 0.8 (0.1) &  0.05 (0.01) & JPL   \\
& F=9/2-7/2 & & & & & & & \\
& F=5-4 & & & & & & & \\
{\medskip}
\textit{c}-C$_{3}$H $ ^{C1}$ & 4$_{1,3}$ -- 3$_{1,2}$  & 216492.6 & 11 & 25   & 17  & 14 & 0.27  & 0.40 (0.01)  & +6.90 (0.13)  & 1.1 (0.3) &  0.05 (0.01) & JPL   \\
& F=9/2-7/2 & & & & & & & \\
& F=4-3 & & & & & & & \\
{\medskip}
D$_{2}$CO  &  8$_{1,7}$ -- 8$_{1,8}$  & 216492.4  &  11  &  111  &  1   & -  & -  & -  & -  &  -  &  -  &  JPL  \\

\textit{c}-C$_{3}$H & 4$_{1,3}$ -- 3$_{1,2}$  & 216641.1 & 11 & 25   & 14  & 11 & 0.27  & 0.04 (0.05)  & +6.58 (0.12)  & 1.4 (0.3) &  0.06 (0.01) & JPL   \\
& F=7/2-5/2 & & & & & & & \\
& F=3-2 & & & & & & & \\
{\medskip}
o-\textit{c}-C$_{3}$H$_{2}$  &  3$_{3,0}$ -- 2$_{2,1}$   &  216278.8  &  11   &  19   &  46   &  12   &  0.27  &  1.25 (0.05)   &  +6.42 (0.01)   & 0.9 (0.1)  &  1.24 (0.01)  &  CDMS   \\
{\medskip}
o-\textit{c}-C$_{3}$H$_{2}$ &  6$_{1,6}$ -- 5$_{0,5}$   &  217822.1  &  11  &  39    &  175   &  13   &  0.27  &  1.75 (0.01)   &  +6.41 (0.01)   &   1.1 (0.1)  &  2.01 (0.01)  &  CDMS   \\ 
{\medskip}
o-\textit{c}-C$_{3}$H$_{2}$  &  5$_{1,4}$ -- 4$_{2,3}$   &  217940.0  &  11  &  35   &  110    &  12   &  0.27  &  1.05 (0.06)   &  +6.41 (0.02)   &  1.1 (0.1)  &  1.16 (0.01)   &  CDMS  \\ 
{\medskip}
p-\textit{c}-C$_{3}$H$_{2}$    &  5$_{2,4}$ -- 4$_{1,3}$   &  218160.5  &  11  &  35   &  37   &  10   &  0.27  &  0.39 (0.02)   &  +6.42 (0.01)   &  1.1 (0.1)   &   0.46 (0.01)  &  CDMS   \\
{\medskip}
o-\textit{c}-C$_{3}$H$_{2}$ $ ^{C2}$ & 8$_{3,6}$ -- 8$_{2,7}$  & 218449.4 & 11 & 87  & 63    & 34 & 0.27  & 0.05 (0.01)  & +6.63 (0.50)  & 2.5 (1.0) &  0.12 (0.04) & CDMS   \\ 
{\medskip}
 p-\textit{c}-C$_{3}$H$_{2}$ &  8$_{2,6}$ -- 8$_{1,7}$  & 218448.8  &  11 & 87  & 21  & - & -  & -  & -  &  -  &  -  &  CDMS  \\
{\medskip}
o-\textit{c}-C$_{3}$H$_{2}$ $ ^{ng}$ &   7$_{1,6}$ -- 7$_{0,7}$   &  218732.7   &   11  &  61   &  33   &  7   &  0.27  &  0.10 (0.01)   &  +6.45 (0.03)  &   1.4 (30.1)   &   0.15 (0.01)   &   CDMS   \\
 {\medskip}
o-H$_{2}$CCO& 11$_{1,11}$ -- 10$_{1,10}$ & 220177.6 & 11 & 76  &  66  & 7 & 0.27  & 0.03 (0.01)  & +6.40 (0.16)  & 2.2 (0.7) &  0.04 (0.06) & CDMS   \\
 {\medskip}
p-H$_{2}$CCO $^{ng}$ & 11$_{0,11}$ -- 10$_{0,10}$ & 222197.6 & 11 & 64  &  22  & 3 & 0.53  & 0.02 (0.01)  & +6.15 (0.14)  & 2.5 (0.3) &  0.05 (0.01) & CDMS   \\
{\medskip}
 %%%%%%%%left here %%%%
 CH$_{3}$OH & 6$_{1,6}$ -- 7$_{2,6}$A & 215302.2 & 11 & 374  &  19  & 8 & 0.27  & 0.04 (0.01)  & +8.98 (0.05)  & 0.6 (0.2) &  0.03 (0.01) & CDMS   \\
&  v$_{t}$=1 & & & & & & & \\
{\medskip}
CH$_{3}$OH $ ^{ng}$ & 5$_{-1,4}$ -- 4$_{-2,3}$E & 216945.5 & 11 & 56  &  4  & 9  & 0.54 & 0.05 (0.01)  & +8.80 (0.16)  & 1.9 (0.5) &  0.08 (0.03) & CDMS   \\
{\medskip}
CH$_{3}$OH $^{e}$      &4$_{-2,3}$ -- 3$_{-1,2}$E &218440.1 & 11 & 45 & 14 & 9& 0.27 & 0.18 (0.01) &+6.83 (0.06) &3.3 (0.2)& 0.35 (0.01) & CDMS \\
{\medskip}
CH$_{3}$OH & 8$_{-1,8}$ -- 7$_{-1,6}$E & 220078.6 & 11 & 97  &  14  & 6  & 0.53 & 0.06 (0.01)  & +8.68 (0.09)  & 2.7 (0.5) &  0.11 (0.01) & CDMS   \\

\end{tabular}

\begin{tablenotes}
\item[a] Frequencies and spectroscopic parameters have been extracted from the Jet Propulsion Laboratory (JPL) molecular database \citep{Pickett1998} and the Cologne Database for Molecular Spectroscopy \citep{Muller2005}. 
{\smallskip}
\item[b] Gaussian fit. 
{\smallskip}
\item[ng] Non-Gaussian profile. The line intensity is obtained by integrating the Zero Power Full Width ($ZPFW$) range.
{\smallskip}
\item[h7] The detected 3 -- 2 line consists of 40 hyperfine components with $S\mu^2$ $\leq$ 312 D$^2$ in a 3 MHz frequency interval. The values are taken from L. Dore in private communication. The line with the highest $S\mu^2$ line is reported (see e.g. Figures \ref{L1551_Spectra}).
{\smallskip}
\item[h8] The detected 7$_{1,7}$ -- 6$_{1,6}$ line consists of 3 hyperfine components with $S\mu^2$ $\leq$ 67 D$^2$ \citep{Muller2005} in a 2 MHz frequency interval. The line with the highest $S\mu^2$ line is reported (see e.g. Figures \ref{L1551_Spectra}).
{\smallskip}
\item[C1] Line blended at the present spectral resolution (0.2 MHz). Possible contamination due to D$_{2}$CO emission at 216492.4 MHz.
{\smallskip}
\item[C2] Line blended at the present spectral resolution (0.2 MHz). Possible contamination due to p-\textit{c}-C$_{3}$H$_{2}$ emission at 218448.8 MHz. 
{\smallskip}

\end{tablenotes} 

    \end{threeparttable} }
\end{table*}
\addtocounter{table}{-1}
\begin{table*}
\centering
\caption{Continued}
%{
\resizebox{19cm}{!}{%
\begin{threeparttable}
 
\begin{tabular}{llccccccccccc}
\hline \hline \noalign {\medskip}

Species & Transition $^{a}$& $\nu$ $^{a}$ & $HPBW$ & $E_{\rm up}$ $^{a}$ & $S\mu^2$ $^{a}$ & rms $^{b}$ & ${\delta} v$ $ ^{b}$ & $T_{peak}$ $^{b}$ & $V_{\rm peak}$ $^{b}$ & $FWHM$ $^{b}$ & $I_{int}$ $^{b}$ & Database \\ 
& & (MHz) &$(\arcsec$)&(K) & (D$^{2}$) & (mK) &(km s$^{-1}$)& (K) & (km s$^{-1}$) & (km s$^{-1}$) &(K km s$^{-1}$) \\
\hline \noalign {\medskip}
{\medskip}
CH$_{3}$OH $ ^{C3}$ & 10$_{2,9}$-9$_{3,6}$A  &  231281.1  &  11  &  165  & 11  &  8  & 0.53  & 0.03(0.01)  & +9.36(0.21)  & 4.0(0.7) &  0.12 (0.02) & CDMS   \\
{\medskip}
 CH$_{3}$CHO   & 12$_{6,6}$ -- 11$_{6,5}$E & 231278.5 & 11 & 153  &  113  & -  & - & -  & -  & - &  - & JPL   \\
 {\medskip}
CH$_{3}$OH  & 10$_{2,8}$ -- 9$_{3,7}$A  &  232418.5  &  11  &  165  & 11  &  10  & 0.25  & 0.05 (0.01)  & +8.81 (0.09)  & 1.5 (0.2) &  0.08 (0.01) & CDMS   \\
{\medskip}
CH$_{3}$OH $ ^{ng}$ & 10$_{3,7}$-11$_{2,9}$E & 232945.8 & 11 & 190  &  12 & 8  & 0.25  & 0.04 (0.01)  & +8.93 (0.13)  & 2.2 (0.4) &  0.06 (0.03) & CDMS   \\
{\medskip}
 CH$_{3}$OH & 4$_{2,3}$ -- 5$_{1,4}$A & 234683.4 & 10 & 61  &  4  & 12  & 0.25  & 0.05 (0.01)  & +8.78 (0.08)  & 0.8 (0.2) &  0.04 (0.01) & CDMS   \\
 {\medskip}
CH$_{3}$OH & 5$_{4,2}$ -- 6$_{3,3}$E  &  234698.5  &  10  &  123  & 2  &  9  & 0.25  & 0.03 (0.01)  & +9.61 (0.13)  & 1.4 (0.4) &  0.04 (0.01) & CDMS   \\
 {\medskip}
CH$_{2}$DOH & 3$_{1,3}$ -- 2$_{0,2}$, e$_{0}$  & 214701.7  & 11 & 17  & 2  & 10  & 0.54  & 0.03 (0.01)  & +9.43 (0.16)  & 1.7 (0.4) &  0.04 (0.01) & JPL   \\
{\medskip}
CH$_{3}$CN &  12$_{4}$ -- 11$_{4}$E & 220679.3  &  11 &  183 & 226 & 8 & 0.26  & 0.03 (0.01)  & +8.07 (0.14)  & 4.3 (0.6) &  0.03 (0.01) & JPL   \\
{\medskip}
CH$_{3}$CN &  12$_{3}$ -- 11$_{3}$A & 220709.3  &  11 &  133 & 477 & 8 & 0.26  & 0.05 (0.01)  & +7.89 (0.20)  & 4.5 (0.4) &  0.06 (0.01) & JPL   \\
{\medskip}
CH$_{3}$CN &  12$_{2}$ -- 11$_{2}$E & 220730.3  &  11 &  97 & 247 & 8 & 0.26  & 0.04 (0.01)  & +7.89 (0.23)  & 3.6 (0.6) &  0.04 (0.01) & JPL   \\
{\medskip}
CH$_{3}$CN &  12$_{1}$ -- 11$_{1}$E & 220743.0  &  11 &  76 & 252 & 11 & 0.53  & 0.03 (0.01)  & +8.55 (0.26)  & 1.3 (0.4) &  0.04 (0.01) & JPL   \\
{\medskip}
CH$_{3}$CN &  12$_{0}$ -- 11$_{0}$A & 220747.3  &  11 &  69 & 254 & 10 & 0.53  & 0.04 (0.01)  & +8.05 (0.12)  & 3.2 (0.7) &  0.07 (0.01) & JPL   \\
 {\medskip}
 CH$_{3}$CCH   $ ^{C4}$  & 13$_{3}$ -- 12$_{3}$  & 222128.8 & 11 & 140  & 9  & 7 & 0.26  & 0.03 (0.01)  & +7.05 (0.24)  & 3.0 (0.9) &  0.05 (0.01) & CDMS   \\ 
 {\medskip}
 HCOOCH$_{3}$ & 8$_{5,4}$ -- 7$_{4,3}$A  &  222128.2  &  11  &  226  & 2  &  -  & -  & -  & -  & - &  - & JPL   \\
{\medskip}
CH$_{3}$CCH  & 13$_{2}$ -- 12$_{2}$  & 222150.0 & 11 & 103  & 5  & 6 & 0.26  & 0.02 (0.01)  & +6.85 (0.12)  & 1.8 (0.4) &  0.05 (0.01) & CDMS   \\ 
{\medskip}
CH$_{3}$CCH  & 13$_{1}$ -- 12$_{1}$  & 222162.7 & 11 & 82  & 5  & 19 & 0.26  & 0.05 (0.01)  & +6.45 (0.17)  & 1.4 (0.5) &  0.07 (0.02) & CDMS   \\ 
{\medskip}
CH$_{3}$CCH  & 13$_{0}$ -- 12$_{0}$  & 222166.9 & 11 & 75  & 5  & 15 & 0.26  & 0.07 (0.01)  & +6.50 (0.10)  & 1.7 (0.2) &  0.11 (0.01) & CDMS   \\ 
{\medskip}
CH$_{3}$CHO $^{ng}$ & 12$_{7,6}$ -- 11$_{7,5}$E & 231268.4 & 11 & 182  &  100  & 15   & 0.51 & 0.03 (0.01)  & +8.15 (0.20)  & 1.4 (0.6) &  0.04 (0.01) & JPL   \\
{\medskip}
CH$_{3}$CHO $^{ng}$ & 3$_{3,1}$ -- 3$_{2,2}$E & 234469.3 & 11 & 26  &  2  & 8  & 0.50 & 0.03(0.01)  & +7.27 (0.16)  & 2.1 (0.3) &  0.06 (0.01) & JPL   \\ 
{\medskip}
CH$_{3}$CHO $^{ng}$ & 12$_{1,11}$ -- 11$_{1,10}$E & 235996.2 & 10 & 76  &  151  & 8   & 0.50 & 0.03 (0.01)  & +8.43 (0.17)  & 3.0 (0.4) &  0.09 (0.01) & JPL   \\
{\medskip}
%% FIRST METHYL FORMATE
%%%% Main
HCOOCH$_{3}$ $ ^{C5}$ & 20$_{1,20}$ -- 19$_{1,19}$A  &  216965.9  &  11  &  111  & 53  &  8  & 0.53   & -  & -  & - &  $\geq$0.24 & JPL   \\
{\medskip}
%%%%%% BLENDED LINES
HCOOCH$_{3}$   & 20$_{0,20}$ -- 19$_{1,19}$E  &  216963.0  &  11  &  111  & 8  &  -  & -  & -  & -  & - &  - & JPL   \\
{\medskip}
HCOOCH$_{3}$   & 20$_{0,20}$ -- 19$_{1,19}$A  &  216964.2  &  11  &  111  & 8 &  -  & -  & -  & -  & - &  - & JPL   \\
{\medskip}
HCOOCH$_{3}$   & 20$_{1,20}$ -- 19$_{1,19}$E  &  216964.8  &  11  &  111  & 53 &  -  & -  & -  & -  & - &  - & JPL   \\
{\medskip}
HCOOCH$_{3}$   & 20$_{0,20}$ -- 19$_{0,19}$E  &  216966.2  &  11  &  111  & 53  &  -  & -  & -  & -  & - &  - & JPL   \\
{\medskip}
HCOOCH$_{3}$   & 20$_{0,20}$ -- 19$_{0,19}$A  &  216967.4  &  11  &  111  & 53  &  -  & -  & -  & -  & - &  - & JPL   \\
{\medskip}
HCOOCH$_{3}$   & 20$_{1,20}$ -- 19$_{0,19}$E  &  216968.0  &  11  &  111  & 8  &  -  & -  & -  & -  & - &  - & JPL   \\
{\medskip}
HCOOCH$_{3}$   & 20$_{1,20}$ -- 19$_{0,19}$A  &  216969.2  &  11  &  8  & 53  &  -  & -  & -  & -  & - &  - & JPL   \\
{\medskip}
%%%%%%%%%%%%%
%%SECOND METHYL FORMATE
 %%%%Main
HCOOCH$_{3}$ $ ^{C6}$ & 18$_{13,5}$ -- 17$_{13,4}$E  &  221139.8  &  11  &  213  & 23  &  6  & 0.53   & -  & -  & - &  $\geq$0.04 & JPL   \\
{\medskip}
%%%%%% BLENDED LINES
HCOOCH$_{3}$   & 18$_{13,5}$ -- 17$_{13,4}$A  &  221141.1  &  11  &  213  & 23   &  -  & -  & -  & -  & - &  - & JPL   \\
{\medskip}
HCOOCH$_{3}$    & 18$_{13,6}$ -- 17$_{13,5}$A  &  221141.1  &  11  &  213  & 23   &  -  & -  & -  & -  & - &  - & JPL   \\
%%%%%%%
%%THIRD METHYL FORMATE
%%%%Main
{\medskip}
HCOOCH$_{3}$ $ ^{C7}$  & 18$_{4,15}$ -- 17$_{4,14}$E  &  221660.5  &  11  &  112  & 45  &  7  & 0.53  & -  & -  & - &  $\geq$0.11 & JPL   \\
%%%%%% BLENDED LINES
{\medskip}
HCOOCH$_{3}$   & 18$_{10,9}$ -- 17$_{10,8}$A  &  221661.0  &  11  &  167  & 33  &  -  & -  & -  & -  & - &  - & JPL   \\
{\medskip}
HCOOCH$_{3}$    & 18$_{10,8}$ -- 17$_{10,8}$A  &  221661.0  &  11  &  167  & 33  &  -  & -  & -  & -  & - &  - & JPL   \\
%%%%%%%
%\hline \noalign {\larskip}
%\hline \noalign {\medskip}
\end{tabular}
    
\begin{tablenotes}
\item[a] Frequencies and spectroscopic parameters have been extracted from the Jet Propulsion Laboratory (JPL) molecular database \citep{Pickett1998} and the Cologne Database for Molecular Spectroscopy \citep{Muller2005}. 
{\smallskip}
\item[b] Gaussian fit. 
{\smallskip}
\item[ng] Non-Gaussian profile. The line intensity is obtained by integrating the Zero Power Full Width ($ZPFW$) range.
{\smallskip}
\item[C3] Line blended at the present spectral resolution (0.2 MHz). Possible contamination due to CH$_{3}$OH emission at 231278.5 MHz. 
{\smallskip}
\item[C4] Line blended at the present spectral resolution (0.2 MHz). Possible contamination due to HCOOCH$_{3}$ emission at 222128.2 MHz.
{\smallskip}
\item[C5] Line blended at the present spectral resolution (0.2 MHz). Possible contamination due to HCOOCH$_{3}$ emissions at 216963.0--216969.2 MHz.
{\smallskip}
\item[C6] Line blended at the present spectral resolution (0.2 MHz). Possible contamination due to HCOOCH$_{3}$ emission at 221141.1 MHz.
{\smallskip}
\item[C7] Line blended at the present spectral resolution (0.2 MHz). Possible contamination due to HCOOCH$_{3}$ emission at 221661.0 MHz.

\end{tablenotes} 

    \end{threeparttable} }
\end{table*}

\section{Rotational diagrams and derived column densities}

\begin{figure*}
\centering
\includegraphics[scale=0.45]{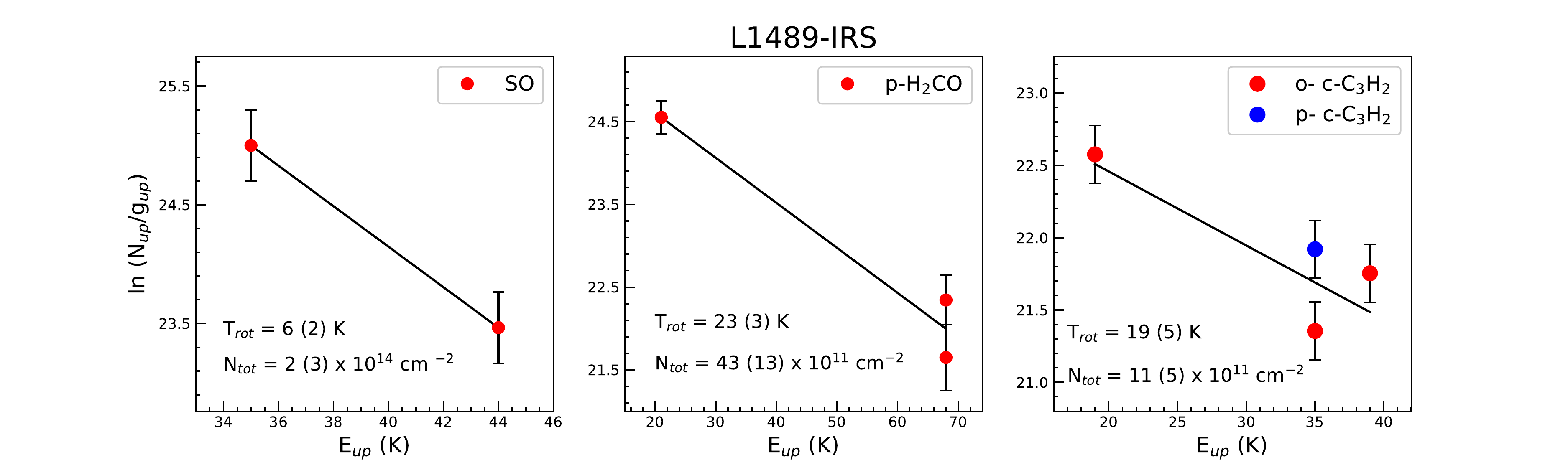}
\caption{Rotational diagrams of SO, H$_{2}$CO, c-C$_{3}$H$_{2}$ in L1489-IRS. The parameters $N_{\rm up}$, $g_{\rm up}$, and $E_{\rm up}$ are the column density, the degeneracy and the energy (with respect to the ground state of each symmetry) of the upper level, respectively. Error bars include the errors on the flux as stated in Table \ref{L1489-list} and the uncertainty of the calibration of \%20.}
\label{RD-L1489}
\end{figure*}

We report in Figures \ref{RD-L1489}--\ref{RD-L1551} (one per source) the rotational diagrams obtained from the whole emitting velocity range as described in Sect. \ref{results}. When applicable, ortho-- and para-- species as well as A-- and E-- types
have been analyzed separately. Tables \ref{L1489-CD}--\ref{L1551-CD} report the beam-averaged column densities derived from rotational diagrams. In addition they include column densities towards envelopes, outflows and hot corinos. Source averaged column densities can be derived by applying the filling factors reported in Table \ref{LVGList} for c-C$_{3}$H$_{2}$ as an envelope tracer and for CH$_{3}$OH and CH$_{3}$CN as hot corino tracers. For single lines, column densities are derived using temperature mentioned in Sect. \ref{LineProfiles}.

%\input{RD_ALL_Source}

%%%%%%%%%%%%%%%%%%%%%%%%%%%%%%%%%%% 
%%% ROTATIONAL DIAGRAMS IN L1489%%%
%%%%%%%%%%%%%%%%%%%%%%%%%%%%%%%%%%%
\clearpage

%%%%%%%%%%%%%%%%%%%%%%%%%%%%%%%%%%% 
%%% ROTATIONAL DIAGRAMS IN B5 %%%
%%%%%%%%%%%%%%%%%%%%%%%%%%%%%%%%%%%

\begin{figure*}
\centering
\includegraphics[scale=0.45]{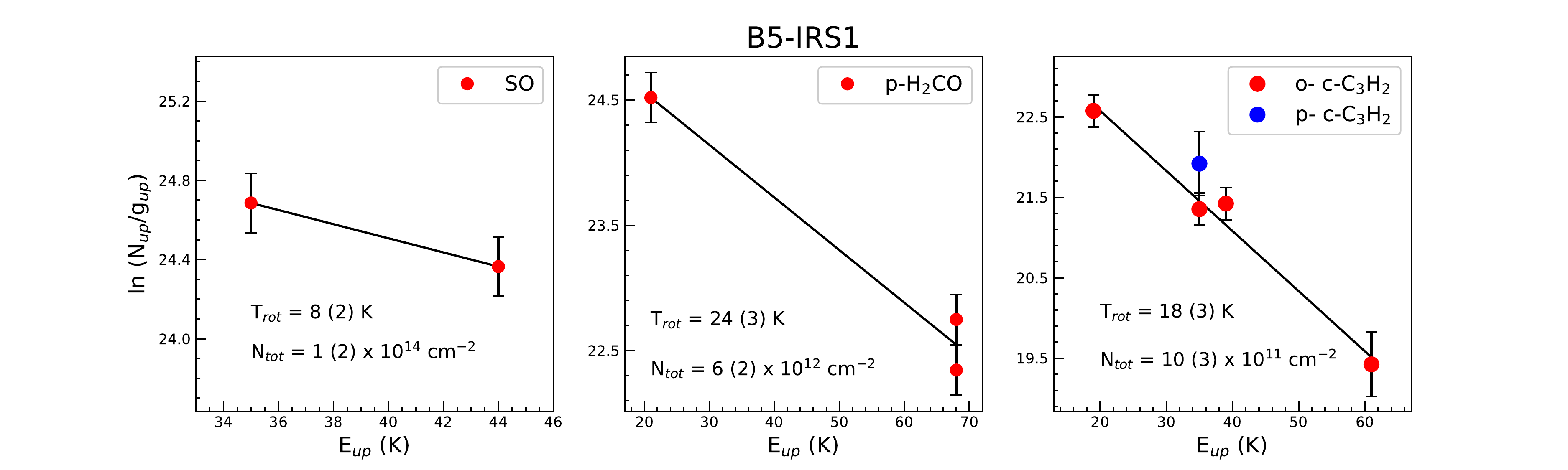}
\caption{Rotational diagrams of SO, H$_{2}$CO, and c-C$_{3}$H$_{2}$ in B5-IRS1. The parameters $N_{\rm up}$, $g_{\rm up}$, and $E_{\rm up}$ are the column density, the degeneracy and the energy (with respect to the ground state of each symmetry) of the upper level, respectively. Error bars include the errors on the flux as stated in Table \ref{B5-list} and the uncertainty of the calibration of \%20.}
%\label{RD-B5}
\end{figure*}

%%%%%%%%%%%%%%%%%%%%%%%%%%%%%%%%%%% 
%%% ROTATIONAL DIAGRAMS IN L1455 %%%
%%%%%%%%%%%%%%%%%%%%%%%%%%%%%%%%%%%
\begin{figure*}
\centering
\includegraphics[scale=0.5]{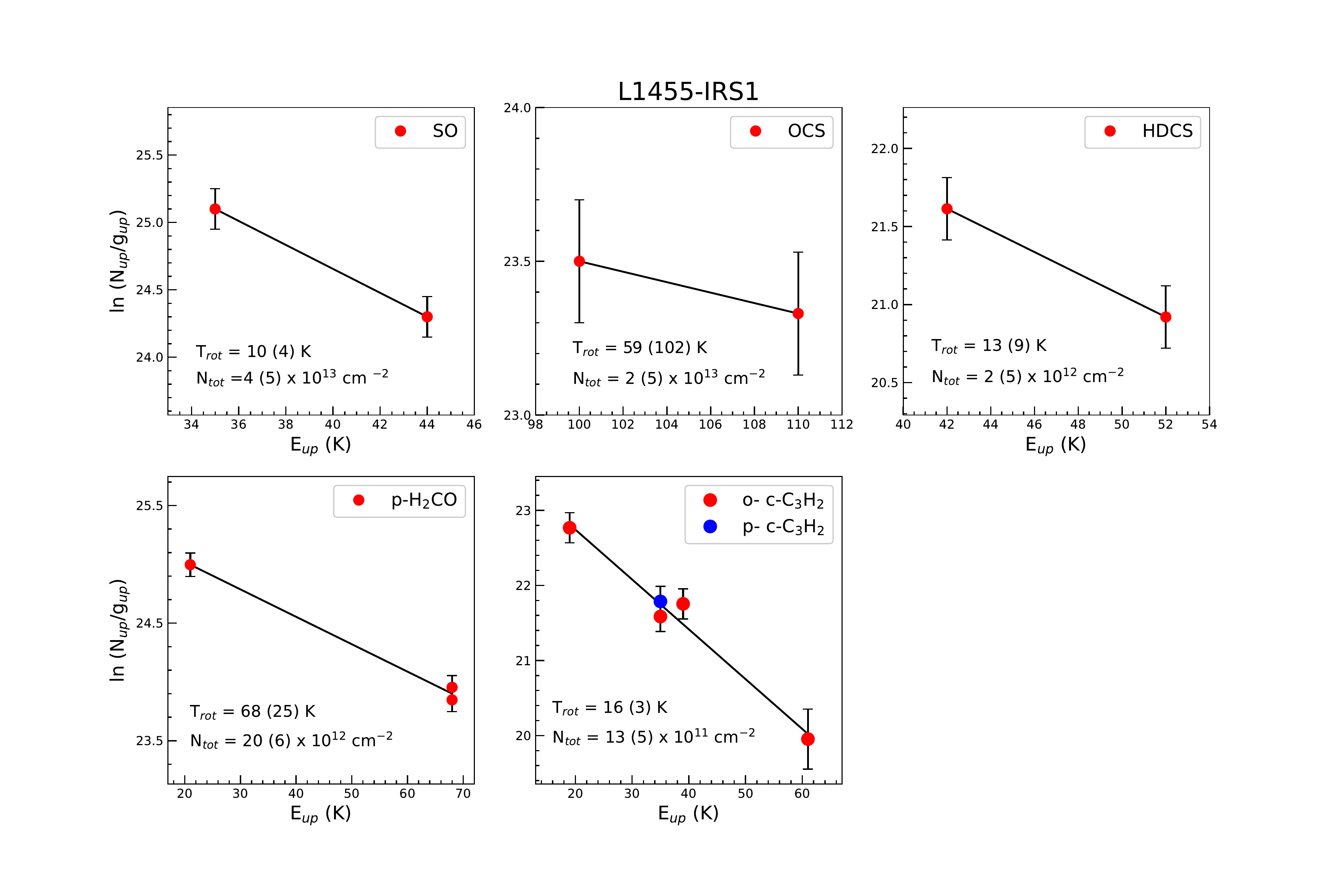}
%\hspace{-1cm}
\caption{Rotational diagrams of SO, OCS, HDCS, H$_{2}$CO and c-C$_{3}$H$_{2}$ in L1455-IRS1. The parameters $N_{\rm up}$, $g_{\rm up}$, and $E_{\rm up}$ are the column density, the degeneracy and the energy (with respect to the ground state of each symmetry) of the upper level, respectively. Error bars include the errors on the flux as stated in Table \ref{L1455-list} and the uncertainty of the calibration of \%20.
}
\label{RD-L1455}

\end{figure*}

%%%%%%%%%%%%%%%%%%%%%%%%%%%%%%%%%%% 
%%% ROTATIONAL DIAGRAMS IN L1551 %%
%%%%%%%%%%%%%%%%%%%%%%%%%%%%%%%%%%%
\begin{figure*}
\centering
\includegraphics[scale=0.5]{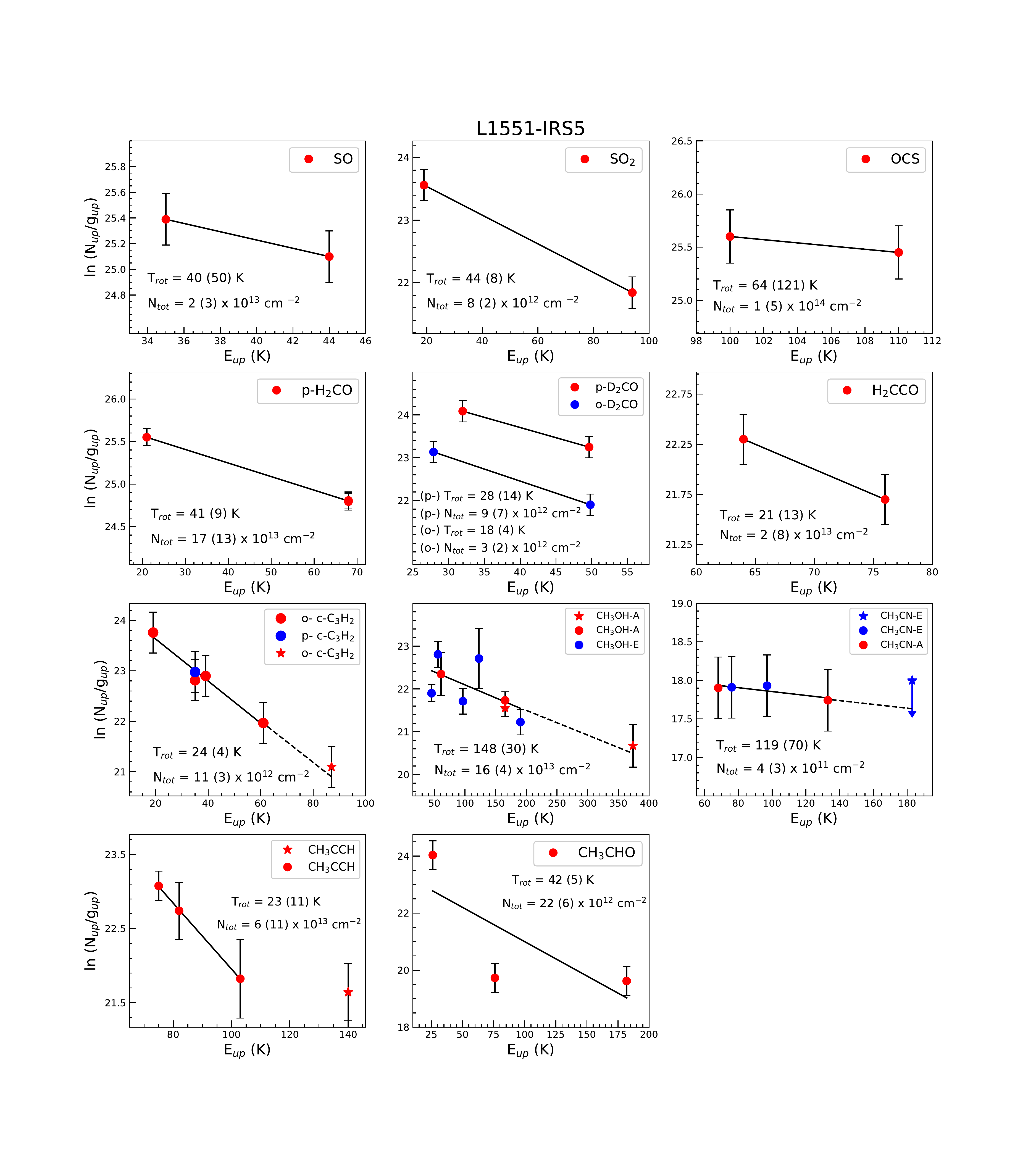}

\caption{Rotational diagrams of SO, SO$_{2}$, OCS, H$_{2}$CO, D$_{2}$CO, H$_{2}$CCO, c-C$_{3}$H$_{2}$, CH$_{3}$OH, CH$_{3}$CN, CH$_{3}$CCH, and CH$_{3}$CHO. The parameters $N_{\rm up}$, $g_{\rm up}$, and $E_{\rm up}$ are the column density, the degeneracy and the energy (with respect to the ground state of each symmetry) of the upper level, respectively. The derived values of the rotational temperature and the column density are reported in each panel for each species. For D$_{2}$CO, blue and red circles indicate ortho- and para- transitions, respectively. For CH$_{3}$OH, blue and red circles indicate E and A transitions, respectively. Red stars in the c-C$_{3}$H$_{2}$ (8$_{3,6}$ -- 8$_{2,7}$) and CH$_{3}$OH (10$_{2,9}$ -- 9$_{3,6}$) indicate transitions not used in the fit due to contamination by other molecules. The blue arrow in the CH$_{3}$CN panel indicates 12$_{4}$ -- 11$_{4}$ transition, which was not used in the fit as upper limit. Error bars include the errors on the flux as stated in Table \ref{L1551-list} and the uncertainty of the calibration of 20\%. }
\label{RD-L1551}
\end{figure*}
%%%%%%%%%%%
%% L1489 %%
%%%%%%%%%%%
\clearpage

\begin{table}[ht]
  \caption{L1489-IRS: Temperatures and beam averaged column densities derived for the envelope (see text).}
\centering
\setlength{\tabcolsep}{15pt}
\label{L1489-CD}    
 { \begin{tabular}{lcc}
 \hline
Species    & $T_{rot}$     &  $N_{tot}$ $^{c}$  \\
 & (K) & (cm$^{-2}$) \\
 \hline
 \hline
\multicolumn{3}{c}{Envelope}\\ 
\hline
SO & 20 (13)$^{a}$&  $2 (3)\times10^{12}$\\
p-H$_{2}$CO & 23 (4)$^{a}$&  $19 (6)\times10^{11}$ \\
c-C$_{3}$H$_{2}$ & 19 (5)$^{a}$&  $5 (2)\times10^{11}$ \\
$^{13}$CO & $20-35^{b}$ & $2-3\times10^{15}$\\
C$^{18}$O & $20-35^{b}$ & $3-4\times10^{14}$ \\
C$^{15}$N & $20-35^{b}$ & $23-27\times10^{10}$ \\
CCD & $20-35^{b}$ & $17-18\times10^{11}$ \\
DCN$^{c}$ & $20-35^{b}$ & $90-98\times10^{9}$ \\
DCO$^{+}$ & $20-35^{b}$ & $3-4\times10^{10}$\\
o-D$_{2}$CO & $20-35^{b}$ & $6-7\times10^{10}$\\

\hline
\end{tabular}}

\tablefoot{
($^a$) Temperatures and column densities as derived from rotational diagrams (see Table \ref{RD}).

($^b$)Temperature range assumed from the c-C$_{3}$H$_{2}$ analysis of the envelope emission (see text). 
($^c$) Column density is corrected for the opacity as derived by fitting hyperfine line pattern (see Table \ref{Hyp}).
($^c$) Source averaged column densities can be derived by applying the filling factor reported in Table \ref{LVGList} for c-C$_{3}$H$_{2}$ as an envelope tracer.

}
   
\end{table}

%%%%%%%%%%%
%% B5-IRS1 %%
%%%%%%%%%%%
\begin{table}[ht]
  \caption{B5-IRS1: Temperatures and beam averaged column densities for different physical components (envelope, outflow, see text).}
\centering
\setlength{\tabcolsep}{15pt}
 \label{B5-CD}   
 { \begin{tabular}{lcc}
\hline
Species    & $T_{rot}$     &  $N_{tot}$ $^{f}$  \\
 & (K) & (cm$^{-2}$) \\
 \hline
 \hline
\multicolumn{3}{c}{Envelope}\\ 
\hline
SO & 22 (15)$^{a}$ & $8(10)\times10^{12}$\\
p-H$_{2}$CO & 19 (4)$^{a}$ & $7(2)\times10^{12}$ \\
c-C$_{3}$H$_{2}$ & 18 (3)$^{a}$ & $10(4)\times10^{11}$ \\
$^{13}$CO & $20-35^{b}$ & $1-2\times10^{15}$\\
C$^{18}$O & $20-35^{b}$ & $6-7\times10^{14}$ \\
$^{13}$CN & $20-35^{b}$ & $36-43\times10^{10}$ \\
$^{13}$CS$^{c}$ & $20-35^{b}$ & $13-16\times10^{11}$ \\
$^{34}$SO & $20-35^{b}$ & $9-10\times10^{11}$ \\
CCD & $20-35^{b}$ & $29-32\times10^{12}$ \\
DCN & $20-35^{b}$ & $83-90\times10^{9}$ \\
N$_{2}$D$^{+}$ & $20-35^{b}$ & $48-51\times10^{9}$ \\
DCO$^{+}$ & $20-35^{b}$ & $1-2\times10^{11}$\\
SO$_{2}$ &$20-35^{b}$ & $4-6\times10^{12}$ \\
o-H$_{2}^{13}$CO & $20-35^{b}$ & $1-10\times10^{10}$ \\
o-D$_{2}$CO & $20-35^{b}$ & $10-12\times10^{11}$ \\
p-D$_{2}$CO & $20-35^{b}$ &$5-15\times10^{11}$ \\
o-H$_{2}$CS & $20-35^{b}$& $1-20\times10^{12}$ \\
CH$_{3}$OH$^{d}$ &$20-35^{b}$  & $1-2\times10^{13}$ \\
CH$_{2}$DOH &$20-35^{b}$  & $4-7\times10^{12}$ \\
\hline
\multicolumn{3}{c}{Outflow}\\ 
\hline
SO & 20 (28)$^{a}$ & $1(2)\times10^{12}$  \\
p-H$_{2}$CO & 23 (12)$^{a}$  & $147(99)\times10^{10}$ \\
$^{13}$CO & $50-70^{e}$ & $1-2\times10^{15}$ \\
C$^{18}$O & $50-70^{e}$ & $8-10\times10^{13}$\\
SO$_{2}$ &$50-70^{e}$ & $2-3\times10^{12}$  \\
\hline
\end{tabular}}

\tablefoot{
($^a$) Temperatures and column densities as derived from rotational diagrams (see Table \ref{RD}).
($^b$) Temperature range assumed from the c-C$_{3}$H$_{2}$ analysis of the envelope emission (see text). 
($^{c}$) Column density is corrected for the opacity as derived by fitting hyperfine line pattern (see Table \ref{Hyp}).
($^d$) Column density refer to the sum of A and E species.
($^e$) Assumed according to the p-H$_{2}$CO analysis of the outflow component (see text). 
($^f$) Source averaged column densities can be derived by applying the filling factor reported in Table \ref{LVGList} for c-C$_{3}$H$_{2}$ as an envelope tracer.
}
\end{table}

%%%%%%%%%%%
%% L1455-IRS1 %%
%%%%%%%%%%%
\begin{table}[ht]
  \caption{L1455-IRS1: Temperatures and beam averaged column densities for different physical components (envelope, outflow, (see text).}
\setlength{\tabcolsep}{15pt}
 \label{L1455-CD}   
  
 { \begin{tabular}{lcc}
\hline
Species    & $T_{rot}$     &  $N_{tot}$ $^h$  \\
 & (K) & (cm$^{-2}$) \\
 \hline
 \hline
\multicolumn{3}{c}{Envelope}\\ 
\hline
SO & 18 (10)$^{a}$ &$7(9)\times10^{12}$ \\
p-H$_{2}$CO & 33 (6)$^{a}$ & $10(3)\times10^{12}$  \\
HDCS  & 13 (9)$^{a}$ & $2(5)\times10^{12}$ \\
c-C$_{3}$H$_{2}$ & 16 (3)$^{a}$ & $13(5)\times10^{11}$\\
$^{13}$CN$^{e}$ & $20-35^{b}$ & $40-49\times10^{11}$\\
$^{13}$CS & $20-35^{b}$ & $28-33\times10^{10}$ \\
SiO &$20-35^{b}$ & $1-2\times10^{11}$  \\
CCD & $20-35^{b}$ & $18-19\times10^{11}$ \\
DCN & $20-35^{b}$ & $10-11\times10^{10}$ \\
N$_{2}$D$^{+}$ & $20-35^{b}$ & $11-12\times10^{10}$ \\
DCO$^{+}$ & $20-35^{b}$ &$1-2\times10^{11}$ \\
SO$_{2}$ &$20-35^{b}$ & $1-2\times10^{12}$ \\
CCS & $20-35^{b}$ & $9-40\times10^{11}$  \\
o-H$_{2}^{13}$CO & $20-35^{b}$ & $3-4\times10^{11}$ \\
o-D$_{2}$CO & $20-35^{b}$ &$1-2\times10^{12}$ \\
p-D$_{2}$CO & $20-35^{b}$ & $1-2\times10^{12}$ \\
o-H$_{2}$CS & $20-35^{b}$& $5-8\times10^{12}$ \\
HNCO & $20-35^{b}$ & $1-2\times10^{12}$ \\
CH$_{3}$OH$^{d}$ &$20-35^{b}$  & $7-8\times10^{13}$\\
CH$_{2}$DOH &$20-35^{b}$  &$3-5\times10^{12}$
\\
\hline
\multicolumn{3}{c}{Outflow}\\ 
\hline
p-H$_{2}$CO &66 (34)$^{a}$& $12(4)\times10^{12}$ \\
SO &$50-70^{c}$   & $2-6\times10^{12}$ \\
$^{13}$CO & $50-70^{c}$ &$3-4\times10^{15}$ \\
C$^{18}$O & $50-70^{c}$ & $2-3\times10^{14}$\\
$^{13}$CS & $50-70^{c}$ & $13-15\times10^{10}$ \\
DCN & $50-70^{c}$ & $80-99\times10^{9}$ \\
SO$_{2}$ &$50-70^{c}$ &$2-3\times10^{12}$  \\
H$_{2}$S & $50-70^{c}$ & $11-12\times10^{9}$ \\
\hline
\multicolumn{3}{c}{Hot corino${^f}$}\\
\hline
OCS & $59(102)$ & $2(5)\times10^{13}$ \\
H$_{2}$S & $60-150{^g}$ & $1-2\times10^{13}$ \\
\hline
\end{tabular}}

\tablefoot{
($^a$) Temperatures and column densities as derived from rotational diagrams (see Table \ref{RD}).
($^b$) Temperature range assumed from the c-C$_{3}$H$_{2}$ analysis of the envelope emission (see text). 
($^c$) Assumed according to the p-H$_{2}$CO analysis of the outflow component (see text). 
($^d$) Column density refer to the sum of A and E species.
($^{e}$) Column density is corrected for the opacity as derived by fitting hyperfine line pattern (see Table \ref{Hyp}).
($^{f}$) Assumed as hot corino discussed in Section \ref{hot corino}
($^{g}$) Temperature range assumed from the CH$_{3}$OH and OCS rotational diagrams in L1551-IRS5.
($^h$) Source averaged column densities can be derived by applying the filling factor reported in Table \ref{LVGList} for c-C$_{3}$H$_{2}$ as an envelope tracer.}

\end{table}

%%%%%%%%%%%
%% L1551 %%
%%%%%%%%%%%
       
\begin{table}[ht]
  \caption{L1551-IRS5: temperatures and beam averaged column densities for different physical components (envelope, outflow, and hot corino, see text).}
\centering
\label{L1551-CD}
\setlength{\tabcolsep}{15pt}
 { \begin{tabular}{lcc}
\hline
Species    & $T_{rot}$     &  $N_{tot}$ $^{h}$ \\
 & (K) & (cm$^{-2}$) \\
 \hline
 \hline
\multicolumn{3}{c}{Envelope}\\ 
\hline
SO & 35 (40)$^{a}$&  $1 (2)\times10^{13}$\\
SO$_{2}$ & 44 (8)$^{a}$& $8 (2)\times10^{12}$ \\
p-H$_{2}$CO & 30 (5)$^{a}$&  $22 (7)\times10^{12}$ \\
o-D$_{2}$CO &  18 (4)$^{a}$ & $3 (2)\times10^{12}$\\
p-D$_{2}$CO &  28 (14)$^{a}$ & $9 (7)\times10^{12}$\\
c-C$_{3}$H$_{2}$ & 24 (4)$^{a}$& $11 (3)\times10^{12}$ \\
H$_{2}$CCO & 21 (13)$^{a}$ & $2 (8)\times10^{13}$\\
CH$_{3}$CCH &  23 (11) & $6(13)\times10^{13}$\\

C$^{18}$O & $20-35$$^{b}$ & $3-4\times10^{15}$\\
$^{13}$CN$^{f}$ &$20-35$$^{b}$ & $45-54\times10^{11}$ \\
C$^{15}$N$^{f}$ &$20-35$$^{b}$ &  $59-71\times10^{10}$ \\
$^{13}$CS &$20-35$$^{b}$ & $13-16\times10^{10}$ \\
$^{34}$SO & $20-35$$^{b}$ & $1-2\times10^{12}$ \\
CCD &$20-35$$^{b}$ & $10-11\times10^{12}$ \\
DCN & $20-35$$^{b}$ & $23-25\times10^{10}$  \\
N$_{2}$D$^{+}$$^{f}$  & $20-35$$^{b}$ & $9-10\times10^{10}$  \\
DCO$^{+}$ & $20-35$$^{b}$ & $3-4\times10^{11}$\\
CCS &$20-35$$^{b}$ & $1-5\times10^{12}$ \\
o-H$_{2}$CS$^{g}$ & $20-35$$^{b}$ & $2-3\times10^{14}$ \\
o-H$_{2}$C$^{33}$S & $20-35$$^{b}$ & $6-7\times10^{11}$ \\
HDCS& $20-35$$^{b}$ & $8-10\times10^{11}$ \\
HNCO & $20-35$$^{b}$ & $5-8\times10^{11}$ \\
c-C$_{3}$H &  $20-35$$^{b}$ & $14-17\times10^{11}$ \\
CH$_{3}$OH$^{e}$ &$20-35^{b}$  & $5-7\times10^{13}$\\

\hline

\multicolumn{3}{c}{Outflow}\\ 
\hline
SO &33 (37)$^{a}$ & $11(11)\times10^{12}$ \\
p-H$_{2}$CO &58 (21)$^{a}$  &$5(2)\times10^{12}$ \\
$^{13}$CO & $50-70^{c}$  & $3-4\times10^{15}$\\
C$^{18}$O & $50-70$$^{c}$ & $9-10\times10^{14}$\\
$^{13}$CS &$50-70^{c}$ & $9-10\times10^{10}$ \\
DCN & $50-70^{c}$  & $16-20\times10^{10}$  \\
DCO$^{+}$ & $50-70^{c}$  & $8-30\times10^{9}$\\
o-H$_{2}$CS & $50-70^{c}$ & $1-2\times10^{12}$ \\
\hline
\multicolumn{3}{c}{Hot corino}\\
\hline
OCS $^{g}$& 64 (121)$^{a}$ & $2(5)\times10^{15}$\\
CH$_{3}$OH$^{e}$ &148 (30)$^{a}$ & $16(4)\times10^{13}$ \\
CH$_{3}$CN   &133 (81)$^{a}$ & $1(5)\times10^{12}$ \\
CH$_{3}$CHO$^{e}$  &42 (5)$^{a}$ & $22(6)\times10^{12}$ \\
O$^{13}$CS & $60-150{^d}$ & $3-4\times10^{12}$\\
H$_{2}$S & $60-150{^d}$ & $4-5\times10^{13}$\\
CH$_{2}$DOH & $60-150{^d}$ & $1-3\times10^{13}$ \\
HCOOCH$_{3}$ & $60-150{^d}$ &  $3-5\times10^{13}$\\

\hline
\end{tabular}

     \tablefoot{
($^a$) Temperatures and column densities as derived from rotational diagrams (see Table \ref{RD}).
($^b$) Temperature range assumed from the D$_{2}$CO and c-C$_{3}$H$_{2}$ analysis of the envelope emission (see text). 
($^c$) Assumed according to the p-H$_{2}$CO analysis of the outflow component (see text). 
($^d$) Temperature range assumed from the CH$_{3}$OH and OCS rotational diagrams
($^e$) Column density refer to the sum of A and E species.
($^f$) Column density is corrected for the opacity as derived by fitting hyperfine line pattern (see Table \ref{Hyp}). 
($^g$) Column density is corrected for the opacity as mentioned in subsection \ref{iso}.
($^h$) Source averaged column densities can be derived by applying the filling factor reported in Table \ref{LVGList} for c-C$_{3}$H$_{2}$ as an envelope tracer and for CH$_{3}$OH and CH$_{3}$CN as hot corino tracers.}
} 
\end{table}

\end{document}